\documentclass{aa}

\usepackage{euclid}
\usepackage{graphicx}
\usepackage{natbib}
\usepackage{scalerel}
\usepackage{fontawesome5}
\usepackage{txfonts}
\usepackage[pdfencoding=auto,psdextra]{hyperref}
\hypersetup{
    colorlinks=true,
    linkcolor=blue,
    filecolor=magenta,      
    urlcolor=blue,
    citecolor=blue
}
\makeatletter
\renewcommand*\aa@pageof{, page \thepage{} of \pageref*{LastPage}}
\makeatother

\usepackage[utf8]{inputenc}

\usepackage[switch, modulo]{lineno}

\usepackage{amsmath}	
\usepackage{dsfont}
\usepackage{physics}
\usepackage{newtxtext,newtxmath}
\usepackage[export]{adjustbox}
\usepackage{tikz}
\usetikzlibrary{angles, quotes}
\usetikzlibrary{shapes.arrows,calc}
\usepackage{multirow}

\usepackage[margin=10pt,font=small,labelfont=bf]{caption} 

\makeatletter
\newcommand{\github}[1]{%
   \href{#1}{\faGithubSquare}%
}
\makeatother

\newcommand{\figref}[1]{Fig. \ref{#1}} 
\newcommand{\secref}[1]{Sect. \ref{#1}}
\newcommand{\appref}[1]{Appendix \ref{#1}}
\newcommand{\Eq}[1]{Eq. (\ref{#1})}

\renewcommand{\ee}{\mathrm{e}}
\newcommand{\btheta}{\boldsymbol{\theta}}
\newcommand{\bvartheta}{\boldsymbol{\vartheta}}
\newcommand{\ii}{\mathrm{i}}

\newcommand{\Omegam}{\Omega_{\mathrm{m}}}

\newcommand{\ttheta}[1]{\pmb{\theta}}

\newcommand{\proj}{\boldsymbol{\mathcal{P}}}
\newcommand{\gammac}{\gamma_{\mathrm{c}}}
\newcommand{\gammaci}[1]{\gamma_{\mathrm{c}{#1}}}

\newcommand{\Norm}[2]{\mathcal{N}}

\newcommand{\gammatttt}{\gamma_{\mathrm{t}\mathrm{t}\mathrm{t}\mathrm{t}}}
\newcommand{\gammaxttt}{\gamma_{\times\mathrm{t}\mathrm{t}\mathrm{t}}}
\newcommand{\gammatxtt}{\gamma_{\mathrm{t}\times\mathrm{t}\mathrm{t}}}
\newcommand{\gammattxt}{\gamma_{\mathrm{t}\mathrm{t}\times\mathrm{t}}}
\newcommand{\gammatttx}{\gamma_{\mathrm{t}\mathrm{t}\mathrm{t}\times}}
\newcommand{\gammaxxtt}{\gamma_{\times\times\mathrm{t}\mathrm{t}}}
\newcommand{\gammaxtxt}{\gamma_{\times\mathrm{t}\times\mathrm{t}}}
\newcommand{\gammaxttx}{\gamma_{\times\mathrm{t}\mathrm{t}\times}}
\newcommand{\gammatxxt}{\gamma_{\mathrm{t}\times\times\mathrm{t}}}
\newcommand{\gammatxtx}{\gamma_{\mathrm{t}\times\mathrm{t}\times}}
\newcommand{\gammattxx}{\gamma_{\mathrm{t}\mathrm{t}\times\times}}
\newcommand{\gammatxxx}{\gamma_{\mathrm{t}\times\times\times}}
\newcommand{\gammaxtxx}{\gamma_{\times\mathrm{t}\times\times}}
\newcommand{\gammaxxtx}{\gamma_{\times\times\mathrm{t}\times}}

\newcommand{\gammaxxxx}{\gamma_{\times\times\times\times}}

\newcommand{\gammat}{\gamma_{\mathrm{t}}}
\newcommand{\map}{\mathcal{M}_{\mathrm{ap}}}
\newcommand{\mperp}{\mathcal{M}_{\times}}
\newcommand{\maptwo}{\map^2}
\newcommand{\mxtwo}{\mperp^2}
\newcommand{\mapthree}{\map^3}

\newcommand{\mxthree}{\mperp^3}

\newcommand{\mapthreeens}{\left\langle\mapthree\right\rangle}

\newcommand{\mapfour}{\map^4}
\newcommand{\mapfourc}{\mathcal{M}_{\mathrm{ap,c}}^4}

\newcommand{\mapfourens}{\left\langle\mapfour\right\rangle}
\newcommand{\mapfourcens}{\left\langle\mapfour\right\rangle_\mathrm{c}}
\newcommand{\mapfourhat}{\hat{\mathcal{M}}_{\mathrm{ap}}^{4}}

\newcommand{\StoN}{\ensuremath{\mathrm{S/N}}}
\newcommand{\binfunc}{\mathcal{B}}
\newcommand{\Ngal}{N_{\mathrm{gal}}}

\newcommand{\npcf}{$N$PCF}
\newcommand{\desyt}{DES Y3 }

\newcommand*\circled[1]{\tikz[baseline=(char.base)]{
            \node[shape=circle,draw,inner sep=1pt] (char) {#1};}}

\tikzset{Pfeil/.style=%
{to path={let \p1 = ($(\tikztotarget)-(\tikztostart)$),
            \n1 = {int(mod(scalar(atan2(\y1,\x1))+360, 360))}, 
            \n2 = {veclen(\x1,\y1)}
        in \pgfextra{\typeout{\n1,\n2,\x1,\y1}}
         (\tikztotarget)
        node[draw,single arrow,
             minimum height=\n2-\pgflinewidth,
             inner sep=1ex,
             single arrow head extend=1ex,
             rotate=\n1, 
             anchor=tip, 
             ]{}
         }}}

\begin{document}

\defcitealias{Schneideretal2005}{SKL05}
\defcitealias{Asgarietal2021}{A21}
\defcitealias{Porthetal2024}{P24}
\defcitealias{SilvestreRoselloetal24}{SR25}

%
%
    \title{Towards an application of fourth-order shear statistics II}
   \subtitle{Efficient estimation of fourth-order shear correlation functions \\ and an application to the DES Y3 data.}

\newcommand{\orcid}[1]{} 
\author{
Lucas Porth,$^{1}$\thanks{E-mail: lporth@astro.uni-bonn.de}
Elena Silvestre-Rosello,$^{2,1}$
Peter Schneider$^{1}$,
Martina Larma$^{1}$}
\institute{$^{1}$University of Bonn, Argelander-Institut f\"ur Astronomie, Auf dem H\"ugel 71, 53121 Bonn, Germany\\
$^{2}$Universit\"at Innsbruck, Institut für Astro- und Teilchenphysik, Technikerstr. 25/8, 6020 Innsbruck, Austria}
%
%
\abstract{
\textit{Context.} Higher-order lensing statistics contain a wealth of cosmological information that is not captured by second-order statistics. Stage-III lensing surveys have sufficient statistical power to significantly detect cumulant-based statistics up to fourth order.\\
\textit{Aims.} We derive and validate an efficient estimation procedure for the four-point correlation function (4PCF) of polar fields such as weak lensing shear. We then use our approach to measure the shear 4PCF and the fourth-order aperture mass statistics on the \desyt survey.\\
\textit{Methods.} We construct an efficient estimator for fourth-order shear statistics which builds on the multipole decomposition of the shear 4PCF. We then validate our estimator on mock ellipticity catalogues obtained from Gaussian random fields and on realistic $N$-body simulations. Finally, we apply our estimator to the \desyt data and present a measurement of the fourth-order aperture statistics in a non-tomographic setup.\\
\textit{Results.} Due to its quadratic scaling, our estimator provides a significant speed-up over hypothetical brute force or tree-based estimation methods of the shear 4PCF. We report a significant detection of the connected part of the fourth-order aperture mass in the \desyt data. We find the sampling distribution of the fourth-order aperture mass to be significantly skewed. We make our estimator code available on GitHub as part of the {\sc orpheus} package  
\github{https://github.com/lporth93/orpheus}.}
%
%
\keywords{gravitational lensing: weak – methods: numerical – large-scale structure of Universe}
%
%
   \titlerunning{Efficient Estimation of the Shear 4PCF}
   \authorrunning{L. Porth et al.}
   \maketitle
%
%
%
%
   
\section{Introduction}

Weak gravitational lensing (WL) is a powerful probe for constraining the cosmological parameters and distinguishing between competing Universe models \citep{Blandfordetal1991,Kaiser1992}, and more than two decades have passed since the pioneering measurements from the shapes of observed galaxies \citep{Baconetal2000,Kaiseretal2000,vanWaerbekeetal2000,Wittmanetal2000}. During this period, cosmic shear has become a primary observational probe for the statistical investigation of the Universe's large-scale structure. Looking ahead, forthcoming and ongoing stage IV surveys, such as the Vera C. Rubin Observatory's Legacy Survey of Space and Time \citep{Ivezicetal2019} and \textit{Euclid} \citep{ECMellieretal2025}, are expected to deliver shape measurements for billions of galaxies, promising an unprecedented volume of data. 

While previous and current surveys like KiDS, DES, and HSC have predominantly focused their cosmic shear analyses on second-order shear statistics, such as the shear two-point correlation functions and the shear angular power spectrum \citep{Asgarietal2021,Amonetal2022,Seccoetal2022a,Dalaletal2023,Lietal2023,Wrightetal2025}, these methods capture only a fraction of the available cosmological information. This limitation arises because second-order statistics are sensitive solely to the Gaussian component of the underlying field. To obtain tighter parameter constraints, the incorporation of higher-order statistics (HOS) becomes essential. Although HOS have been extensively explored in theoretical frameworks and simulated data \citep{Jainetal2000,Schneideretal2005,Barthelemyetal2020,Kratochviletal2012,ECAjanietal2023}, their practical application to real observational data is more recent \citep{Heydenreichetal2022b,Seccoetal2022,Thieleetal2023,Burgeretal2024,HarnoisDerapsetal2024}. This delay is mainly due to their inherently lower signal-to-noise detection significance and the considerably higher computational cost associated with their estimation. 

One prominent HOS are the shear $N$-point shear correlation functions (\npcf) and compressions thereof, such as the aperture mass statistics. The latter, in particular, can be recast as an integral over the \npcf \, that naturally separates the shear signal in its $E$- and $B$-modes. With the second- and higher-order aperture mass moments being dependent on different orders of the WL polyspectra, they carry complementary information such that a joint analysis might yield tight parameter constraints due to degeneracy breaking; this effect has been explicitly observed in analyses of stage II and stage III surveys \citep{Fuetal2014,Burgeretal2024}. The unprecedented data volume of ongoing stage IV surveys will allow for a significant detection of even higher-order aperture mass measures that, when combined with second- and third-order statistics, could extract additional information from the same data.

Estimating higher-order correlation functions by counting $N$-tuplets of galaxies rapidly becomes computationally untractable for $N \geq 3$. A more efficient estimator has been developed for the multipole components of the \npcf \ of scalar fields, which, after estimation, can readily be transformed to the traditional \npcf \ components \citep{ChenSzapudi2005, SlepianEisenstein2016, Philcoxetal2022a}. Recently, this approach has been used for utilizing 4PCFs to investigate potential cosmological parity violation from the three-dimensional distribution of galaxies and the cosmic microwave background \citep{Houetal2023,Philcoxetal2024,Philcox2025}.  

This work builds upon the formalism introduced in \citealt{Porthetal2024} (herafter P24), who applied the multipole decomposition to the natural components of the 3PCF associated with spin-2 polar fields, such as cosmic shear. We extend this formalism to correlation functions of arbitrary order and provide an efficient implementation of the estimator up to fourth order, making it feasible to compute those statistics on stage III and stage IV survey data.  

This work has a companion paper (\citealp[][hereafter SR25)]{SilvestreRoselloetal24}, which gives a detailed derivation of the connection between the shear 4PCF and the fourth-order aperture measures, provides a numerical setup for an efficient and reliable integration using a bin-averaged 4PCF, and explores the cosmological information contained in the fourth-order aperture mass statistics. We use the same notation and conventions as \citetalias{SilvestreRoselloetal24} and refer for additional details in \secref{sec:FourthOrderMeasuresOfCosmicShear} and \secref{sec:ApertureMass} to this work.

This paper is structured as follows: In \secref{sec:FourthOrderMeasuresOfCosmicShear}, we introduce the shear 4PCF and its natural compoments, $\Gamma_\mu$. In \secref{sec:Estimator}, we derive a multipole-based estimator for the shear 4PCF and discuss some approximations. In \secref{sec:Validation}, we validate the estimator on Gaussian random fields, for which the theoretical 4PCF is known. In \secref{sec:ApertureMass}, we review the connection between the shear 4PCF and the fourth-order aperture mass measures and validate our implementation using a large suite of $N$-body simulations, the SLICS ensemble. In \secref{sec:DESY3Measurement}, we apply the estimator to the first data release of the Dark Energy Survey, hereafter \desyt\!. After obtaining an empirical estimate of the expected covariance matrix for the \desyt data from a large suite of mock simulations and assessing higher-order effects such as reduced shear and source clustering, we present our measurement of the fourth-order aperture statistics in the \desyt data. We conclude in \secref{sec:Conclusions}.

Readers familiar with the basics of WL who are mainly interested in the application of our methods to data may skip the theoretical part of this work and directly jump to \secref{sec:DESY3Measurement}.

\section{Fourth-order measures of cosmic shear}\label{sec:FourthOrderMeasuresOfCosmicShear}
We introduce the relevant equations of gravitational lensing that are necessary to describe fourth-order shear statistics. We use a similar presentation as \citetalias{Porthetal2024}. For extensive reviews of weak gravitational lensing see e.g. \citet{BartelmannSchneider2001}, \citet{Kilbinger2015}, \citet{Dodelson2017} and \citet{Mandelbaum2018}.

\subsection{Basics of WL}
Cosmological WL investigates how large-scale structure influences the shape of light bundles propagating through spacetime. To leading order, this effect is characterised by two fundamental quantities: the convergence, $\kappa$, and the shear, $\gamma$, which describe the isotropic stretching and the distortion experienced by a bundle. The convergence can be written as a weighted projection of the density contrast, $\delta$, along the line of sight, up until  the limiting comoving distance, $\chi_{\mathrm{lim}}$, of the galaxy sample:
\begin{align}\label{eq:kappadef}
    \kappa(\btheta) = \int_0^{\chi_{\mathrm{lim}}} \dd \chi' W(\chi') \  \delta\left[f_K(\chi')\btheta;\chi'\right] \ ,
\end{align}
where the associated projection kernel, $W$, is defined as 
\begin{align}
    W(\chi) \equiv \frac{3\Omegam H_0^2}{2c^2} \frac{f_K(\chi)}{a(\chi)} \int_{z(\chi)}^{z(\chi_{\mathrm{lim}})} \dd z' n(z')  \frac{f_K[\chi(z')-\chi]}{f_K[\chi(z')]} 
,\end{align}
and $f_K[\chi(z)]$ denotes the comoving angular diameter distance of the comoving radial distance, $\chi$, at redshift, $z$. We introduced the dimensionless matter density parameter, $\Omegam$, and the Hubble constant, $H_0$, as well as the scale factor, $a$, and the source redshift distribution, $n(z)$. For the remainder of this work, we assume a flat universe for which $f_K(\chi) = \chi$.

The components of the complex shear field, $\gamma$, can be characterised with respect to a Cartesian coordinate frame, $\gammac \equiv \gamma_1 + \ii \gamma_2$. To evaluate the shear at position $\btheta$ in a reference frame, which is rotated by an angle, $\zeta$, with respect to the Cartesian basis, one has
\begin{align}\label{eq:gammatx_def}
    \gamma(\btheta;\zeta) \equiv \gammat(\btheta;\zeta) + \ii \gamma_\times(\btheta;\zeta) \equiv -\gammac(\btheta) \ \ee^{-2\ii\zeta} \ ,
\end{align}
where we introduced the tangential $(\gammat)$ and cross-components $(\gamma_\times)$ of the shear with respect to the projection direction, $\zeta$.

\begin{figure}
  \centering
  \includegraphics[width=\linewidth]{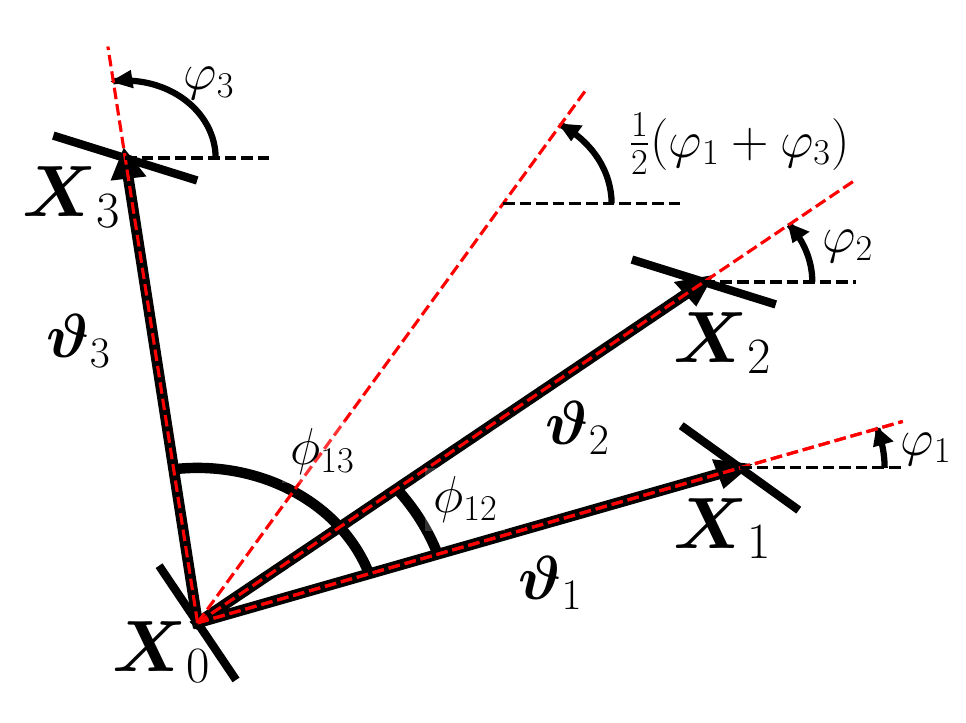}
  \captionsetup{width=\linewidth}
  \caption{Parametrization of a quadruplet of shears used in this work. For some shear at position $\mathbf{X}_0$, we denote the connecting lines to the other shears at positions $\mathbf{X}_i$ as $\boldsymbol{\vartheta}_i$ and the enclosing angles as $\phi_{12}$ and $\phi_{13}$. The red dashed lines show the directions of the $\times$-projection \Eq{eq:xprojection} for which the three projection axes intersect in $\mathbf
{X}_0$}\label{fig:Xprojection}
\end{figure}
\subsection{The shear 4PCF and its natural components}
Due to the polar nature of the shear, there are 16 different real-valued components for its four-point correlator. We follow the prescription outlined in \citet{SchneiderLombardi2003} and regroup those components into eight complex-valued natural components, $\Gamma_\mu$, that do not mix and only transform by some phase factor under rotations of the corresponding quadruplet configuration. Parametrizing a quadrilateral as depicted in Fig. 1 and choosing some arbitrary projection, $\proj$, in which angles, $\zeta_\mu^{\proj}$, rotate the individual shear components, we define the natural
components of the shear 4PCF as\footnote{We slightly adapt the notation of \citetalias{Porthetal2024} in order to have a similar indexing of quantities appearing in expressions related to the 4PCF and to the fourth-order aperture mass.}

{\allowdisplaybreaks
\begin{align}
    \label{eq:Gamma0_def}
    \Gamma_0^{\proj} (\vartheta_1, \vartheta_2, \vartheta_3, \phi_{12}, \phi_{13}) 
    \nonumber \\ &\hspace{-3cm}=
    \left\langle \ 
    \gamma\left(\mathbf{X_0};\zeta_0^{\proj}\right) \ \gamma\left(\mathbf{X_1};\zeta_1^{\proj}\right) \ 
    \gamma\left(\mathbf{X_3};\zeta_2^{\proj}\right) \ 
    \gamma\left(\mathbf{X_3};\zeta_3^{\proj}\right)
    \ \right\rangle \ , \\
    \label{eq:Gamma1_def}
    \Gamma_1^{\proj} (\vartheta_1, \vartheta_2, \vartheta_3, \phi_{12}, \phi_{13}) 
    \nonumber \\ &\hspace*{-3cm}= 
    \left\langle \ \gamma^*\hspace{-.1cm}\left(\mathbf{X_0};\zeta_0^{\proj}\right) \ \gamma\left(\mathbf{X_1};\zeta_1^{\proj}\right) \ 
    \gamma\left(\mathbf{X_2};\zeta_2^{\proj}\right) \ 
    \gamma\left(\mathbf{X_3};\zeta_3^{\proj}\right)
    \ \right\rangle \ , \\
    \Gamma_2^{\proj} (\vartheta_1, \vartheta_2, \vartheta_3, \phi_{12}, \phi_{13}) 
    \nonumber \\ &\hspace*{-3cm}= 
    \left\langle \ 
    \gamma\left(\mathbf{X_0};\zeta_0^{\proj}\right) \ \gamma^*\hspace{-.1cm}\left(\mathbf{X_1};\zeta_1^{\proj}\right) \ \gamma\left(\mathbf{X_2};\zeta_2^{\proj}\right) \ 
    \gamma\left(\mathbf{X_3};\zeta_3^{\proj}\right)  \ 
    \right\rangle \ , \\
    \label{eq:Gamma3_def}
    \Gamma_3^{ \boldsymbol{\mathcal{P}}} (\vartheta_1, \vartheta_2, \vartheta_3, \phi_{12}, \phi_{13}) 
    \nonumber \\ &\hspace*{-3cm}= 
    \left\langle \ 
    \gamma\left(\mathbf{X_0};\zeta_0^{\proj}\right) \  \gamma \left(\mathbf{X_1};\zeta_1^{\proj}\right) \ \gamma^*\hspace{-.1cm} \left(\mathbf{X_2};\zeta_2^{\proj}\right) \ 
    \gamma\left(\mathbf{X_3};\zeta_3^{\proj}\right)
    \ \right\rangle \ , \\
    \label{eq:Gamma4_def}
    \Gamma_4^{ \boldsymbol{\mathcal{P}}} (\vartheta_1, \vartheta_2, \vartheta_3, \phi_{12}, \phi_{13}) 
    \nonumber \\ &\hspace*{-3cm}= 
    \left\langle \ 
    \gamma\left(\mathbf{X_0};\zeta_0^{\proj}\right) \  \gamma \left(\mathbf{X_1};\zeta_2^{\proj}\right) \ \gamma \left(\mathbf{X_2};\zeta_2^{\proj}\right) \ 
    \gamma^*\hspace{-.1cm} \left(\mathbf{X_3};\zeta_3^{\proj}\right)
    \ \right\rangle \ , \\
    \label{eq:Gamma5_def}
    \Gamma_5^{ \boldsymbol{\mathcal{P}}} (\vartheta_1, \vartheta_2, \vartheta_3, \phi_{12}, \phi_{13}) 
    \nonumber \\ &\hspace*{-3cm}= 
    \left\langle \ 
    \gamma^*\hspace{-.1cm} \left(\mathbf{X_0};\zeta_0^{\proj}\right) \  \gamma^*\hspace{-.1cm} \left(\mathbf{X_1};\zeta_1^{\proj}\right) \ \gamma \left(\mathbf{X_2};\zeta_2^{\proj}\right) \ 
    \gamma \left(\mathbf{X_3};\zeta_3^{\proj}\right)
    \ \right\rangle \ , \\
    \label{eq:Gamma6_def}
    \Gamma_6^{ \boldsymbol{\mathcal{P}}} (\vartheta_1, \vartheta_2, \vartheta_3, \phi_{12}, \phi_{13}) 
    \nonumber \\ &\hspace*{-3cm}= 
    \left\langle \ 
    \gamma^*\hspace{-.1cm} \left(\mathbf{X_0};\zeta_0^{\proj}\right) \  \gamma \left(\mathbf{X_1};\zeta_1^{\proj}\right) \ \gamma^*\hspace{-.1cm} \left(\mathbf{X_2};\zeta_2^{\proj}\right) \ 
    \gamma\left(\mathbf{X_3};\zeta_3^{\proj}\right)
    \ \right\rangle \ , \\
    \label{eq:Gamma7_def}
    \Gamma_7^{ \boldsymbol{\mathcal{P}}} (\vartheta_1, \vartheta_2, \vartheta_3, \phi_{12}, \phi_{13}) 
    \nonumber \\ &\hspace*{-3cm}= 
    \left\langle \ 
    \gamma^*\hspace{-.1cm} \left(\mathbf{X_0};\zeta_0^{\proj}\right) \  \gamma \left(\mathbf{X_1};\zeta_1^{\proj}\right) \ \gamma \left(\mathbf{X_2};\zeta_2^{\proj}\right) \ 
    \gamma^*\hspace{-.1cm} \left(\mathbf{X_3};\zeta_3^{\proj}\right)
    \ \right\rangle \ , 
\end{align}
}
where we have $\mathbf{X}_1=\mathbf{X}_0 + \boldsymbol{\vartheta}_1$, $\mathbf{X}_2=\mathbf{X}_0 + \boldsymbol{\vartheta}_2$ and $\mathbf{X}_3=\mathbf{X}_0 + \boldsymbol{\vartheta}_3$, the separation vectors $\boldsymbol{\vartheta}_i$ have an angle $\varphi_i$ with the $x$--axis and we have defined the inner angles as $\phi_{ij}\equiv\varphi_j-\varphi_i$, as well as the projection directions $\zeta_\mu^{\proj}$. This implies that the natural components are invariant under rotations of the quadrilateral such that a parametrization in terms of five variables is sufficient. As noted earlier, any of the natural components can be written in terms of 16 real-valued basis components that we write as the sequence
\begin{align}
    \label{eq:4PCF_rawcomponents}
     (
    &\gammatttt, 
    \gammaxttt, \gammatxtt, \gammattxt, \gammatttx, 
    \gammattxx, \gammatxtx, \gammatxxt, \gammaxttx, 
    \nonumber \\ &\gammaxtxt, \gammaxxtt, \gammatxxx, \gammaxtxx, \gammaxxtx, \gammaxxtx, \gammaxxxx
    ) \ ,
\end{align}
where we introduced shorthand notation like $\gammatttt \equiv \left\langle\gammat\gammat\gammat\gammat\right\rangle$ for the various quadruple products of shears \citep{SchneiderLombardi2003}.
Inverting the relations \eqref{eq:Gamma0_def}--\eqref{eq:Gamma7_def}, one can relate the components in \Eq{eq:4PCF_rawcomponents} to the natural components.

In this work, we mainly use the 4PCF in an adapted version of the $\times$--projection of \citetalias{Porthetal2024}:
\begin{align}\label{eq:xprojection}
    \zeta_0^\times = \frac{1}{2} 
    \left(\varphi_1+\varphi_3\right)
    \ \ \ , \ \ \ 
    \zeta_1^\times = \varphi_1
    \ \ \ , \ \ \ 
    \zeta_2^\times = \varphi_2\
    \ \ \ , \ \ \ 
    \zeta_3^\times = \varphi_3\
     \ .
\end{align}
The corresponding projection axes are depicted as the red dashed lines in \figref{fig:Xprojection} and we use \Eq{eq:gammatx_def} to convert between the Cartesian projections and the $\times$-projection.

\subsection{Bin-averaged 4PCF}
When estimating the $\Gamma_\mu^{\mathcal{P}}$ from a finite set of galaxy ellipticities, one does not have direct access to every possible quadrilateral shape but instead collects the point quadruplets into radial and angular bins. Such a measurement then provides an estimator of the bin-averaged shear 4PCF, $\overline{\Gamma}_\mu^{\mathcal{P}}$:
\begin{align}\label{eq:Gamma0_binned}
    \overline{\Gamma}_\mu^{\mathcal{P}}(&\Theta_i,\Theta_j,\Theta_k,\Phi_{m},\Phi_{n})  
   \equiv \int_{\Theta_i}\frac{\dd \vartheta_1 \ \vartheta_1}{A_i} \int_{\Theta_j}\frac{\dd \vartheta_2 \ \vartheta_2}
   {A_j} \int_{\Theta_k}\frac{\dd \vartheta_3 \ \vartheta_3} {A_k}
   \nonumber \\ &\times 
   \int_{\Phi_{m}}\frac{\dd \phi_{12}}{|\Delta\Phi_{m}|} \ 
   \int_{\Phi_{n}}\frac{\dd \phi_{13}}{|\Delta\Phi_{n}|} \ 
\Gamma_\mu^{\mathcal{P}}(\vartheta_1,\vartheta_2,\vartheta_3,\phi_{12},\phi_{13}) \ ,
\end{align}
where for each radial bin, $\Theta_i$, we have $\vartheta \in [\Theta_{\mathrm{low},i},\Theta_{\mathrm{up},i}]$, and where we defined $A_i\equiv (\Theta_{\mathrm{up},i}^2-\Theta_{\mathrm{low},i}^2)/2$. Similarly, for the angular bins we have $\phi \in [\Phi_{\mathrm{low},m},\Phi_{\mathrm{up},m}]$ and $|\Delta\Phi_m|\equiv \Phi_{\mathrm{up},m}-\Phi_{\mathrm{low},m}$.

\section{An efficient estimator for the shear 4PCF}\label{sec:Estimator}

Given a discrete set of $\Ngal$ galaxies at positions $\btheta_i$, estimated ellipticities, $\gammaci{,i}$, and weights, $w_i$, the traditional estimator of the bin-averaged natural components of the shear 4PCF assigns all galaxy quadruplets to their corresponding quadrilateral configuration bin and then averages over those. In particular, for the $\mu$th natural component in the $\times$-projection \Eq{eq:xprojection}, the estimator is given by the ratio
\begin{align}\label{eq:4PCF_quadrupletestimator}
\hat{\overline{\Gamma}}^{\mathcal{P}}_\mu(\Theta_1,\Theta_2,\Theta_3,\Phi_{12},\Phi_{13})
\equiv \frac{\Upsilon_\mu^{\mathcal{P}}(\Theta_1,\Theta_2,\Theta_3,\Phi_{12},\Phi_{13}) }{\mathcal{N}(\Theta_1,\Theta_2,\Theta_3,\Phi_{12},\Phi_{13}) } \ ,
\end{align}
where, e.g. for the zeroth natural component in the $\times$-projection we have
\begin{align}
\label{eq:QuadrupletUpsilon}
\Upsilon_0^{\times}(\Theta_1,\Theta_2,\Theta_3,\Phi_{12},\Phi_{13})  
    &\nonumber \\ &\hspace{-3cm}
    \equiv \sum_{i,j,k,l=1}^{\Ngal}w_i\gammaci{,i} \ w_j\gammaci{,j}\  w_k\gammaci{,k} \ w_l\gammaci{,l} \ 
    \ee^{-\ii(3\varphi_{ij}+2\varphi_{ik}+3\varphi_{il})}
    \nonumber \\ &\hspace{-2.5cm}\times \binfunc\left(\theta_{ij}\in\Theta_1\right) \ \binfunc\left(\theta_{ik}\in\Theta_2\right) \binfunc\left(\theta_{il}\in\Theta_3\right) 
    \nonumber \\ &\hspace{-2.5cm}\times \binfunc\left(\phi_{ijk}\in\Phi_{12}\right) \binfunc\left(\phi_{ijl}\in\Phi_{13}\right) \ ,
\end{align}
\begin{align}\label{eq:TripletNorm}
\mathcal{N}(\Theta_1,\Theta_2,\Theta_3,\Phi_{12},\Phi_{13})  
    &\equiv  
    \sum_{i,j,k,l=1}^{\Ngal }w_i \, w_j \, w_k \, w_l
    &\nonumber \\ &\hspace{-2.33cm} \times \binfunc\left(\theta_{ij}\in\Theta_1\right) \ \binfunc\left(\theta_{ik}\in\Theta_2\right) \binfunc\left(\theta_{il}\in\Theta_3\right) 
    \nonumber \\ &\hspace{-2.33cm}\times \binfunc\left(\phi_{ijk}\in\Phi_{12}\right) \binfunc\left(\phi_{ijl}\in\Phi_{13}\right) \ .
\end{align}
As in \citetalias{Porthetal2024} we further defined $\theta_{ij} \equiv |\btheta_{ij}| \equiv |\btheta_j-\btheta_i|$ and $\phi_{ijk}\equiv\varphi_{ik}-\varphi_{ij}$, with $\varphi_{ij}$ denoting the polar angle of $\btheta_{ij}$ and introduced the bin selection function, $\mathcal{B}(x \in X)$, which is unity if $x\in X$ and zero otherwise. The different combinations of the bin selection functions then specify the various quadrilateral configurations.

\subsection{Multipole decomposition}\label{ssec:EstimatorMultipoles}
To accelerate the estimation of the shear 4PCF, we used its representation in the multipole basis, where the angular arguments of the 4PCF are expanded in complex exponentials. Adopting the notation of \citetalias{Porthetal2024}, we denote the 4PCF multipoles as $\Upsilon^\times_{\mu,\mathbf{n}}$, where $\mathbf{n} \equiv \left(n_2, n_3\right)$. In general, the two bases are related as
\begin{align}\label{eq:Multipole2RealConversion}
    \Upsilon^{\mathcal{P}}_{\mu}&\left(\Theta_1,\Theta_2,\Theta_3,\phi_{12},\phi_{13}\right) 
     \nonumber \\ &\equiv
     \frac{1}{(2\pi)^2} \sum_{n_2,n_3=-\infty}^\infty \Upsilon^{\mathcal{P}}_{\mu,\mathbf{n}}(\Theta_1,\Theta_2,\Theta_3) \ \ee^{\ii n_2\phi_{12}}\ee^{\ii n_{3}\phi_{13}}
\end{align}
and, again specifying to the zeroth component and to the $\times$-projection, the multipole components, $\Upsilon^\times_{0,\mathbf{n}}$, are computed as 
{\allowdisplaybreaks
\begin{align}\label{eq:4PCFUpsilon0_Derivation}
    \Upsilon^\times_{0,\mathbf{n}}&\left(\Theta_1,\Theta_2,\Theta_3\right)  \\ &=
    \int_0^{2\pi} \dd \phi_{12} \int_0^{2\pi} \dd \phi_{13} \ \ee^{-\ii n_2\phi_{12}} \, \ee^{-\ii n_{3}\phi_{13}}
    \nonumber \\ &\hspace{1cm}
    \Upsilon^\times_{0} \left(\Theta_1,\Theta_2,\Theta_3,\phi_{12},\phi_{13}\right)
    \nonumber \\ &= 
    \int_0^{2\pi} \dd \phi_{12} \int_0^{2\pi} \dd \phi_{13} \ \ee^{-\ii n_2\phi_{12}}  \, \ee^{-\ii n_{3}\phi_{13}}
    \nonumber \\ &\hspace{1cm} 
    \sum_{i,j,k,l} w_i\gammaci{,i} \, w_j\gammaci{,j} \, w_k\gammaci{,k} \, w_l\gammaci{,l} \  \ee^{-\ii\left(3\varphi_{ij}+2\varphi_{ik}+3\varphi_{il}\right)} 
    \nonumber \\ &\hspace{2cm} 
    \binfunc\left(\theta_{ij}\in\Theta_1\right)
    \binfunc\left(\theta_{ik}\in\Theta_2\right)
    \binfunc\left(\theta_{il}\in\Theta_3\right)
    \nonumber \\ &\hspace{2cm} 
    \binfunc\left(\phi_{12}\in\{\phi_{ijk}\}\right)
    \binfunc\left(\phi_{13}\in\{\phi_{ijl}\}\right)
    \nonumber \\ &= 
    \sum_i^{\Ngal} w_i\gammaci{,i} \ 
    \sum_j^{\Ngal} w_j\gammaci{,j} \ \ee^{\ii\left(n_2+n_{3}-3\right)\,\varphi_{ij}} \ \binfunc\left(\theta_{ij}\in\Theta_1\right)
    \nonumber \\ &\hspace{2cm} 
    \sum_k^{\Ngal} w_k\gammaci{,k} \ \ee^{-\ii\left(n_2+2\right)\,\varphi_{ik}} \ \binfunc\left(\theta_{ik}\in\Theta_2\right)
    \nonumber \\ &\hspace{2cm} 
    \sum_l^{\Ngal} w_l\gammaci{,l} \ \ee^{-\ii\left(n_{3}+3\right)\,\varphi_{il}} \ \binfunc\left(\theta_{il}\in\Theta_3\right)
    \nonumber \\ &\equiv
    \sum_i^{\Ngal} w_i\gammaci{,i} \ 
    G^{\mathrm{disc}}_{n_2+n_{3}-3}(\btheta_i,\Theta_1) 
    \nonumber \\ &\hspace{2cm} 
    G^{\mathrm{disc}}_{-\left(n_2+2\right)}(\btheta_i,\Theta_2) \ G^{\mathrm{disc}}_{-\left(n_{3}+3\right)}(\btheta_i,\Theta_3) \ , 
    \label{eq:4PCFUpsilon0}
\end{align}
}
where we have defined the $G_n^{\mathrm{disc}}$ as in \citetalias{Porthetal2024}. 
Similarly, we computed the other seven multipoles as
{\allowdisplaybreaks
\begin{align}
    \Upsilon^\times_{1,\mathbf{n}}\left(\Theta_1,\Theta_2,\Theta_3\right)  &= \sum_i^{\Ngal} w_i\gammaci{,i}^* \ G^{\mathrm{disc}}_{n_2+n_{3}-1}(\btheta_i,\Theta_1) 
    \nonumber \\&\hspace{-1cm}
    G^{\mathrm{disc}}_{-\left(n_2+2\right)}(\btheta_i,\Theta_2) \ G^{\mathrm{disc}}_{-\left(n_{3}+1\right)}(\btheta_i,\Theta_3)
    \ , \\
    \Upsilon^\times_{2,\mathbf{n}}\left(\Theta_1,\Theta_2,\Theta_3\right)  
    &= \sum_i^{\Ngal} w_i\gammaci{,i} \ 
    \left(G^{\mathrm{disc}}_{-\left(n_2+n_{3}+1\right)}(\btheta_i,\Theta_1)\right)^* 
    \nonumber \\&\hspace{-1cm}
    G^{\mathrm{disc}}_{-\left(n_2+2\right)}(\btheta_i,\Theta_2) \ G^{\mathrm{disc}}_{-\left(n_{3}+3\right)}(\btheta_i,\Theta_3)
    \ , \\
    \Upsilon^\times_{3,\mathbf{n}}\left(\Theta_1,\Theta_2,\Theta_3\right)  
    &= \sum_i^{\Ngal} w_i\gammaci{,i} \ 
    G^{\mathrm{disc}}_{n_2+n_{3}-3}(\btheta_i,\Theta_1) 
    \nonumber \\&\hspace{-1cm}
    \left(G^{\mathrm{disc}}_{n_2-2}(\btheta_i,\Theta_2)\right)^* \ G^{\mathrm{disc}}_{-\left(n_{3}+3\right)}(\btheta_i,\Theta_3)
    \ , \\
    \Upsilon^\times_{4,\mathbf{n}}\left(\Theta_1,\Theta_2,\Theta_3\right)  
    &= \sum_i^{\Ngal} w_i\gammaci{,i} \ 
    G^{\mathrm{disc}}_{n_2+n_{3}-3}(\btheta_i,\Theta_1) 
    \nonumber \\&\hspace{-1cm}
    G^{\mathrm{disc}}_{-\left(n_2+2\right)}(\btheta_i,\Theta_2) \ \left(G^{\mathrm{disc}}_{n_{3}-1}(\btheta_i,\Theta_3) \right)^*
    \ , \\
    \Upsilon^\times_{5,\mathbf{n}}\left(\Theta_1,\Theta_2,\Theta_3\right)  
    &= \sum_i^{\Ngal} w_i\gammaci{,i}^* \ 
    \left(G^{\mathrm{disc}}_{-\left(n_2+n_{3}+3\right)}(\btheta_i,\Theta_1)\right)^* 
    \nonumber \\&\hspace{-1cm}
    G^{\mathrm{disc}}_{-\left(n_2+2\right)}(\btheta_i,\Theta_2) \ G^{\mathrm{disc}}_{-\left(n_{3}+1\right)}(\btheta_i,\Theta_3)
    \ , \\
    \Upsilon^\times_{6,\mathbf{n}}\left(\Theta_1,\Theta_2,\Theta_3\right)  
    &= \sum_i^{\Ngal} w_i\gammaci{,i}^* \ 
    G^{\mathrm{disc}}_{n_2+n_{3}-1}(\btheta_i,\Theta_1) 
    \nonumber \\&\hspace{-1cm}
    \left(G^{\mathrm{disc}}_{n_2-2}(\btheta_i,\Theta_2)\right)^* \ G^{\mathrm{disc}}_{-\left(n_{3}+1\right)}(\btheta_i,\Theta_3)
    \ , \\
    \Upsilon^\times_{7,\mathbf{n}}\left(\Theta_1,\Theta_2,\Theta_3\right)  
    &= \sum_i^{\Ngal} w_i\gammaci{,i}^* \ 
    G^{\mathrm{disc}}_{n_2+n_{3}-1}(\btheta_i,\Theta_1) 
    \nonumber \\&\hspace{-1cm}
    G^{\mathrm{disc}}_{-\left(n_2+2\right)}(\btheta_i,\Theta_2) \ \left(G^{\mathrm{disc}}_{n_{3}-3}(\btheta_i,\Theta_3) \right)^*
    \ .
\end{align}
}

Finally, we obtained for the normalization, $\mathcal{N}$, see also \citet{Sunserietal2023},
\begin{align}
    \label{eq:Nnmultipoles}\mathcal{N}_{\mathbf{n}}\left(\Theta_1,\Theta_2,\Theta_3\right)  &= \sum_i^{\Ngal} w_i \ W^{\mathrm{disc}}_{n_2+n_{3}}(\btheta_i,\Theta_1) 
    \nonumber \\&\hspace{-1cm}
    W^{\mathrm{disc}}_{-n_2}(\btheta_i,\Theta_2) \ W^{\mathrm{disc}}_{-n_{3}}(\btheta_i,\Theta_3)
    \ ,
\end{align}
where the $W^{\mathrm{disc}}_{n}$ are defined in \citetalias{Porthetal2024}.

We note that the $\Upsilon^\times_{0,\mathbf{n}}$ and the $\mathcal{N}_{\mathbf{n}}$ are not independent of each other and can be related to each other by permuting the three radial bins. As an example, letting $(\Theta_1,\Theta_2,\Theta_3) \rightarrow (\Theta_2,\Theta_3,\Theta_1)$ in $\Upsilon^\times_{\mu,\mathbf{n}}$, the contribution of the $i$th galaxy in \Eq{eq:4PCFUpsilon0} becomes
\begin{align}
    &G_{n'_2+n'_3-3}(\btheta_i,\Theta_2) \, G_{-n'_2-2}(\btheta_i,\Theta_3) \, G_{-n'_3-3}(\btheta_i,\Theta_1)
    \nonumber \\ &\hspace{.3cm}\overset{!}{=}
    G_{n_2+n_3-3}(\btheta_i,\Theta_1) \, G_{-n_2-2}(\btheta_i,\Theta_2) \, G_{-n_3-3}(\btheta_i,\Theta_3) \ ,
\end{align}
which is fulfilled for $(n'_2,n'_3) = (-n_2-n_3,n_3+1)$. Extending this relation to each summand in \Eq{eq:4PCFUpsilon0} we conclude that 
\begin{align}
\Upsilon^\times_{0,\left(n_2,n_3\right)} \left(\Theta_1,\Theta_2,\Theta_3\right) = \Upsilon^\times_{0,\left(n_3+1,-n_2-n_3\right)} \left(\Theta_2,\Theta_3,\Theta_1\right) \ .
\end{align}
The remaining interrelations are given in Table \ref{tab:4PCFSymm}. Combining the results for all possible such permutations implies that it is sufficient to compute the multipoles for the radial bins with $\Theta_1 \leq \Theta_2 \leq \Theta_3$.\footnote{This is a natural extension of the symmetries of the multipoles of the 3PCF, for which it is sufficient to compute all radial bins with $\Theta_1 \leq \Theta_2$; in \citetalias{Porthetal2024} this symmetry was rephrased as only requiring positive values for the multipoles to be allocated.} 

We further note that to restrict ourselves to correlators in which all galaxies in the quadruplet are distinct from each other, one needs to subtract the contributions for which the above requirement is not met, leading to double- and triple-counting corrections, for which we defer the explicit expressions to \appref{app:MCCorr}. 

Finally, we note that we can extend the above construction of multipole-based estimators of the shear 4PCF to arbitrary order. For the explicit expressions, we refer to \appref{app:NthOrderEstimator}

\subsection{Practical implementation}
Similar to the case of the 3PCF introduced in Sect. 3.2--3.3 of \citetalias{Porthetal2024}, the multipole-based estimator of the 4PCF can be accelerated using tree-based methods. While in \citetalias{Porthetal2024}, the speedup had been achieved by distributing the catalogue into a series of hierarchical grids, ${\mathcal{G}(\Delta_d)}$, from which the $G_n^{(\Delta_d)}$ could be readily computed using Fast Fourier Transforms (FFTs), we modified the approach such that the $G_n^{(\Delta_d)}$ are allocated using the standard formalism. The main reasons for this change were the high memory consumption of the FFT-based implementation (i.e. all $G_n^{(\Delta_d)}$ of the footprint need to be kept in memory -- see Appendix A3 in \citetalias{Porthetal2024}) and the constraint that due to the regularity of the FFT, the cell-averaged shear was always placed at the pixels' centre. 
In our updated implementation, we continued to use a series of hierarchical grids of resolutions $\Delta_{d} = 2^{d-1} \, \Delta_1$ and placed the cell-averaged shear at its corresponding centre of mass, thereby constructing a hierarchy of `reduced' catalogues. 
To control the level of angular resolution in the $G_n^{(\Delta_d)}$ we introduced the parameter $r_{\mathrm{min},\Delta}$, which fixes the choice of the grid resolution for an angular bin, $\Theta$. In particular, we set the resolution for $\Theta$ as the coarsest resolution level, $d'$, for which $\Theta_{\mathrm{low}}/\Delta_{d'} \geq r_{\mathrm{min},\Delta}$; if this condition could not be fulfilled, we used the discrete estimator. For a more detailed description and a complexity analysis of our implemented 4PCF algorithm, we refer to \appref{app:Est_LowMemImplementation}. For a more detailed explanation of the approximation schemes used within our public implementation, \textsc{orpheus}, we refer to \appref{app:NthOrderImplementation}.

\section{Validation of the estimator on Gaussian random fields}\label{sec:Validation}
\begin{figure*}
  \centering
  \includegraphics[width=.999\textwidth,valign=t]{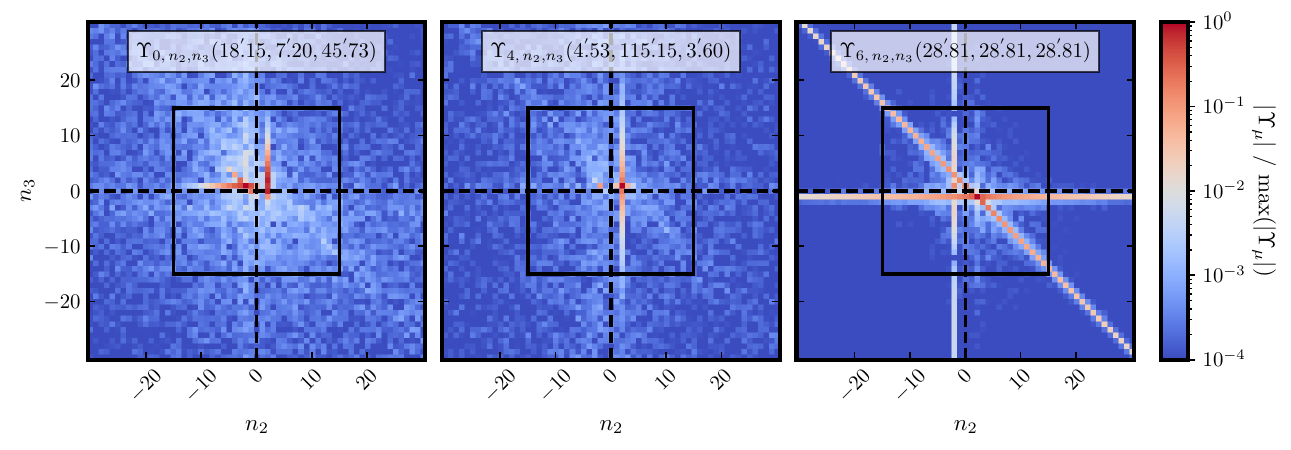}\\
  \includegraphics[width=.999\textwidth,valign=t]{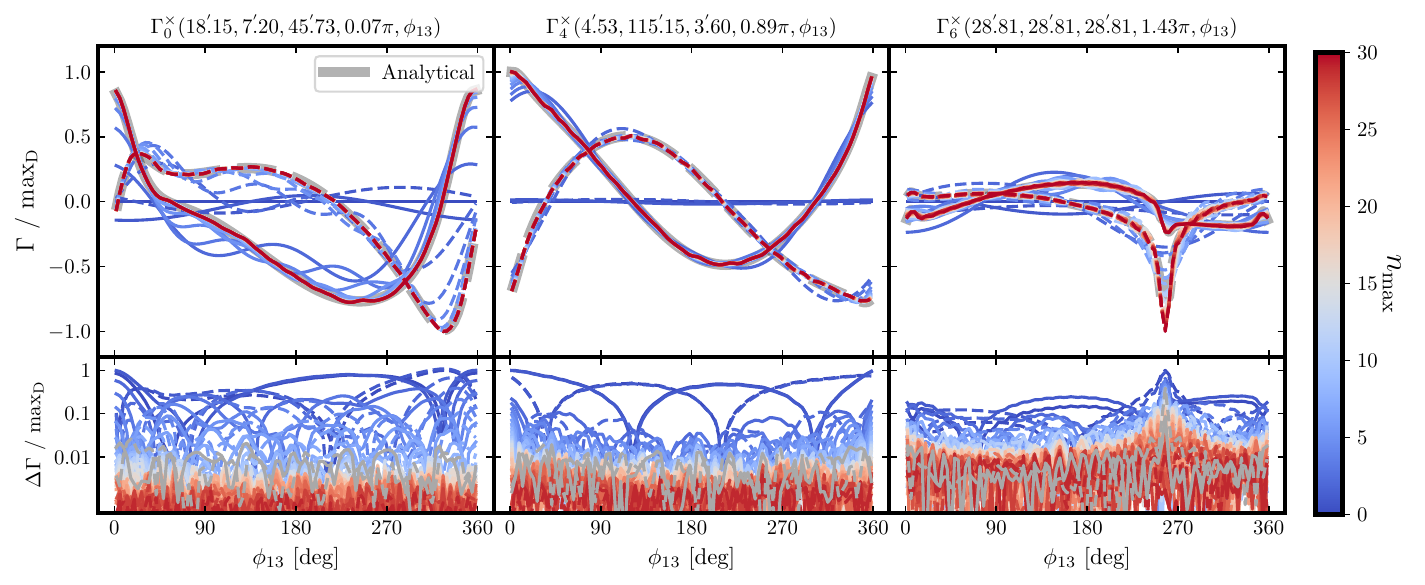}\\
  \includegraphics[width=.999\textwidth,valign=t]{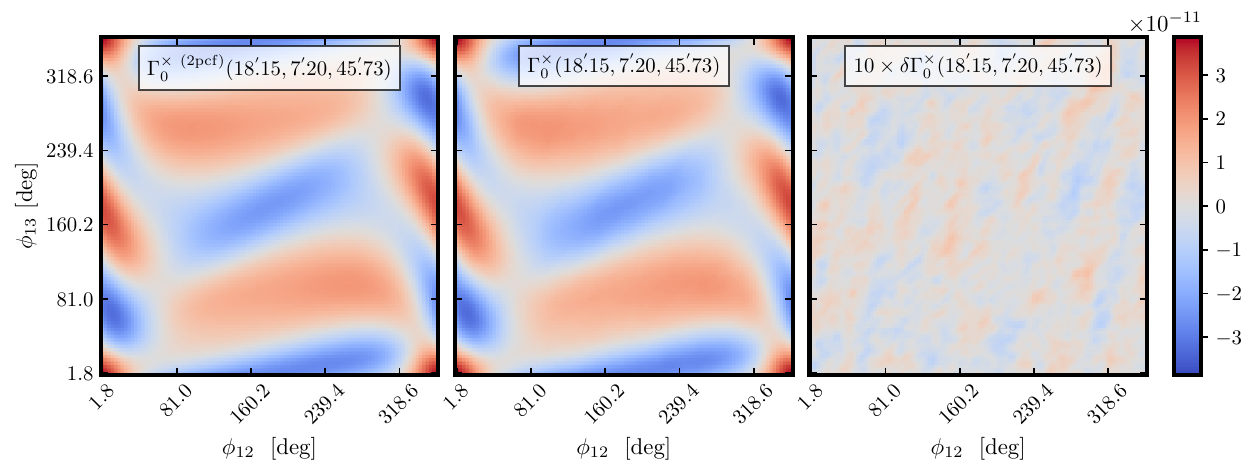}
\captionsetup{width=\linewidth}
\caption{Estimator validation using the shear 4PCF from an ensemble of GRFs. \textit{Top row:} Absolute value of 4PCF multipoles for three different natural components and different radial configurations, normalised by their largest value. The square surrounding the region with $|n_i|\leq15$ indicates the multipole cuts used for our default analysis. \textit{Middle row:} Convergence of the 4PCF in real space for the same natural components and radial configurations as in the top row, evaluated at some fixed $\phi_{12}$. Solid lines indicate the real part while dashed lines show the imaginary part. The lines are colour-coded by the largest multipole, $n_{\rm max}$, used in the conversion \Eq{eq:Multipole2RealConversion}. The thick grey lines denote the disconnected part of the 4PCF when reconstructed from the measured 2PCF. In the upper sub-panels, the signal is normalised by its largest value, $\rm{max}_D$, while the lower sub-panels display the difference, $\Delta \Gamma \equiv \left|\Gamma_{n_{\rm{max}}}-\Gamma_{n_{\rm{max}}=30}\right|$, normalised by $\rm{max}_D$.
\textit{Bottom row:} The angular dependence of the real part of the zeroth natural component of the 4PCF when reconstructed from the measured 2PCF (left) or the measured 4PCF (middle). The right panel displays the difference between the two measurements, multiplied by a factor of ten.}
\label{fig:Gaussfield_4PCFcompare}
\end{figure*}

To validate the implementation of the estimator, we applied it to an ensemble of Gaussian random fields (GRFs). For such fields, Wicks' theorem guarantees that the components of the 4PCF are solely built from the second-order theory, i.e., the $\xi_{\pm}$, which themselves are related to the convergence power spectrum used to generate the mock \citep{Schneideretal2002}. For the explicit expressions, we refer to Appendix C.1 of \citetalias{SilvestreRoselloetal24}.

As the 4PCF consists of eight complex-valued components, each dependent on five variables, exhibiting nontrivial structure, and spanning multiple orders of magnitude, it is not straightforward to define a single sufficient convergence measure. We therefore opted to first check the convergence on different subsamples of the 4PCF and then to assess the overall estimator convergence by virtue of the fourth-order aperture mass statistics. 

\subsection{Mock generation}\label{ssec:Gaussmock_generation}
For our tests, we generated an ensemble of 100 GRFs using a convergence power spectrum using the best-fit parameters of the $\Lambda$CDM-\textit{Optimised} analysis reported in \cite{Amonetal2022} and the cumulative $n(z)$ reported in \cite{Mylesetal2021}. Each mock was on a square footprint with $10\deg$ side length and a pixel resolution of $\approx 0 \farcm 2$. After having generated the convergence GRF and having obtained its corresponding shear GRF using the method of \cite{KaiserSquires1993}, we randomly selected $360\,000$ unique pixels, assigned their shear to a hypothetical galaxy located at the pixels' centre, and then applied a random displacement within the pixel to this galaxy. We note that to circumvent finite-field and pixel resolution effects, we used the measured second-order shear correlation functions (2PCF) to construct any fourth-order statistics that was compared to the measured statistics derived from the 4PCF multipoles in the following validation tests.

\subsection{Validation of shear 4PCF estimator}\label{ssec:Gaussfield_4PCFcompare}
For assessing the largest required multipole, $n_{\rm max}$, that allows for an accurate reconstruction of the angular dependence of the 4PCF in real-space, we chose to compute its multipole components for $(n_2, n_3) \in \{-30, \cdots, 30\}^2$. To keep the memory footprint manageable we used 20 logarithmically spaced radial bins covering the interval $\Theta \in [1\farcm 0, 128\farcm 0]$.

In the top row of \figref{fig:Gaussfield_4PCFcompare} we show the absolute value of the 4PCF multipoles for three different natural components and three different configurations of radial bins, normalised by its maximum, $\max_{\{\mathbf{n}\}}\left(|\Upsilon_{\mu,\mathbf{n}}(\Theta_1,\Theta_2,\Theta_3)|\right)$. We see that the largest values are generally obtained for small values of both $n_2$ and $n_3$, and that there are three linear combinations of the multipoles for which the values are enhanced.\footnote{We note that this structure is tied to the field being Gaussian, and we refer to \appref{app:DisconnectedMultipoles} for an explicit derivation of the structure visible in the top right panel in \figref{fig:Gaussfield_4PCFcompare}. As can be seen in \appref{app:SLICS_4PCF}, more multipole components become significant for a field with a non-vanishing disconnected four-point function.}

The second row of \figref{fig:Gaussfield_4PCFcompare} displays the 4PCF in the real space basis when the transformation \Eq{eq:Multipole2RealConversion} is truncated at order $n_{\rm max} \leq 30$. We see that including the first few multipoles recovers the shape of the 4PCF for most configurations at the level of a few percent and we further see that the convergence is slowest in the regimes in which sharp peaks in the 4PCF are visible. The signal contained in the highest-order multipoles does not visibly affect the convergence and therefore we chose $n_{\rm max} \equiv 15$ as the default setup for future analyses in this work. We also compared the estimated 4PCF to the 4PCF that is obtained from the measured 2PCF in the mocks.\footnote{We chose $\vartheta \in [0\farcm 1, 256']$, used a logarithmic bin width of $0.05$ and set the 2PCF to zero outside those bounds.} To ensure a reasonable matching of the five-dimensional bins, we bin-averaged the estimated 2PCF over $4^3$ sub-bins in the radial integrations, linearly interpolating the 2PCF at the corresponding bin centres. We found agreement between both methods at the percent level.

In the bottom row of \figref{fig:Gaussfield_4PCFcompare} we compare the full angular dependence of $\Gamma_0$ for a fixed radial configuration to the Gaussian prediction. While there are small visible differences, those do not appear to have a structure and we attribute them to noise originating from the truncation of the multipole expansion and to the slight mismatching of the bins in both computations.
In \appref{app:SLICS_4PCF} we repeated the same analysis on the SLICS ensemble, finding similar convergence properties and, as expected, non-vanishing contributions of the connected part of the 4PCF that manifest themselves both, in the real space and the multipole space basis.

\section{Application to fourth-order aperture mass measures}\label{sec:ApertureMass}

\subsection{The aperture-mass statistic} 
\label{ssec:ApertureMass}
To compress the information of the 4PCF we make use of higher-order aperture mass measures. Originally, the aperture mass introduced in \cite{Kaiser1995} and \citet{Schneider1996} is a measure for the projected overdensity within a circular region of radius $\theta$ around a particular location, $\bvartheta$:
\begin{align}\label{eq:MapDefinition}
    \map(\bvartheta;\theta) = \int \dd^2 \bvartheta' \ U_\theta\left(|\bvartheta'|\right) \ \kappa(\bvartheta+\bvartheta') \ ,
\end{align}
where $U_\theta$ is a compensated filter function. Within the flat-sky approximation, one can define the aperture mass as the real part of a complex aperture measure that is written in terms of the shear,
\begin{align}\label{eq:MapGamma}
   \mathcal{M}(\bvartheta;\theta) 
    &=
   \map(\bvartheta;\theta) + \ii \mperp(\bvartheta;\theta) \nonumber \\
    &=
    \int \dd^2 \bvartheta' \ Q_\theta\left(|\bvartheta'|\right) \ \gamma(\bvartheta+\bvartheta';\varphi') \ ,
\end{align}
in which $Q_\theta$ is another filter function uniquely related to the choice of the $U_\theta$ filter, and $\varphi'$ denotes the polar angle of the separation vector, $\bvartheta'$. With the aperture measure being able to separate the shear signal into its $E$- and $B$-modes, it can be used for both extracting cosmological information and for pinning down potential systematics. For this work, we used the exponential filter introduced in \cite{Crittendenetal2002},
\begin{align}
    \label{eq:CrittendenQ}
    Q_\theta(\vartheta) = \frac{\vartheta^2}{4\pi\theta^4}\exp[\left(-\frac{\vartheta^2}{2\theta^2}\right)] \ .
\end{align}
By taking moments of the complex aperture measure and applying an average over cosmological ensembles, one can connect those moments to higher-order correlators of the underlying field. In particular, by combining four aperture measures one can relate them to a filtered version of the shear trispectrum or of the shear 4PCF. Focusing on the latter, the corresponding equations structurally read
\begin{align}\label{eq:4PCF2M4}
&\hspace*{0cm} \expval{\mathcal{M}^{(*)}_{4,\mu}}(\theta) 
\nonumber \\ &\hspace*{0.5cm}= \int_0^\infty \dd \vartheta_1 \int_0^\infty \dd \vartheta_2 \int_0^\infty \dd \vartheta_3 \int_0^{2\pi}\dd \phi_{12} \int_0^{2\pi}\dd \phi_{13} 
\nonumber \\ &\hspace*{1cm} 
\times \Gamma_{\mu}^\times(\vartheta_1,\vartheta_2,\vartheta_3,\phi_{12},\phi_{13}) \ F^{(4,\times)}_\mu(\theta;\vartheta_1,\vartheta_2,\vartheta_3,\phi_{12},\phi_{13})\ ,
\end{align}
where we ordered the correlators as 
\begin{align}\label{eq:M4Ordering}
    \expval{\mathcal{M}^{(*)}_4}
    &\equiv 
    \left[
    \expval{\mathcal{M}\mathcal{M}\mathcal{M}\mathcal{M}},
    \expval{\mathcal{M}^*\mathcal{M} \mathcal{M} \mathcal{M}}, 
    \right. \nonumber \\ &\hspace*{-1cm} \left.
    \expval{\mathcal{M}\mathcal{M}^* \mathcal{M} \mathcal{M}}, 
    \expval{\mathcal{M}\mathcal{M} \mathcal{M}^* \mathcal{M}}, 
    \expval{\mathcal{M}\mathcal{M} \mathcal{M} \mathcal{M}^*}, 
    \right. \nonumber \\ &\hspace*{-1cm} \left.
    \expval{\mathcal{M}^*\mathcal{M}^* \mathcal{M} \mathcal{M}}, 
    \expval{\mathcal{M}^*\mathcal{M} \mathcal{M}^* \mathcal{M}}, 
    \expval{\mathcal{M}^*\mathcal{M} \mathcal{M} \mathcal{M}^*} 
    \right] \ 
\end{align}
and we refer to the explicit expressions for the filter functions, $F^{(4,\times)}_{\mu}$ 
 and for a generalization to arbitrary orders to \citetalias{SilvestreRoselloetal24}. Using \Eq{eq:MapGamma}, the correlators $\expval{\mathcal{M}^{(*)}_4}$ can then be transformed to the $E$/$B$-decomposed aperture-mass statistics, $\expval{\mathcal{M}^{(\rm{ap})}_4}$, which consists of $16$ real-valued components that we ordered as the sequence \Eq{eq:4PCF_rawcomponents}. The pure $E$-mode contribution is contained in $\expval{\mathcal{M}_{\rm{ap}}^4}$ while the remaining components contain $B$-modes, parity-violating modes, and mixtures thereof, that are all expected to vanish at lowest order and in the absence of systematic and astrophysical effects.

\subsection{Estimators of fourth-order aperture measures}
\subsubsection{Estimation via shear 4PCF}
As established in the previous subsection one can estimate the fourth-order aperture mass measures by integrating over the 4PCF in the $\times$-basis. In practice we first used the multipole-based estimator introduced in \secref
{sec:Estimator} to obtain the shear 4PCF in the multipole-space basis, then transformed it to the real-space basis via \Eq{eq:Multipole2RealConversion} and \Eq{eq:4PCF_quadrupletestimator}, and finally performed the numerical integrals contained in \Eq{eq:4PCF2M4}. Using GRFs, our companion paper \citetalias{SilvestreRoselloetal24} investigated the level of accuracy to which the 4PCF needs to be computed in order to guarantee a stable integration and we based our binning choices on those results.

\subsubsection{Estimation via the direct estimator}
On an unmasked survey, the aperture statistics can alternatively be estimated using a higher-order extension of the direct method of \citet{Schneideretal1998} that averages moments of the convolution \Eq{eq:MapGamma} over the survey field, using the ellipticities at the galaxy positions. In particular, one samples an ensemble of apertures over the survey field, and for each aperture at a particular angular position, $\bvartheta$, computes
\begin{align}\label{eq:MapDirectEstimator}
    \mapfourhat(&\bvartheta; \theta)  \nonumber \\
    &\hspace{-.5cm}=\frac{\sum_{i\neq j \neq k \neq l}^{N_{\mathrm{gal}}} 
    w_i w_j w_k w_l \
    Q_{\theta,i} Q_{\theta,j}Q_{\theta,k}Q_{\theta,l} \ 
    \gamma_{\mathrm{t},i} 
    \gamma_{\mathrm{t},j} 
    \gamma_{\mathrm{t},k}
    \gamma_{\mathrm{t},kl}}
    {\sum_{i\neq j \neq k \neq l}^{N_{\mathrm{gal}}}
    w_i w_j w_k w_l \
     \ } \ ,
\end{align}
where we abbreviated $Q_{\theta,i}\equiv Q_{\theta}(|\bvartheta_i-\bvartheta|)$ and $\gamma_{\mathrm{t},i} \equiv \gammat (\bvartheta_i; \zeta_i)$, with $\bvartheta_i$ being the angular position of the $i$th galaxy and $\zeta_i$ denoting the direction, $\bvartheta_i-\bvartheta$.
In our implementation we used the accelerated estimator introduced in \cite{Porthetal2021} that decomposes the nested sums appearing in \Eq{eq:MapDirectEstimator} in a linear-order expression and we applied inverse shot noise weights when averaging the apertures over the footprint. To keep the support of the $Q$ filter, \Eq{eq:CrittendenQ}, finite, we imposed a hard cut at $4\theta$ within which almost all the weight of the filter is contained \citep{Heydenreichetal2022}. To prevent the leakage of $B$-modes due to the finite extent of the mocks we only sampled apertures that are separated from the fields' boundary by at least $4\theta$.

\subsection{Comparison of both estimators}
\label{ssec:EstimatorCompare}
\begin{figure*}
    \centering
    \includegraphics[width=.99\textwidth]{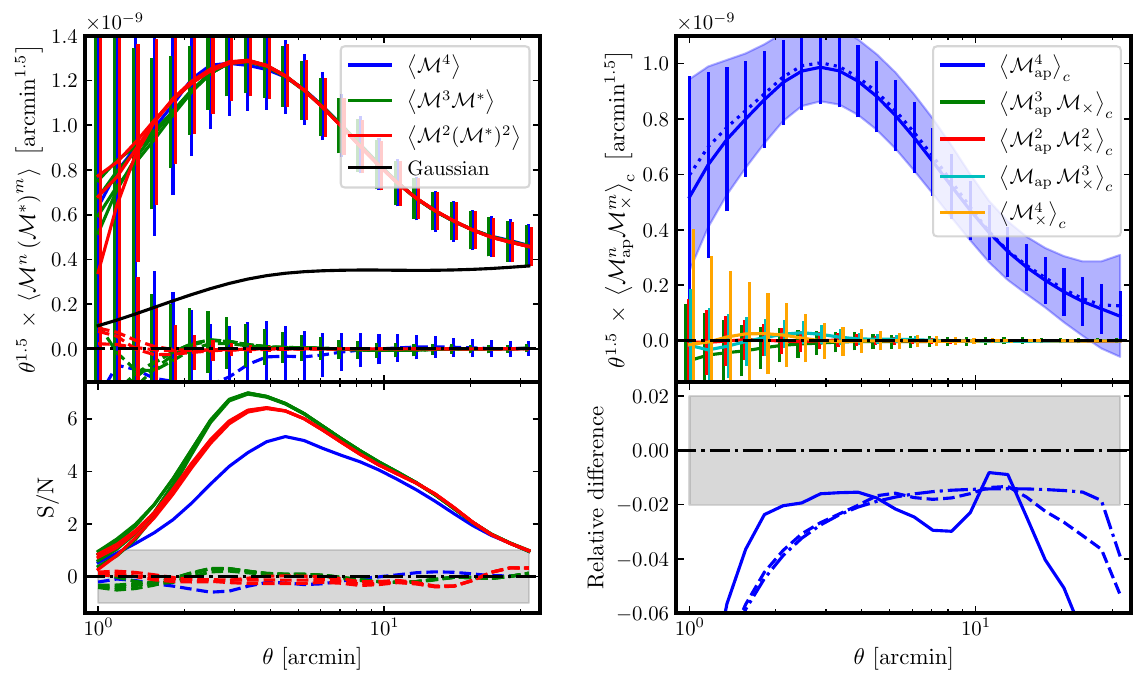}
    \captionsetup{width=\linewidth}
    \caption{Validation of the fourth-order aperture statistics on the SLICS simulation suite.  
    \textit{Left}: The independent aperture measures \eqref{eq:M4Ordering} obtained by applying the transformations \eqref{eq:4PCF2M4} to the estimated shear 4PCF. In the top panel, the real (imaginary) parts of the aperture measures are displayed as solid (dashed) lines which are colour-coded according to the structurally different 4PCF components. The black line shows the disconnected part of $\mapfour$ estimated from the shear 2PCF. In the bottom panel, we show the signal-to-noise of the connected part of the measurement when rescaled to an area of $4000 \, \rm{deg}^2$ and we indicate the regime with less than unity signal-to-noise as the gray shaded region.
    \textit{Right:} Accuracy of the estimated fourth-order aperture statistics. In the top panel, we show the $E$/$B$-decomposed aperture mass statistics as measured from the 4PCF (solid lines) as well as the pure $E$-mode component from the direct method (blue dotted line). The statistical uncertainties of the estimators are given by the error bars (4PCF) and the blue error band (direct). In the bottom panel, we show the relative difference between the two $\mapfourcens$ estimators (solid line), as well as the expected integration bias for the disconnected part when comparing with the direct estimate (dashed) and the correlation-function estimate (dash-dotted). In all panels, the zero line is shown as a black dash-dotted line. For visual purposes, the errorbars of the different measurements are slightly shifted with respect to their corresponding lines.}
    \label{fig:SLICS_Map4compare}
\end{figure*}
As a validation of both, the convergence of the multipole-based 4PCF estimator and the binning setup for the conversion \Eq{eq:4PCF2M4}, we computed both estimators on the SLICS simulation suite \citep{HarnoisDerapsetal2018}; see \citetalias{Porthetal2024} for a description of the simulations. To make some connection with the DES-Y3 data while keeping the compute time feasible, we subselected galaxies such that each mock had a source density of $\overline{n}_{\rm{SLICS}}=0.3 \, \mathrm{arcmin}^{-2}$ and we adjusted the shapenoise-level to $\sigma_{\epsilon,\rm{SLICS}}^2 = \frac{\overline{n}_{\rm{DES}}}{\overline{n}_{\rm{SLICS}}} \, \sigma_{\epsilon,\mathrm{DES}}^2$, where $\overline{n}_{\rm{DES}} = 5.592 \, \mathrm{arcmin}^{-2}$ and $\sigma_{\epsilon,\mathrm{DES}}=0.261$ are the corresponding values reported in \citet{Gattietal2021}.

For the multipole-based estimator we chose $n_{\rm{max}}=15$ and evaluated the 4PCF using $65$ logarithmically-spaced bins in the interval $\Theta \in [0\farcm 25, 166\farcm 29]$. We then transformed the 4PCF multipoles in the real-space basis using $129$ linearly-spaced angular bins to which finally the transformations \eqref{eq:4PCF2M4} were applied\footnote{Due to the large memory footprint of the 4PCF, about 38 GB (658 GB) in the multipole-space (real-space) basis, we never stored the 4PCF but updated the transformation equations on-the-fly; see \appref{app:Est_LowMemImplementation} for additional details.} for 24 logarithmically-spaced aperture radii, $\theta \in [1\farcm 0, 32\farcm 0]$. Such a radial binning setup for the 4PCF was found to be accurate at the $2\%$-level for the disconnected part of $\mapfour$ for $\theta \gtrsim 4'$ in \citetalias{SilvestreRoselloetal24}. We estimated $\mapfour$ from the direct method for the same set of aperture radii. To obtain the connected part of the fourth-order aperture statistics we subtracted off the disconnected contribution, which for the 4PCF estimator was estimated from the $\xi_{\pm}$ and for the direct method was obtained from the directly estimated $\maptwo$.

In \figref{fig:SLICS_Map4compare} we show the results from 800 lines-of-sight from the SLICS ensemble. To compare the estimator accuracy to a rough estimate of the expected measurement uncertainty in a \desyt like survey, we area-rescaled the sample covariance to match a survey observing $4000 \, \rm{deg}^2$ on the celestial sphere.
Looking at the eight aperture correlators we see that their real parts are all consistent with their mean, $\mapfour$, while their imaginary parts are well within their $1\sigma$ uncertainty region around zero. We further see that for large aperture radii the signal-to-noise of the correlators is comparable while when choosing a small aperture radius, the signal-to-noise of $\left\langle\mathcal{M}^4\right\rangle$ is significantly lower compared to the other components. We attribute this to the fact that the covariance of large aperture radii is dominated by the cosmic variance contribution while for gradually smaller scales the shapenoise contribution becomes more important. In particular, the latter is related to the number of independent galaxy quadruplets within an aperture, which for $\left\langle\mathcal{M}^4\right\rangle$ is proportional to $N_{\rm{gal,ap}}^4/24$ while for $\left\langle\mathcal{M}^3\mathcal{M}^*\right\rangle$ $\left(\left\langle\mathcal{M}^2(\mathcal{M}^*)^2\right\rangle\right)$ this number increases by a factor of four (three). 

In the right panel of \figref{fig:SLICS_Map4compare} we show the results for the fourth-order aperture mass statistics. We again find the $B$-modes and the parity-violating modes to be consistent with zero while the $E$-mode contains a significant signal. Comparing the $E$-mode to the results obtained using the direct method we find good agreement at the level of a few percent. The increasing bias for small aperture radii is expected due to the finite radial cutoff of the estimated shear 4PCF while for larger scales we attribute the bias to the different weighting of the data when using the direct estimator and the 4PCF estimator. We further compared the measured integration bias to the expected integration bias for the disconnected part of $\mapfourens$. The latter was obtained by allocating the disconnected 4PCF from the measured $\xi_\pm$ and by transforming it to $\mapfourens$ using the same integration setup that was used for the full 4PCF. Finding good agreement between the curves we conclude that the employed estimation setup produces a sufficient estimate for $\mapfourcens$ and further does not artificially produce significant $B$- and parity-violating modes. 

\section{Fourth-order shear statistics in the \desyt data}\label{sec:DESY3Measurement}
\subsection{The \desyt data}
We use the \text{\small{\textsc{metacalibration}}} shape catalogue derived from the first three years of data from the Dark Energy Survey (\desyt\!\!) for our analysis \citep{DESOverview2005,Flaugheretal2015,DESOverview2016}. The catalogue contains over 100 million galaxies spread over 4143 $\mathrm{deg}^2$ on the sky resulting in a mean source redshift of $\overline{z}_{\rm{DES}}=0.63$ and a weighted source number density of $\overline{n}_{\rm{DES}}=5.592 \, \mathrm{arcmin}^{-2}$. It has been validated in \citet{Gattietal2021} for the \desyt analysis. While for the shear two-point analysis of \citet{Amonetal2022} and \citet{Seccoetal2022a} the shape catalogue had been divided into four tomographic redshift bins \citep{Mylesetal2021} we did not perform such a split for our measurement of fourth-order shear statistics. We made this choice due to computational constraints (the multipole-based 4PCF estimator scales quartic with the number of tomographic bins) and quantification of the expected signal-to-noise of different tomographic bin combinations for the fourth-order aperture mass on a suite of mock catalogues (see \secref{ssec:T17CovMatrix}). For the latter, we found the non-tomographic setup to yield the largest \StoN \ amongst each possible tomographic split.

\subsection{Measurement setup}
We measured the 4PCF following the procedure outlined in \citetalias{Porthetal2024}. In particular, we first decomposed the catalogue into a set of 100 overlapping patches for which the flat-sky approximation holds and then obtained the estimate for the 4PCF multipole correlators $\Upsilon_{\mu,\mathbf{n}}$ and $\mathcal{N}_{\mathbf{n}}$ by summing over their measurement on the individual patches.

The binning setup for the shear 4PCF multipoles was chosen in accordance with the findings from \citetalias{SilvestreRoselloetal24} and from \secref{ssec:EstimatorCompare}. In particular, we computed the 4PCF multipoles up to $n_{\rm{max}}=15$ using $65$ logarithmically-spaced bins in the interval $\Theta \in [0\farcm 25, 166\farcm 29]$. To reduce the runtime we applied the tree-based approximation using a hierarchical mesh with resolutions $\Delta_{d} \in  \{\textrm{0\farcm 5},\textrm{1\farcm 0},\textrm{2\farcm 0}\}$ and we set $r_{\mathrm{min},\Delta}\equiv 40$. 
\subsection{Covariance matrix}\label{ssec:T17CovMatrix}
For constructing estimates of the covariance matrix of $\mapfourcens$ we made use of a suite of full-sky gravitational lensing simulations introduced in \citet{Takahashietal2017}, hereafter the T17 ensemble. The underlying
dark matter-only $N$-body simulations were run on a $\Lambda$CDM cosmology with $\Omegam = 0.279$, $\Omega_{\Lambda}=1-\Omegam$, $h = 0.7$, $\sigma_8 = 0.82$ and spectral index $n_{\rm s} = 0.97$. 

\begin{figure*}
    \centering
    \includegraphics[width=.98\textwidth]{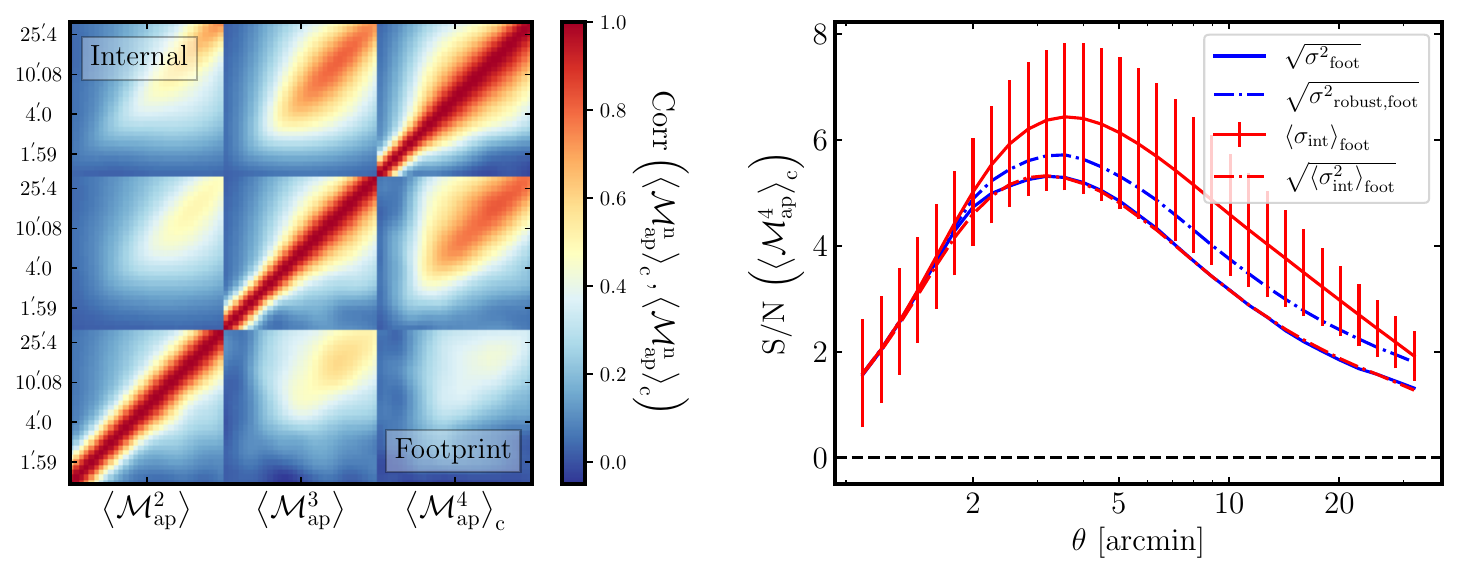}
    \captionsetup{width=\linewidth}
    \caption{\textit{Left:}  Joint correlation matrix of the second, third, and fourth aperture mass cumulants in the T17 ensemble obtained from the footprints (lower triangle) or from the internal covariance estimates averaged over the footprints (upper triangle). \textit{Right:} Signal-to-noise of $\mapfourcens$ in the T17 ensemble when choosing different definitions of the standard deviation. The blue solid (red dash-dotted) line shows the results when a global standard deviation is computed as in the lower (upper) triangle from the plot on the left. The red line shows the mean and the standard deviation of the patch-based signal-to-noise ratio when computed for each footprint individually. The blue dash-dotted line shows the MCD covariance matrix estimate using 855 out of the available 864 footprints.} 
    \label{fig:t17mocks_corrcoef}
\end{figure*}
\subsubsection{Mock generation}

For constructing the mocks we used the convergence and shear maps from the T17 at a resolution of $\text{\small{\textsc{nside}}}=8192$. To have a reasonable number of $864$ independent mock footprints we first split the full-sky into octants and then cut off stripes across the two great circles dividing the octants such that each footprint had an area of $4143 \, \deg^2$. 

To mimic some properties of the \desyt catalogue in the mocks we distributed its per-object intrinsic shape and weight values on each footprint in two steps. In the first step, we constructed a base catalog: 
\begin{enumerate}
    \item We Poisson-distributed $N_{\rm{gal}}^{\rm{DES}}$ discrete points in the footprint located within the first octant.
    \item For each lens plane we obtained the total weight per redshift bin on that plane given the \desyt redshift distribution.
    \item We randomly selected sources of the corresponding redshift bin until their cumulative weight reached the target weight.
    \item We saved the resulting base catalogue containing the randomised positions, the galaxy weight, the intrinsic shape, the tomographic redshift bin, and the index of the lens plane.
\end{enumerate}
By applying the above algorithm we ensured that the weighted $n(z)$ and the distribution of intrinsic shapes across individual lens planes was equal across all mocks and further that the connection between galaxy weights and intrinsic noise in the base catalogue matches the one from the \desyt shape catalogue. To obtain a mock catalogue for a specific footprint we proceeded as follows:
\begin{enumerate}
    \item We translated the positions of the base catalogue to the corresponding octant.
    \item We generated random directions for the intrinsic ellipticity.
    \item For each lens plane we retrieved the T17 convergence and shear map values at the pixels corresponding to the galaxy positions.
    \item We applied the diagonal components of the \text{\small{\textsc{metacalibaration}}} response matrix, $R_i$, to the intrinsic ellipticity as $\epsilon'^{\text{S}} \equiv \epsilon^{\text{S}}/R_i$, and, following \citet{SeitzSchneider1997}, used the latter quantity to define the
    per-object observed ellipticity,  $\epsilon^{\rm{obs}} = \frac{\epsilon'^{\text{S}}+g}{1+\epsilon'^{\text{S}}g^*}$, where $g \equiv \frac{\gamma}{1-\kappa}$ denotes the reduced shear.
\end{enumerate}
We note that one pitfall of the resulting mock catalogues is their simpler geometry which might result in a slight misestimation of the correlation structure for large aperture scales. Our choice of octants above the realistic \desyt footprint was first, to be able to place eight instead of four non-overlapping, contiguous footprints on a full-sky and second, to be able to apply the direct estimator on the mocks without having to cut out a significant area. We expect the geometry-related effects on the covariance to be small as the \desyt footprint consists of a single contiguous area with the local thickness being larger than any considered aperture radius for most points.\vspace*{.2cm}
\subsubsection{Estimated covariance matrices}
\label{ssec:T17Cov}
To keep the estimation time feasible we employed the direct estimator to measure the second- and fourth-order aperture statistics in $31$ logarithmically spaced bins between $1\farcm 0$ and $32\farcm 0$. For circumventing full-sky effects we followed the same outline as for the \desyt data and decomposed each mock footprint into $100$ overlapping patches.\footnote{Due to the equal spatial distribution of the sources and the weights across all mocks, also the patches will be the same across all mocks.} We estimated the aperture statistics on each patch by sampling apertures within its interior and obtained the aperture statistics on the footprint by a weighted sum over the statistics on each patch with the weights being the cumulative inverse-shot noise-derived aperture weights on the patches.   

Due to the similar average patch size in the T17 mocks and in the decomposed \desyt footprint, we estimated the covariance of the aperture statistics in two different ways. 
\begin{enumerate}[(i)]
    \item The sample covariance, $\mathrm{cov}_{\mathrm{foot}}$, estimated from the $864$ mock footprints.
    \item The average $\left\langle\mathrm{cov}_{\mathrm{int}}\right\rangle_{\mathrm{foot}}$ of the internally estimated sample covariances per footprint $\mathrm{cov}_{\mathrm{int}}$.
\end{enumerate}
We show the correlation matrices of the second, third, and fourth aperture mass cumulant\footnote{For both estimates we construct the fourth-order cumulant by subtracting off the disconnected fourth-order contribution at the footprint level.} computed by the two approaches in the left panel of \figref{fig:t17mocks_corrcoef}. We see that both estimates result in a similar correlation structure, but note that there are some visible fluctuations in the regions of low correlation in $\mathrm{cov}_{\mathrm{foot}}$, implying that this estimate is not fully converged yet. Generally, we observe that for increasing cumulant order and aperture scales outside the shape noise-dominated limit, there is a tendency for larger cross-correlations between different aperture scales and further an increase of the cross-correlation between cumulants of adjacent order. This implies that only a few equal-scale apertures are needed to capture the full information contained in these statistics, but also hints towards the necessity of including multi-scale apertures to access the bulk of information contained in the 4PCF. The need for considering multi-scale apertures can also be motivated by looking at the formulation of $\left\langle\map^n\right\rangle_\mathrm{c}$ in terms of the convergence polyspectrum (see Eq. 17 in \citetalias{SilvestreRoselloetal24}), where due to the localised structure of the $\Hat{U}$-filter the fraction of $\boldsymbol{\ell}$-multiplets significantly contributing to the equal-scale statistics decreases with $n$. We leave a more thorough investigation of this note to future work. 

While the correlation structure appears similar between both approaches, this is not the case for the estimated standard deviation. As shown in the right panel of \figref{fig:t17mocks_corrcoef}, using an internally estimated covariance tends to underestimate the diagonal elements of the covariance matrix and therefore provides an apparent increase in signal-to-noise. While part of this effect could arise due to neglecting the mutual information shared between different Jackknife samples, the main contributing factor is the non-Gaussian sampling distribution of $\mapfourc$, see \figref{fig:t17mocks_mapnsamplingdistribution_notomo} for a visualization. In general, the sampling distribution becomes more skewed for larger aperture radii and, due to the central limit theorem, this effect is expected to be enhanced on the individual Jackknife patches as compared to the footprint. When internally estimating the covariance matrix on each footprint, most of the measurements are expected to stem from the bulk of the sampling distribution, indicating an underestimated covariance matrix; this effect is leveraged, however, when also taking into account the patches containing the tails of the sampling distribution (see the good agreement between the solid blue and the red dash-dotted lines in \figref{fig:t17mocks_corrcoef}). As in a realistic survey one does not have access to the ensemble but only to a single footprint, internally computing the standard deviation per footprint will result in overconfident error bars. This is because taking the square root operation before averaging over the patches will reduce the impact of the tails of the sampling distribution of the Jackknife covariance as compared to taking the square root after having averaged the internally estimated covariances over the full ensemble. In \figref{fig:t17mocks_corrcoef}, this effect can be seen by the red solid line that systematically overpredicts the \StoN \ for aperture radii in which the sampling distribution of $\mapfourens$ is not dominated by shape noise. We introduce a heuristic way to correct for this bias in which one multiplies the \StoN \ of the internally estimated covariance by a correction factor $r_{\rm {std,foot}} \equiv \sqrt{\sigma^2_{\rm foot}}\, / \, \left\langle \sigma_{\rm int} \right\rangle_{\rm foot}$.

To further highlight the impact of the tails of the sampling distribution of $\mapfourcens$ at the footprint level we used an estimate of a robust covariance matrix as the basis for computing the \StoN\footnote{We applied the {\sc scikit-learn} implementation of the Minimal Covariance Determinant (MCD) estimator \citep{Rousseeuw1999} using the set of 855 out of the 864 available observations whose sample covariance has the smallest determinant.}. As expected, we found the relative effect of the tails of the sampling distribution to increase with aperture radius

\subsection{Measurement results}
\begin{figure*}
    \centering
    \includegraphics[width=.98\textwidth]{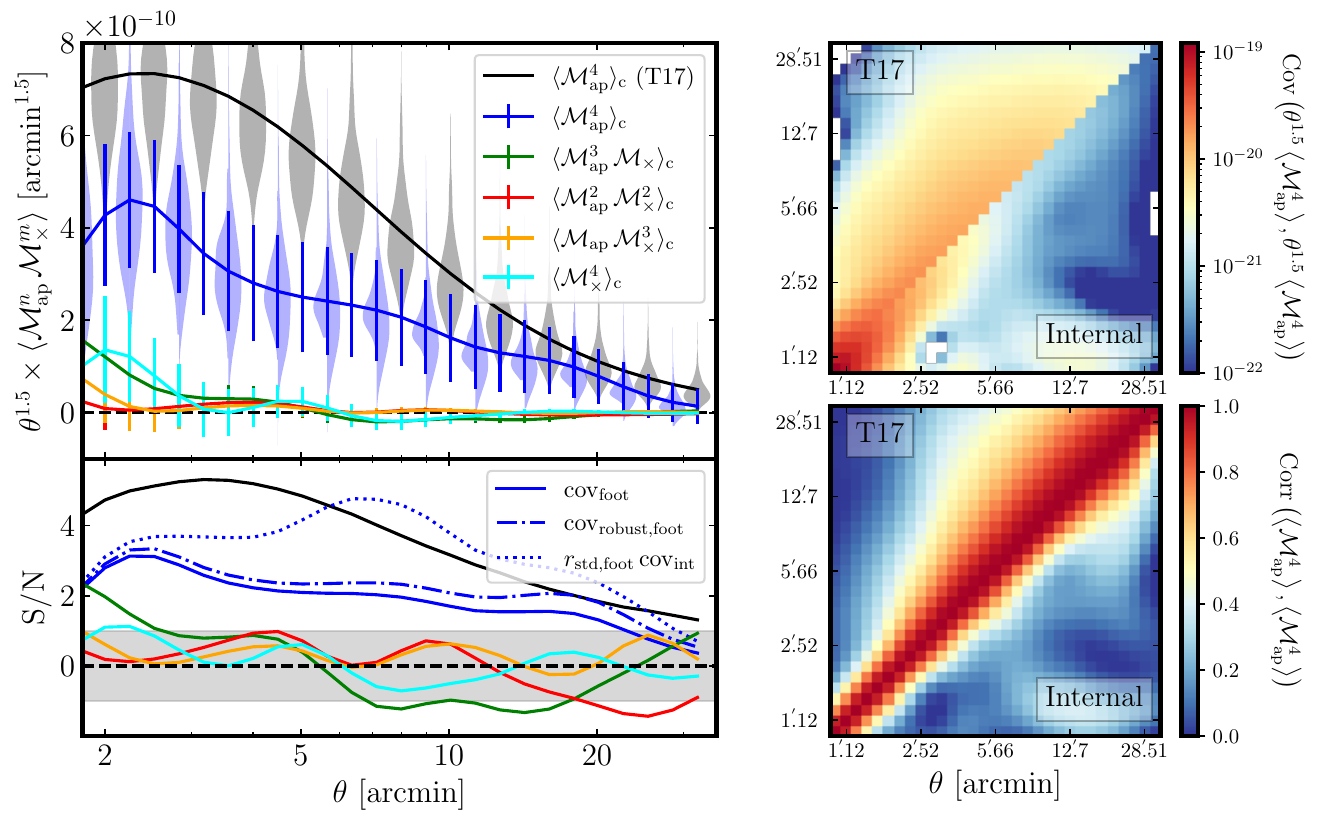}
    \captionsetup{width=\linewidth}
    \caption{\textit{Left:} The fourth-order aperture statistics in the T17 ensemble (black) and in the \desyt data (other). In the upper panel, the sampling distribution of the $E$-mode is shown as a violin shape, where for the \desyt data the left part of each violin shows the area-rescaled distribution on the \desyt patches, whereas the right part shows the distribution in the T17 ensemble. The lower panel shows the signal-to-noise for each of the statistics, where for $\mapfourcens$ we include the different error estimates introduced in \secref{ssec:T17Cov}.  \textit{Right:} Comparison of the covariance and correlation matrices of $\mapfourens$ when estimated from the T17 ensemble (upper triangles) and internally from the \desyt data (lower triangles).}
    \label{fig:desy3res_map4signal}
\end{figure*}
We show the measurement of the fourth-order aperture statistics in the \desyt data in \figref{fig:desy3res_map4signal}. Before discussing the pure $E$-mode signal, we assess the significance of the other four modes that would indicate systematics. Assuming a Gaussian sampling distribution and using the internally estimated covariance matrix, we find for their $p$-values: 
$p(\langle \mapthree \mperp \rangle)=0.0554$, $p(\langle \maptwo \mxtwo \rangle)=0.638$, 
$p(\langle \map \mxthree \rangle)=0.399$, 
$p(\langle \mperp^4 \rangle)=0.0444$, indicating those modes not to be highly significant. In contrast, testing the null-hypothesis `no cosmological connected $E$-mode signal', using the internally estimated covariance matrix of $\langle\mperp^4\rangle$, yields $p=4.09\times 10^{-137}$, strongly rejecting the null-hypothesis.

Given the highly non-Gaussian sampling distribution of $\mapfourcens$, we do not attempt a thorough analysis of the measurement, but aim to make a few qualitative statements on the two main observations in \figref{fig:desy3res_map4signal}, namely that
\begin{enumerate}
    \item The overall amplitude in the \desyt is significantly smaller than in the T17.
    \item The internally estimated error bars appear to be overconfident.
\end{enumerate}

Although the underlying cosmological parameters of the T17 mocks and the \desyt best-fit cosmology are inherently different, their inferred values of $S_8$ appear to be too similar to explain the observed difference. We have cross-checked our measurement against the direct estimator, which we found to be in agreement with the \npcf-based estimate at the ten-percent level. As a second check, in \appref{app:PSFTests} we used the $\rho$-statistics \citep{Rowe2010,Jarvisetal2016} to assess the relevance of PSF-induced systematics; again finding no significant contribution. However, we also note that our results are in qualitative agreement with the findings of \citet{Seccoetal2022} who reported a significantly reduced amplitude of $\mapthreeens$ for aperture scales $\lesssim 10'$ with respect to a mock constructed from the T17 simulations. We reproduced their results using our suite of T17 mocks and by repeating the third-order measurement using the {\sc{orpheus}} code. Noting the high correlation between $\mapthreeens$ and $\mapfourcens$ in the T17 mocks and assuming the signal to be dominated by cosmology, makes the measurement appear more plausible. Finally, a further hint towards a reduced amplitude of the fourth-order aperture mass cumulant can be found in Fig. 6 of \citet{Anbajaganeetal2023}, which shows the fourth-order cumulant of a smoothed convergence map reconstructed from the data in the highest tomographic bin of the \desyt data. While they perform their analysis on a regular grid, such that the impact of shape noise dominates their measurement, they also find a reduced amplitude of the fourth-order cumulant when comparing it to measurements on their suite of reference simulations.  

The second observation can be understood as a combination of the small amplitude of the measured signal and the intrinsic bias when internally estimating the covariance. In particular, from the lower panel on the left hand side in \figref{fig:desy3res_map4signal} we see that the internally estimated \StoN \ traces the mean obtained from the T17 well once adjusted by the expected bias, $r_{\mathrm{std,foot}}$ and assuming the \StoN \ to be independent of the the amplitude of the signal. We have validated the latter assumption on the T17, finding a small negative correlation between the standardised signal amplitude and the internally estimated \StoN. 

\section{Conclusions}\label{sec:Conclusions}
In this work, we presented and validated an efficient estimator for the
natural components of the fourth-order shear correlation functions
and applied it to the \desyt data.

Our work directly extended the multipole-based methods introduced in \citet{SlepianEisenstein2016} and \citetalias{Porthetal2024}, allowing the 4PCF to be computed in quadratic time complexity, which we further reduced by utilizing tree-based methods. To validate our implementation of the estimator, we created a suite of Gaussian shear fields and compared the measured 4PCF to the theoretically computed one, finding good agreement if the 4PCF multipoles are allocated up until $n_{\rm max} \gtrsim 15$. After a schematic review of the findings of our companion paper \citetalias{SilvestreRoselloetal24}, namely on how to convert the shear 4PCF to the aperture statistics, we tested our implementation against a direct estimation method on the SLICS simulation suite. We found that the binning scheme of the 4PCF recommended in \citetalias{SilvestreRoselloetal24} yields a two-percent accuracy for aperture scale $4\farcm 0 \lesssim \theta \lesssim 30\farcm 0$ and that the relative level of the integration accuracy is similar for the connected and for the disconnected part of $\mapfourens$.  

We proceeded to apply the 4PCF estimator to the \desyt data. For an estimate of the covariance matrix, we used the T17 ray-tracing simulations to create a suite of 864 mock footprints, mimicking the \desyt data. We presented the measurement of the fourth-order aperture measures on the \desyt data, finding no significant systematic modes, but noted the low amplitude of the $E$-mode as compared to the T17 ensemble. While we did not find any indication that this effect was due to observational systematics, we note that it could be driven by astrophysical effects, such as intrinsic alignments or baryonic feedback. We postpone a thorough analysis of such possibilities to future work.

Using multipole-based methods, we have demonstrated that the measurement of the shear 4PCF is feasible for stage III and stage IV weak lensing surveys. We have further shown that the fourth-order aperture statistics can serve as a compression scheme of the 4PCF signal. We note that while this compression has several advantageous properties, such as its implicit $E$-/$B$-mode decomposition, its clear physical interpretation, and the availability of multiple classes of estimators, it also comes at the cost of having to compute the 4PCF in a very fine radial and angular binning, significantly increasing the computation time. This limitation could be addressed by using a coarser radial binning or a reduced dynamical range for the 4PCF and by explicitly modelling the induced bias at the aperture statistics level (see the approach of \citealt{Sugiyamaetal2024} and \citealt{Gomesetal2025} at the $\mapthreeens$-level), but this would come at the expense of additionally leaking $B$-modes in the data vector, making the analysis of systematics at the fourth-order level more challenging.

%
%

\begin{acknowledgements}
  The figures in this work were created with {\sc matplotlib} \citep{Hunter2007}
  Some of the results in this paper have been derived using the healpy and HEALPix\footnote{currently http://healpix.sourceforge.net} package \citep{Gorskietal2005, Zonca2019}. We further make use of the {\sc numpy} \citep{Harrisetal2020}, {\sc scipy} \citep{Virtanenetal2020} and {\sc scikit-learn} \citep{Pedregosaetal2011} software packages. LP acknowledges support from the DLR grant 50QE2302. 
  This project used public archival data from the Dark Energy Survey (DES). Funding for the DES Projects has been provided by the U.S. Department of Energy, the U.S. National Science Foundation, the Ministry of Science and Education of Spain, the Science and Technology FacilitiesCouncil of the United Kingdom, the Higher Education Funding Council for England, the National Center for Supercomputing Applications at the University of Illinois at Urbana-Champaign, the Kavli Institute of Cosmological Physics at the University of Chicago, the Center for Cosmology and Astro-Particle Physics at the Ohio State University, the Mitchell Institute for Fundamental Physics and Astronomy at Texas A\&M University, Financiadora de Estudos e Projetos, Funda{\c c}{\~a}o Carlos Chagas Filho de Amparo {\`a} Pesquisa do Estado do Rio de Janeiro, Conselho Nacional de Desenvolvimento Cient{\'i}fico e Tecnol{\'o}gico and the Minist{\'e}rio da Ci{\^e}ncia, Tecnologia e Inova{\c c}{\~a}o, the Deutsche Forschungsgemeinschaft, and the Collaborating Institutions in the Dark Energy Survey
The Collaborating Institutions are Argonne National Laboratory, the University of California at Santa Cruz, the University of Cambridge, Centro de Investigaciones Energ{\'e}ticas, Medioambientales y Tecnol{\'o}gicas-Madrid, the University of Chicago, University College London, the DES-Brazil Consortium, the University of Edinburgh, the Eidgen{\"o}ssische Technische Hochschule (ETH) Z{\"u}rich,  Fermi National Accelerator Laboratory, the University of Illinois at Urbana-Champaign, the Institut de Ci{\`e}ncies de l'Espai (IEEC/CSIC), the Institut de F{\'i}sica d'Altes Energies, Lawrence Berkeley National Laboratory, the Ludwig-Maximilians Universit{\"a}t M{\"u}nchen and the associated Excellence Cluster Universe, the University of Michigan, the National Optical Astronomy Observatory, the University of Nottingham, The Ohio State University, the OzDES Membership Consortium, the University of Pennsylvania, the University of Portsmouth, SLAC National Accelerator Laboratory, Stanford University, the University of Sussex, and Texas A\&M University.
Based in part on observations at Cerro Tololo Inter-American Observatory, National Optical Astronomy Observatory, which is operated by the Association of Universities for Research in Astronomy (AURA) under a cooperative agreement with the National Science Foundation.
  
\end{acknowledgements}

%
%

\bibliography{bibliography.bib}

\begin{thebibliography}{67}
\expandafter\ifx\csname natexlab\endcsname\relax\def\natexlab#1{#1}\fi

\bibitem[{{Amon} {et~al.}(2022){Amon}, {Gruen}, {Troxel}, {MacCrann}, {Dodelson}, {Choi}, {Doux}, {Secco}, {Samuroff}, {Krause}, {Cordero}, {Myles}, {DeRose}, {Wechsler}, {Gatti}, {Navarro-Alsina}, {Bernstein}, {Jain}, {Blazek}, {Alarcon}, {Fert{\'e}}, {Lemos}, {Raveri}, {Campos}, {Prat}, {S{\'a}nchez}, {Jarvis}, {Alves}, {Andrade-Oliveira}, {Baxter}, {Bechtol}, {Becker}, {Bridle}, {Camacho}, {Carnero Rosell}, {Carrasco Kind}, {Cawthon}, {Chang}, {Chen}, {Chintalapati}, {Crocce}, {Davis}, {Diehl}, {Drlica-Wagner}, {Eckert}, {Eifler}, {Elvin-Poole}, {Everett}, {Fang}, {Fosalba}, {Friedrich}, {Gaztanaga}, {Giannini}, {Gruendl}, {Harrison}, {Hartley}, {Herner}, {Huang}, {Huff}, {Huterer}, {Kuropatkin}, {Leget}, {Liddle}, {McCullough}, {Muir}, {Pandey}, {Park}, {Porredon}, {Refregier}, {Rollins}, {Roodman}, {Rosenfeld}, {Ross}, {Rykoff}, {Sanchez}, {Sevilla-Noarbe}, {Sheldon}, {Shin}, {Troja}, {Tutusaus}, {Tutusaus}, {Varga}, {Weaverdyck}, {Yanny}, {Yin}, {Zhang}, {Zuntz}, {Aguena}, {Allam}, {Annis}, {Bacon},
  {Bertin}, {Bhargava}, {Brooks}, {Buckley-Geer}, {Burke}, {Carretero}, {Costanzi}, {da Costa}, {Pereira}, {De Vicente}, {Desai}, {Dietrich}, {Doel}, {Ferrero}, {Flaugher}, {Frieman}, {Garc{\'\i}a-Bellido}, {Gaztanaga}, {Gerdes}, {Giannantonio}, {Gschwend}, {Gutierrez}, {Hinton}, {Hollowood}, {Honscheid}, {Hoyle}, {James}, {Kron}, {Kuehn}, {Lahav}, {Lima}, {Lin}, {Maia}, {Marshall}, {Martini}, {Melchior}, {Menanteau}, {Miquel}, {Mohr}, {Morgan}, {Ogando}, {Palmese}, {Paz-Chinch{\'o}n}, {Petravick}, {Pieres}, {Romer}, {Sanchez}, {Scarpine}, {Schubnell}, {Serrano}, {Smith}, {Soares-Santos}, {Tarle}, {Thomas}, {To}, {Weller}, \& {DES Collaboration}}]{Amonetal2022}
{Amon}, A., {Gruen}, D., {Troxel}, M.~A., {et~al.} 2022, \prd, 105, 023514

\bibitem[{{Anbajagane} {et~al.}(2023){Anbajagane}, {Chang}, {Banerjee}, {Abel}, {Gatti}, {Ajani}, {Alarcon}, {Amon}, {Baxter}, {Bechtol}, {Becker}, {Bernstein}, {Campos}, {Carnero Rosell}, {Carrasco Kind}, {Chen}, {Choi}, {Davis}, {DeRose}, {Diehl}, {Dodelson}, {Doux}, {Drlica-Wagner}, {Eckert}, {Elvin-Poole}, {Everett}, {Fert{\'e}}, {Gruen}, {Gruendl}, {Harrison}, {Hartley}, {Huff}, {Jain}, {Jarvis}, {Jeffrey}, {Kacprzak}, {Kokron}, {Kuropatkin}, {Leget}, {MacCrann}, {McCullough}, {Myles}, {Navarro-Alsina}, {Pandey}, {Prat}, {Raveri}, {Rollins}, {Roodman}, {Rykoff}, {S{\'a}nchez}, {Secco}, {Sevilla-Noarbe}, {Sheldon}, {Shin}, {Troxel}, {Tutusaus}, {Whiteway}, {Yanny}, {Yin}, {Zhang}, {Abbott}, {Allam}, {Aguena}, {Alves}, {Andrade-Oliveira}, {Annis}, {Bacon}, {Blazek}, {Brooks}, {Cawthon}, {da Costa}, {Pereira}, {Davis}, {Desai}, {Doel}, {Ferrero}, {Frieman}, {Giannini}, {Gutierrez}, {Hinton}, {Hollowood}, {Honscheid}, {James}, {Kuehn}, {Lahav}, {Marshall}, {Mena-Fern{\'a}ndez}, {Menanteau}, {Miquel},
  {Palmese}, {Pieres}, {Plazas Malag{\'o}n}, {Reil}, {Sanchez}, {Smith}, {Swanson}, {Tarle}, {Wiseman}, \& {DES Collaboration}}]{Anbajaganeetal2023}
{Anbajagane}, D., {Chang}, C., {Banerjee}, A., {et~al.} 2023, \mnras, 526, 5530

\bibitem[{{Asgari} {et~al.}(2021){Asgari}, {Lin}, {Joachimi}, {Giblin}, {Heymans}, {Hildebrandt}, {Kannawadi}, {Stölzner}, {Tröster}, {van den Busch}, \& et~al.}]{Asgarietal2021}
{Asgari}, M., {Lin}, C.-A., {Joachimi}, B., {et~al.} 2021, \aap, 645, A104

\bibitem[{{Bacon} {et~al.}(2000){Bacon}, {Refregier}, \& {Ellis}}]{Baconetal2000}
{Bacon}, D.~J., {Refregier}, A.~R., \& {Ellis}, R.~S. 2000, \mnras, 318, 625

\bibitem[{{Bartelmann} \& {Schneider}(2001)}]{BartelmannSchneider2001}
{Bartelmann}, M. \& {Schneider}, P. 2001, \physrep, 340, 291

\bibitem[{{Barthelemy} {et~al.}(2020){Barthelemy}, {Codis}, {Uhlemann}, {Bernardeau}, \& {Gavazzi}}]{Barthelemyetal2020}
{Barthelemy}, A., {Codis}, S., {Uhlemann}, C., {Bernardeau}, F., \& {Gavazzi}, R. 2020, \mnras, 492, 3420

\bibitem[{{Blandford} {et~al.}(1991){Blandford}, {Saust}, {Brainerd}, \& {Villumsen}}]{Blandfordetal1991}
{Blandford}, R.~D., {Saust}, A.~B., {Brainerd}, T.~G., \& {Villumsen}, J.~V. 1991, \mnras, 251, 600

\bibitem[{{Burger} {et~al.}(2024){Burger}, {Porth}, {Heydenreich}, {Linke}, {Wielders}, {Schneider}, {Asgari}, {Castro}, {Dolag}, {Harnois-D{\'e}raps}, {Hildebrandt}, {Kuijken}, \& {Martinet}}]{Burgeretal2024}
{Burger}, P.~A., {Porth}, L., {Heydenreich}, S., {et~al.} 2024, \aap, 683, A103

\bibitem[{{Chen} \& {Szapudi}(2005)}]{ChenSzapudi2005}
{Chen}, G. \& {Szapudi}, I. 2005, \apj, 635, 743

\bibitem[{{Crittenden} {et~al.}(2002){Crittenden}, {Natarajan}, {Pen}, \& {Theuns}}]{Crittendenetal2002}
{Crittenden}, R.~G., {Natarajan}, P., {Pen}, U.-L., \& {Theuns}, T. 2002, \apj, 568, 20

\bibitem[{{Dalal} {et~al.}(2023){Dalal}, {Li}, {Nicola}, {Zuntz}, {Strauss}, {Sugiyama}, {Zhang}, {Rau}, {Mandelbaum}, {Takada}, {More}, {Miyatake}, {Kannawadi}, {Shirasaki}, {Taniguchi}, {Takahashi}, {Osato}, {Hamana}, {Oguri}, {Nishizawa}, {Malag{\'o}n}, {Sunayama}, {Alonso}, {Slosar}, {Luo}, {Armstrong}, {Bosch}, {Hsieh}, {Komiyama}, {Lupton}, {Lust}, {MacArthur}, {Miyazaki}, {Murayama}, {Nishimichi}, {Okura}, {Price}, {Tait}, {Tanaka}, \& {Wang}}]{Dalaletal2023}
{Dalal}, R., {Li}, X., {Nicola}, A., {et~al.} 2023, \prd, 108, 123519

\bibitem[{Dodelson(2017)}]{Dodelson2017}
Dodelson, S. 2017, Gravitational Lensing (Cambridge University Press)

\bibitem[{{Euclid Collaboration: Ajani} {et~al.}(2023){Euclid Collaboration: Ajani}, {Baldi}, {Barthelemy}, {et~al.}}]{ECAjanietal2023}
{Euclid Collaboration: Ajani}, V., {Baldi}, M., {Barthelemy}, A., {et~al.} 2023, \aap, 675, A120

\bibitem[{{Euclid Collaboration: Mellier} {et~al.}(2025){Euclid Collaboration: Mellier}, {Abdurro'uf}, {Acevedo~Barroso}, {et~al.}}]{ECMellieretal2025}
{Euclid Collaboration: Mellier}, Y., {Abdurro'uf}, {Acevedo~Barroso}, J., {et~al.} 2025, A\&A, 697, A1

\bibitem[{{Flaugher} {et~al.}(2015){Flaugher}, {Diehl}, {Honscheid}, {Abbott}, {Alvarez}, {Angstadt}, {Annis}, {Antonik}, {Ballester}, {Beaufore}, {Bernstein}, {Bernstein}, {Bigelow}, {Bonati}, {Boprie}, {Brooks}, {Buckley-Geer}, {Campa}, {Cardiel-Sas}, {Castander}, {Castilla}, {Cease}, {Cela-Ruiz}, {Chappa}, {Chi}, {Cooper}, {da Costa}, {Dede}, {Derylo}, {DePoy}, {de Vicente}, {Doel}, {Drlica-Wagner}, {Eiting}, {Elliott}, {Emes}, {Estrada}, {Fausti Neto}, {Finley}, {Flores}, {Frieman}, {Gerdes}, {Gladders}, {Gregory}, {Gutierrez}, {Hao}, {Holland}, {Holm}, {Huffman}, {Jackson}, {James}, {Jonas}, {Karcher}, {Karliner}, {Kent}, {Kessler}, {Kozlovsky}, {Kron}, {Kubik}, {Kuehn}, {Kuhlmann}, {Kuk}, {Lahav}, {Lathrop}, {Lee}, {Levi}, {Lewis}, {Li}, {Mandrichenko}, {Marshall}, {Martinez}, {Merritt}, {Miquel}, {Mu{\~n}oz}, {Neilsen}, {Nichol}, {Nord}, {Ogando}, {Olsen}, {Palaio}, {Patton}, {Peoples}, {Plazas}, {Rauch}, {Reil}, {Rheault}, {Roe}, {Rogers}, {Roodman}, {Sanchez}, {Scarpine}, {Schindler}, {Schmidt},
  {Schmitt}, {Schubnell}, {Schultz}, {Schurter}, {Scott}, {Serrano}, {Shaw}, {Smith}, {Soares-Santos}, {Stefanik}, {Stuermer}, {Suchyta}, {Sypniewski}, {Tarle}, {Thaler}, {Tighe}, {Tran}, {Tucker}, {Walker}, {Wang}, {Watson}, {Weaverdyck}, {Wester}, {Woods}, {Yanny}, \& {DES Collaboration}}]{Flaugheretal2015}
{Flaugher}, B., {Diehl}, H.~T., {Honscheid}, K., {et~al.} 2015, \aj, 150, 150

\bibitem[{{Fu} {et~al.}(2014){Fu}, {Kilbinger}, {Erben}, {Heymans}, {Hildebrandt}, {Hoekstra}, {Kitching}, {Mellier}, {Miller}, {Semboloni}, {Simon}, {Van Waerbeke}, {Coupon}, {Harnois-D{\'e}raps}, {Hudson}, {Kuijken}, {Rowe}, {Schrabback}, {Vafaei}, \& {Velander}}]{Fuetal2014}
{Fu}, L., {Kilbinger}, M., {Erben}, T., {et~al.} 2014, \mnras, 441, 2725

\bibitem[{{Gatti} {et~al.}(2021){Gatti}, {Sheldon}, {Amon}, {Becker}, {Troxel}, {Choi}, {Doux}, {MacCrann}, {Navarro-Alsina}, {Harrison}, {Gruen}, {Bernstein}, {Jarvis}, {Secco}, {Fert{\'e}}, {Shin}, {McCullough}, {Rollins}, {Chen}, {Chang}, {Pandey}, {Tutusaus}, {Prat}, {Elvin-Poole}, {Sanchez}, {Plazas}, {Roodman}, {Zuntz}, {Abbott}, {Aguena}, {Allam}, {Annis}, {Avila}, {Bacon}, {Bertin}, {Bhargava}, {Brooks}, {Burke}, {Carnero Rosell}, {Carrasco Kind}, {Carretero}, {Castander}, {Conselice}, {Costanzi}, {Crocce}, {da Costa}, {Davis}, {De Vicente}, {Desai}, {Diehl}, {Dietrich}, {Doel}, {Drlica-Wagner}, {Eckert}, {Everett}, {Ferrero}, {Frieman}, {Garc{\'\i}a-Bellido}, {Gerdes}, {Giannantonio}, {Gruendl}, {Gschwend}, {Gutierrez}, {Hartley}, {Hinton}, {Hollowood}, {Honscheid}, {Hoyle}, {Huff}, {Huterer}, {Jain}, {James}, {Jeltema}, {Krause}, {Kron}, {Kuropatkin}, {Lima}, {Maia}, {Marshall}, {Miquel}, {Morgan}, {Myles}, {Palmese}, {Paz-Chinch{\'o}n}, {Rykoff}, {Samuroff}, {Sanchez}, {Scarpine}, {Schubnell},
  {Serrano}, {Sevilla-Noarbe}, {Smith}, {Soares-Santos}, {Suchyta}, {Swanson}, {Tarle}, {Thomas}, {To}, {Tucker}, {Varga}, {Wechsler}, {Weller}, {Wester}, \& {Wilkinson}}]{Gattietal2021}
{Gatti}, M., {Sheldon}, E., {Amon}, A., {et~al.} 2021, \mnras, 504, 4312

\bibitem[{{Gomes} {et~al.}(2025){Gomes}, {Sugiyama}, {Jain}, {Jarvis}, {Anbajagane}, {Gatti}, {Gebauer}, {Gong}, {Halder}, {Marques}, {Pandey}, {Marshall}, {Allam}, {Alves}, {Andrade-Oliveira}, {Bacon}, {Blazek}, {Bocquet}, {Brooks}, {Carnero Rosell}, {Carretero}, {da Costa}, {Doel}, {Doux}, {Everett}, {Flaugher}, {Frieman}, {Garc{\'\i}a-Bellido}, {Gaztanaga}, {Gruen}, {Gruendl}, {Gutierrez}, {Herner}, {Hinton}, {Hollowood}, {Honscheid}, {Huterer}, {James}, {Jeffrey}, {Mena-Fern{\'a}ndez}, {Miquel}, {Muir}, {Ogando}, {Pereira}, {Pieres}, {Plazas Malag{\'o}n}, {Samuroff}, {Sanchez}, {Sanchez Cid}, {Santiago}, {Sevilla-Noarbe}, {Smith}, {Suchyta}, {Swanson}, {Tarle}, {To}, {Vikram}, {Weaverdyck}, \& {Weller}}]{Gomesetal2025}
{Gomes}, R.~C.~H., {Sugiyama}, S., {Jain}, B., {et~al.} 2025, arXiv:2503.03964

\bibitem[{{G{\'o}rski} {et~al.}(2005){G{\'o}rski}, {Hivon}, {Banday}, {Wandelt}, {Hansen}, {Reinecke}, \& {Bartelmann}}]{Gorskietal2005}
{G{\'o}rski}, K.~M., {Hivon}, E., {Banday}, A.~J., {et~al.} 2005, \apj, 622, 759

\bibitem[{{Harnois-D{\'e}raps} {et~al.}(2018){Harnois-D{\'e}raps}, {Amon}, {Choi}, {Demchenko}, {Heymans}, {Kannawadi}, {Nakajima}, {Sirks}, {van Waerbeke}, {Cai}, {Giblin}, {Hildebrandt}, {Hoekstra}, {Miller}, \& {Tr{\"o}ster}}]{HarnoisDerapsetal2018}
{Harnois-D{\'e}raps}, J., {Amon}, A., {Choi}, A., {et~al.} 2018, \mnras, 481, 1337

\bibitem[{{Harnois-D{\'e}raps} {et~al.}(2024){Harnois-D{\'e}raps}, {Heydenreich}, {Giblin}, {Martinet}, {Tr{\"o}ster}, {Asgari}, {Burger}, {Castro}, {Dolag}, {Heymans}, {Hildebrandt}, {Joachimi}, \& {Wright}}]{HarnoisDerapsetal2024}
{Harnois-D{\'e}raps}, J., {Heydenreich}, S., {Giblin}, B., {et~al.} 2024, \mnras, 534, 3305

\bibitem[{Harris {et~al.}(2020)Harris, Millman, van~der Walt, Gommers, Virtanen, Cournapeau, Wieser, Taylor, Berg, Smith, Kern, Picus, Hoyer, van Kerkwijk, Brett, Haldane, del R{\'{i}}o, Wiebe, Peterson, G{\'{e}}rard-Marchant, Sheppard, Reddy, Weckesser, Abbasi, Gohlke, \& Oliphant}]{Harrisetal2020}
Harris, C.~R., Millman, K.~J., van~der Walt, S.~J., {et~al.} 2020, Nature, 585, 357

\bibitem[{{Heydenreich} {et~al.}(2022){Heydenreich}, {Br{\"u}ck}, {Burger}, {Harnois-D{\'e}raps}, {Unruh}, {Castro}, {Dolag}, \& {Martinet}}]{Heydenreichetal2022b}
{Heydenreich}, S., {Br{\"u}ck}, B., {Burger}, P., {et~al.} 2022, \aap, 667, A125

\bibitem[{{Heydenreich} {et~al.}(2023){Heydenreich}, {Linke}, {Burger}, \& {Schneider}}]{Heydenreichetal2022}
{Heydenreich}, S., {Linke}, L., {Burger}, P., \& {Schneider}, P. 2023, \aap, 672, A44

\bibitem[{{Hou} {et~al.}(2023){Hou}, {Slepian}, \& {Cahn}}]{Houetal2023}
{Hou}, J., {Slepian}, Z., \& {Cahn}, R.~N. 2023, \mnras, 522, 5701

\bibitem[{Hunter(2007)}]{Hunter2007}
Hunter, J.~D. 2007, Computing in Science \& Engineering, 9, 90

\bibitem[{{Ivezi{\'c}} {et~al.}(2019){Ivezi{\'c}}, {Kahn}, {Tyson}, {Abel}, {Acosta}, {Allsman}, {Alonso}, {AlSayyad}, {Anderson}, {Andrew}, \& et~al.}]{Ivezicetal2019}
{Ivezi{\'c}}, {\v Z}., {Kahn}, S.~M., {Tyson}, J.~A., {et~al.} 2019, \apj, 873, 111

\bibitem[{{Jain} \& {Van Waerbeke}(2000)}]{Jainetal2000}
{Jain}, B. \& {Van Waerbeke}, L. 2000, \apjl, 530, L1

\bibitem[{{Jarvis} {et~al.}(2016){Jarvis}, {Sheldon}, {Zuntz}, {Kacprzak}, {Bridle}, {Amara}, {Armstrong}, {Becker}, {Bernstein}, {Bonnett}, {Chang}, {Das}, {Dietrich}, {Drlica-Wagner}, {Eifler}, {Gangkofner}, {Gruen}, {Hirsch}, {Huff}, {Jain}, {Kent}, {Kirk}, {MacCrann}, {Melchior}, {Plazas}, {Refregier}, {Rowe}, {Rykoff}, {Samuroff}, {S{\'a}nchez}, {Suchyta}, {Troxel}, {Vikram}, {Abbott}, {Abdalla}, {Allam}, {Annis}, {Benoit-L{\'e}vy}, {Bertin}, {Brooks}, {Buckley-Geer}, {Burke}, {Capozzi}, {Carnero Rosell}, {Carrasco Kind}, {Carretero}, {Castander}, {Clampitt}, {Crocce}, {Cunha}, {D'Andrea}, {da Costa}, {DePoy}, {Desai}, {Diehl}, {Doel}, {Fausti Neto}, {Flaugher}, {Fosalba}, {Frieman}, {Gaztanaga}, {Gerdes}, {Gruendl}, {Gutierrez}, {Honscheid}, {James}, {Kuehn}, {Kuropatkin}, {Lahav}, {Li}, {Lima}, {March}, {Martini}, {Miquel}, {Mohr}, {Neilsen}, {Nord}, {Ogando}, {Reil}, {Romer}, {Roodman}, {Sako}, {Sanchez}, {Scarpine}, {Schubnell}, {Sevilla-Noarbe}, {Smith}, {Soares-Santos}, {Sobreira}, {Swanson},
  {Tarle}, {Thaler}, {Thomas}, {Walker}, \& {Wechsler}}]{Jarvisetal2016}
{Jarvis}, M., {Sheldon}, E., {Zuntz}, J., {et~al.} 2016, \mnras, 460, 2245

\bibitem[{{Kaiser}(1992)}]{Kaiser1992}
{Kaiser}, N. 1992, \apj, 388, 272

\bibitem[{{Kaiser}(1995)}]{Kaiser1995}
{Kaiser}, N. 1995, \apjl, 439, L1

\bibitem[{{Kaiser} \& {Squires}(1993)}]{KaiserSquires1993}
{Kaiser}, N. \& {Squires}, G. 1993, \apj, 404, 441

\bibitem[{{Kaiser} {et~al.}(2000){Kaiser}, {Wilson}, \& {Luppino}}]{Kaiseretal2000}
{Kaiser}, N., {Wilson}, G., \& {Luppino}, G.~A. 2000, arXiv:0003338

\bibitem[{{Kilbinger}(2015)}]{Kilbinger2015}
{Kilbinger}, M. 2015, Reports on Progress in Physics, 78, 086901

\bibitem[{{Kratochvil} {et~al.}(2012){Kratochvil}, {Lim}, {Wang}, {Haiman}, {May}, \& {Huffenberger}}]{Kratochviletal2012}
{Kratochvil}, J.~M., {Lim}, E.~A., {Wang}, S., {et~al.} 2012, \prd, 85, 103513

\bibitem[{{Li} {et~al.}(2023){Li}, {Zhang}, {Sugiyama}, {Dalal}, {Terasawa}, {Rau}, {Mandelbaum}, {Takada}, {More}, {Strauss}, {Miyatake}, {Shirasaki}, {Hamana}, {Oguri}, {Luo}, {Nishizawa}, {Takahashi}, {Nicola}, {Osato}, {Kannawadi}, {Sunayama}, {Armstrong}, {Bosch}, {Komiyama}, {Lupton}, {Lust}, {MacArthur}, {Miyazaki}, {Murayama}, {Nishimichi}, {Okura}, {Price}, {Tait}, {Tanaka}, \& {Wang}}]{Lietal2023}
{Li}, X., {Zhang}, T., {Sugiyama}, S., {et~al.} 2023, \prd, 108, 123518

\bibitem[{{Mandelbaum}(2018)}]{Mandelbaum2018}
{Mandelbaum}, R. 2018, \araa, 56, 393

\bibitem[{{Myles} {et~al.}(2021){Myles}, {Alarcon}, {Amon}, {S{\'a}nchez}, {Everett}, {DeRose}, {McCullough}, {Gruen}, {Bernstein}, {Troxel}, {Dodelson}, {Campos}, {MacCrann}, {Yin}, {Raveri}, {Amara}, {Becker}, {Choi}, {Cordero}, {Eckert}, {Gatti}, {Giannini}, {Gschwend}, {Gruendl}, {Harrison}, {Hartley}, {Huff}, {Kuropatkin}, {Lin}, {Masters}, {Miquel}, {Prat}, {Roodman}, {Rykoff}, {Sevilla-Noarbe}, {Sheldon}, {Wechsler}, {Yanny}, {Abbott}, {Aguena}, {Allam}, {Annis}, {Bacon}, {Bertin}, {Bhargava}, {Bridle}, {Brooks}, {Burke}, {Carnero Rosell}, {Carrasco Kind}, {Carretero}, {Castander}, {Conselice}, {Costanzi}, {Crocce}, {da Costa}, {Pereira}, {Desai}, {Diehl}, {Eifler}, {Elvin-Poole}, {Evrard}, {Ferrero}, {Fert{\'e}}, {Flaugher}, {Fosalba}, {Frieman}, {Garc{\'\i}a-Bellido}, {Gaztanaga}, {Giannantonio}, {Hinton}, {Hollowood}, {Honscheid}, {Hoyle}, {Huterer}, {James}, {Krause}, {Kuehn}, {Lahav}, {Lima}, {Maia}, {Marshall}, {Martini}, {Melchior}, {Menanteau}, {Mohr}, {Morgan}, {Muir}, {Ogando}, {Palmese},
  {Paz-Chinch{\'o}n}, {Plazas}, {Rodriguez-Monroy}, {Samuroff}, {Sanchez}, {Scarpine}, {Secco}, {Serrano}, {Smith}, {Soares-Santos}, {Suchyta}, {Swanson}, {Tarle}, {Thomas}, {To}, {Varga}, {Weller}, \& {Wester}}]{Mylesetal2021}
{Myles}, J., {Alarcon}, A., {Amon}, A., {et~al.} 2021, \mnras, 505, 4249

\bibitem[{Pedregosa {et~al.}(2011)Pedregosa, Varoquaux, Gramfort, Michel, Thirion, Grisel, Blondel, Prettenhofer, Weiss, Dubourg, Vanderplas, Passos, Cournapeau, Brucher, Perrot, \& Duchesnay}]{Pedregosaetal2011}
Pedregosa, F., Varoquaux, G., Gramfort, A., {et~al.} 2011, Journal of Machine Learning Research, 12, 2825

\bibitem[{{Philcox}(2025)}]{Philcox2025}
{Philcox}, O. H.~E. 2025, \prd, 111, 123534

\bibitem[{{Philcox} \& {Ereza}(2025)}]{Philcoxetal2024}
{Philcox}, O.~H.~E. \& {Ereza}, J. 2025, Philosophical Transactions of the Royal Society of London Series A, 383, 20240034

\bibitem[{{Philcox} {et~al.}(2022){Philcox}, {Slepian}, {Hou}, {Warner}, {Cahn}, \& {Eisenstein}}]{Philcoxetal2022a}
{Philcox}, O. H.~E., {Slepian}, Z., {Hou}, J., {et~al.} 2022, \mnras, 509, 2457

\bibitem[{{Porth} {et~al.}(2024){Porth}, {Heydenreich}, {Burger}, {Linke}, \& {Schneider}}]{Porthetal2024}
{Porth}, L., {Heydenreich}, S., {Burger}, P., {Linke}, L., \& {Schneider}, P. 2024, \aap, 689, A227

\bibitem[{{Porth} \& {Smith}(2021)}]{Porthetal2021}
{Porth}, L. \& {Smith}, R.~E. 2021, \mnras, 508, 3474

\bibitem[{{Rousseeuw} \& {van Driessen}(1999)}]{Rousseeuw1999}
{Rousseeuw}, P.~J. \& {van Driessen}, K. 1999, Technometrics, 41, 212

\bibitem[{{Rowe}(2010)}]{Rowe2010}
{Rowe}, B. 2010, \mnras, 404, 350

\bibitem[{{Schneider}(1996)}]{Schneider1996}
{Schneider}, P. 1996, \mnras, 283, 837

\bibitem[{{Schneider} {et~al.}(2005){Schneider}, {Kilbinger}, \& {Lombardi}}]{Schneideretal2005}
{Schneider}, P., {Kilbinger}, M., \& {Lombardi}, M. 2005, \aap, 431, 9

\bibitem[{{Schneider} \& {Lombardi}(2003)}]{SchneiderLombardi2003}
{Schneider}, P. \& {Lombardi}, M. 2003, \aap, 397, 809

\bibitem[{{Schneider} {et~al.}(1998){Schneider}, {van Waerbeke}, {Jain}, \& {Kruse}}]{Schneideretal1998}
{Schneider}, P., {van Waerbeke}, L., {Jain}, B., \& {Kruse}, G. 1998, \mnras, 296, 873

\bibitem[{{Schneider} {et~al.}(2002){Schneider}, {van Waerbeke}, {Kilbinger}, \& {Mellier}}]{Schneideretal2002}
{Schneider}, P., {van Waerbeke}, L., {Kilbinger}, M., \& {Mellier}, Y. 2002, \aap, 396, 1

\bibitem[{{Secco} {et~al.}(2022{\natexlab{a}}){Secco}, {Jarvis}, {Jain}, {Chang}, {Gatti}, {Frieman}, {Adhikari}, {Alarcon}, {Amon}, {Bechtol}, {Becker}, {Bernstein}, {Blazek}, {Campos}, {Carnero Rosell}, {Carrasco Kind}, {Choi}, {Cordero}, {DeRose}, {Dodelson}, {Doux}, {Drlica-Wagner}, {Everett}, {Giannini}, {Gruen}, {Gruendl}, {Harrison}, {Hartley}, {Herner}, {Krause}, {MacCrann}, {McCullough}, {Myles}, {Navarro-Alsina}, {Prat}, {Rollins}, {Samuroff}, {S{\'a}nchez}, {Sevilla-Noarbe}, {Sheldon}, {Troxel}, {Zeurcher}, {Aguena}, {Andrade-Oliveira}, {Annis}, {Bacon}, {Bertin}, {Bocquet}, {Brooks}, {Burke}, {Carretero}, {Castander}, {Crocce}, {da Costa}, {Pereira}, {De Vicente}, {Diehl}, {Doel}, {Eckert}, {Ferrero}, {Flaugher}, {Friedel}, {Garc{\'\i}a-Bellido}, {Gutierrez}, {Hinton}, {Hollowood}, {Honscheid}, {Huterer}, {Kuehn}, {Kuropatkin}, {Maia}, {Marshall}, {Menanteau}, {Miquel}, {Mohr}, {Morgan}, {Muir}, {Paz-Chinch{\'o}n}, {Pieres}, {Plazas Malag{\'o}n}, {Rodriguez-Monroy}, {Roodman}, {Sanchez},
  {Serrano}, {Suchyta}, {Swanson}, {Tarle}, {Thomas}, {To}, {Weller}, \& {DES Collaboration}}]{Seccoetal2022}
{Secco}, L.~F., {Jarvis}, M., {Jain}, B., {et~al.} 2022{\natexlab{a}}, \prd, 105, 103537

\bibitem[{{Secco} {et~al.}(2022{\natexlab{b}}){Secco}, {Samuroff}, {Krause}, {Jain}, {Blazek}, {Raveri}, {Campos}, {Amon}, {Chen}, {Doux}, {Choi}, {Gruen}, {Bernstein}, {Chang}, {DeRose}, {Myles}, {Fert{\'e}}, {Lemos}, {Huterer}, {Prat}, {Troxel}, {MacCrann}, {Liddle}, {Kacprzak}, {Fang}, {S{\'a}nchez}, {Pandey}, {Dodelson}, {Chintalapati}, {Hoffmann}, {Alarcon}, {Alves}, {Andrade-Oliveira}, {Baxter}, {Bechtol}, {Becker}, {Brandao-Souza}, {Camacho}, {Carnero Rosell}, {Carrasco Kind}, {Cawthon}, {Cordero}, {Crocce}, {Davis}, {Di Valentino}, {Drlica-Wagner}, {Eckert}, {Eifler}, {Elidaiana}, {Elsner}, {Elvin-Poole}, {Everett}, {Fosalba}, {Friedrich}, {Gatti}, {Giannini}, {Gruendl}, {Harrison}, {Hartley}, {Herner}, {Huang}, {Huff}, {Jarvis}, {Jeffrey}, {Kuropatkin}, {Leget}, {Muir}, {Mccullough}, {Navarro Alsina}, {Omori}, {Park}, {Porredon}, {Rollins}, {Roodman}, {Rosenfeld}, {Ross}, {Rykoff}, {Sanchez}, {Sevilla-Noarbe}, {Sheldon}, {Shin}, {Troja}, {Tutusaus}, {Varga}, {Weaverdyck}, {Wechsler}, {Yanny}, {Yin},
  {Zhang}, {Zuntz}, {Abbott}, {Aguena}, {Allam}, {Annis}, {Bacon}, {Bertin}, {Bhargava}, {Bridle}, {Brooks}, {Buckley-Geer}, {Burke}, {Carretero}, {Costanzi}, {da Costa}, {De Vicente}, {Diehl}, {Dietrich}, {Doel}, {Ferrero}, {Flaugher}, {Frieman}, {Garc{\'\i}a-Bellido}, {Gaztanaga}, {Gerdes}, {Giannantonio}, {Gschwend}, {Gutierrez}, {Hinton}, {Hollowood}, {Honscheid}, {Hoyle}, {James}, {Jeltema}, {Kuehn}, {Lahav}, {Lima}, {Lin}, {Maia}, {Marshall}, {Martini}, {Melchior}, {Menanteau}, {Miquel}, {Mohr}, {Morgan}, {Ogando}, {Palmese}, {Paz-Chinch{\'o}n}, {Petravick}, {Pieres}, {Plazas Malag{\'o}n}, {Rodriguez-Monroy}, {Romer}, {Sanchez}, {Scarpine}, {Schubnell}, {Scolnic}, {Serrano}, {Smith}, {Soares-Santos}, {Suchyta}, {Swanson}, {Tarle}, {Thomas}, {To}, \& {DES Collaboration}}]{Seccoetal2022a}
{Secco}, L.~F., {Samuroff}, S., {Krause}, E., {et~al.} 2022{\natexlab{b}}, \prd, 105, 023515

\bibitem[{{Seitz} \& {Schneider}(1997)}]{SeitzSchneider1997}
{Seitz}, C. \& {Schneider}, P. 1997, \aap, 318, 687

\bibitem[{{Silvestre-Rosello} {et~al.}(2025){Silvestre-Rosello}, {Porth}, {Linke}, {Kr"uger}, {Grandis}, {Oel}, \& {Schneider}}]{SilvestreRoselloetal24}
{Silvestre-Rosello}, E., {Porth}, L., {Linke}, L., {et~al.} 2025, submitted to \aap, and published on arXiv

\bibitem[{{Slepian} \& {Eisenstein}(2015)}]{SlepianEisenstein2016}
{Slepian}, Z. \& {Eisenstein}, D.~J. 2015, \mnras, 454, 4142

\bibitem[{{Sugiyama} {et~al.}(2024){Sugiyama}, {Gomes}, \& {Jarvis}}]{Sugiyamaetal2024}
{Sugiyama}, S., {Gomes}, R. C.~H., \& {Jarvis}, M. 2024, arXiv:2407.01798

\bibitem[{{Sunseri} {et~al.}(2023){Sunseri}, {Slepian}, {Portillo}, {Hou}, {Kahraman}, \& {Finkbeiner}}]{Sunserietal2023}
{Sunseri}, J., {Slepian}, Z., {Portillo}, S., {et~al.} 2023, RAS Techniques and Instruments, 2, 62

\bibitem[{{Takahashi} {et~al.}(2017){Takahashi}, {Hamana}, {Shirasaki}, {Namikawa}, {Nishimichi}, {Osato}, \& {Shiroyama}}]{Takahashietal2017}
{Takahashi}, R., {Hamana}, T., {Shirasaki}, M., {et~al.} 2017, \apj, 850, 24

\bibitem[{{The Dark Energy Survey Collaboration}(2005)}]{DESOverview2005}
{The Dark Energy Survey Collaboration}. 2005, arXiv:0510346

\bibitem[{{The Dark Energy Survey Collaboration}(2016)}]{DESOverview2016}
{The Dark Energy Survey Collaboration}. 2016, \mnras, 460, 1270

\bibitem[{{Thiele} {et~al.}(2023){Thiele}, {Marques}, {Liu}, \& {Shirasaki}}]{Thieleetal2023}
{Thiele}, L., {Marques}, G.~A., {Liu}, J., \& {Shirasaki}, M. 2023, \prd, 108, 123526

\bibitem[{{Van Waerbeke} {et~al.}(2000){Van Waerbeke}, {Mellier}, {Erben}, {Cuilland re}, {Bernardeau}, {Maoli}, {Bertin}, {McCracken}, {Le F{\`e}vre}, {Fort}, {Dantel-Fort}, {Jain}, \& {Schneider}}]{vanWaerbekeetal2000}
{Van Waerbeke}, L., {Mellier}, Y., {Erben}, T., {et~al.} 2000, \aap, 358, 30

\bibitem[{Virtanen {et~al.}(2020)Virtanen, Gommers, Oliphant, Haberland, Reddy, Cournapeau, Burovski, Peterson, Weckesser, Bright, {van der Walt}, Brett, Wilson, Millman, Mayorov, Nelson, Jones, Kern, Larson, Carey, Polat, Feng, Moore, {VanderPlas}, Laxalde, Perktold, Cimrman, Henriksen, Quintero, Harris, Archibald, Ribeiro, Pedregosa, {van Mulbregt}, \& {SciPy 1.0 Contributors}}]{Virtanenetal2020}
Virtanen, P., Gommers, R., Oliphant, T.~E., {et~al.} 2020, Nature Methods, 17, 261

\bibitem[{{Wittman} {et~al.}(2000){Wittman}, {Tyson}, {Kirkman}, {Dell'Antonio}, \& {Bernstein}}]{Wittmanetal2000}
{Wittman}, D.~M., {Tyson}, J.~A., {Kirkman}, D., {Dell'Antonio}, I., \& {Bernstein}, G. 2000, \nat, 405, 143

\bibitem[{{Wright} {et~al.}(2025){Wright}, {St{\"o}lzner}, {Asgari}, {Bilicki}, {Giblin}, {Heymans}, {Hildebrandt}, {Hoekstra}, {Joachimi}, {Kuijken}, {Li}, {Reischke}, {von Wietersheim-Kramsta}, {Yoon}, {Burger}, {Chisari}, {de Jong}, {Dvornik}, {Georgiou}, {Harnois-D{\'e}raps}, {Jalan}, {William}, {Joudaki}, {Lesci}, {Linke}, {Loureiro}, {Mahony}, {Maturi}, {Miller}, {Moscardini}, {Napolitano}, {Porth}, {Radovich}, {Schneider}, {Tr{\"o}ster}, {Wittje}, {Yan}, \& {Zhang}}]{Wrightetal2025}
{Wright}, A.~H., {St{\"o}lzner}, B., {Asgari}, M., {et~al.} 2025, arXiv:2503.19441

\bibitem[{Zonca {et~al.}(2019)Zonca, Singer, Lenz, Reinecke, Rosset, Hivon, \& Gorski}]{Zonca2019}
Zonca, A., Singer, L., Lenz, D., {et~al.} 2019, Journal of Open Source Software, 4, 1298

\end{thebibliography}

%

\begin{appendix}
\appendix
\section{Additional implementation specifics and scaling}\label{app:EstimatorComplexity}

\subsection{Multiple-counting correction}\label{app:MCCorr}
From \Eq{eq:Nnmultipoles} we see that $\mathcal{N}_0(\Theta_1,\cdots,\Theta_{N-1})$ can be interpreted as the total number of (weighted) multiplets within the combination of annuli $(\Theta_1, \cdots, \Theta_{N-1})$. However, when setting all the weights to unity, letting $\Theta_1=\Theta_2=\cdots =\Theta_{N-1} \equiv \Theta$ and assuming to have $N_{\rm gal}$ galaxies within $\Theta$, we see that $\mathcal{N}_0(\Theta,\cdots,\Theta) = N_{\rm gal}^{N-1}$. In contrast, when considering only the multiplets with $N$ different points, one would expect $N_{\rm gal}(N_{\rm gal}-1)\cdots(N_{\rm gal}-N+1)$ counts.

To correct this multiple-counting effect in the real space basis, we need to explicitly subtract each contribution for which at least two galaxies are equal. As an example, for the scalar 4PCF the corrected quadruplet counts, $\mathcal{N}'$, around some fixed base point at angular position, $\btheta_i$, are computed as 
\begin{align}
    \mathcal{N}'&(\Theta_1,\Theta_2,\Theta_3;\btheta_i)
    \nonumber \\ &=
    \sum_{j\neq l \neq l} w_i \, w_j \, w_k \, w_l
    \nonumber \\ &=
    \sum_{j, k, l} w_i \, w_j \, w_k \, w_l
    \ - \ \left( \sum_{j, k=l} w_i \, w_j \, w_k^2 \ \delta^\mathrm{K}_{\Theta_2,\Theta_3} \   + 2 \ \mathrm{ perm} \right)
    \nonumber \\
    &+2 \sum_{j=k=l} w_i \, w_j^3  \ \delta^\mathrm{K}_{\Theta_1,\Theta_2} \delta^\mathrm{K}_{\Theta_1,\Theta_3} \ , 
\end{align}
which in the multipole basis, and after considering all possible base points for the quadruplets, becomes 
\begin{align}
    \mathcal{N}'_{\mathbf{n}}
    (\Theta_1, \Theta_2,\Theta_3)
     &=
    \mathcal{N}_{\mathbf{n}}
    (\Theta_1, \Theta_2,\Theta_3)
    \nonumber \\
    &-\left( 
    W_{n_2}(\Theta_1) \, W_{\mathbf{n}}^{(2,3)}(\Theta_2) \ \delta^\mathrm{K}_{\Theta_2,\Theta_3}
    + 2 \ \mathrm{ perm}\right)
    \nonumber \\ 
    &+2 \, W_{\mathbf{n}}^{(1,2,3)}(\Theta_1) \ \delta^\mathrm{K}_{\Theta_1,\Theta_2} \delta^\mathrm{K}_{\Theta_1,\Theta_3} \ .
\end{align}
In the preceding equation, the $W_{\mathbf{n}}^{(i_1, \cdots, i_m)}(\btheta,\Theta)$ refer to the multiple-counting corrections:
\begin{align}\label{eq:MCCorr} 
W_{\mathbf{n}}^{(i_1, \cdots, i_m)}(\Theta) \equiv  \sum_{i=1}^{N_{\rm gal}} w_i^m \, g_{a_1 + a_2n_2 + a_3n_3}(\btheta_i,\Theta) \ ,
\end{align}
in which the $a_i$ are integer-valued. In particular, we have  $(a_1,a_2,a_3) \equiv (0,0,1)$ for $W_{\mathbf{n}}^{(2,3)}(\Theta_2)$ and $(a_1,a_2,a_3) \equiv (0,0,0)$ for  $W_{\mathbf{n}}^{(1,2,3)}$. The multiple-counting corrections for the shear correlators are structurally equivalent; one only needs to replace the $m$th power of the weight in \Eq{eq:MCCorr} with the appropriate quantities appearing in the corresponding correlator.
The explicit expressions for the multiple-counting corrections of all correlators in the shear 4PCF are listed in Table \ref{tab:4PCFMCCorr}. 

\begin{table*}[t] 
\centering
\captionsetup{}
\caption{Double- and triple-counting corrections for the multipole components of the shear 4PCF estimator. The quantities listed correspond to the form of the individual summands in \Eq{eq:MCCorr}}
\begin{tabular}{|c|c|c|c|c|}
\hline
 & $\Theta_1=\Theta_2$ & $\Theta_1=\Theta_3$ & $\Theta_2=\Theta_3$ & $\Theta_1=\Theta_2=\Theta_3$ \\
 \hline 
 $\Upsilon_{0;n_2,n_3}^\times$
 & $(w\,e_c)^2 \, g_{n_3-5}$ 
 & $(w\,e_c)^2 \, g_{n_2-6}$ 
 & $(w\,e_c)^2 \, g_{-(n_2+n_3+5)}$ & $(w\,e_c)^3 \, g_{-8}$   
 \\
 $\Upsilon_{1;n_2,n_3}^\times$ 
 & $(w\,e_c)^2 \, g_{n_3-3}$ 
 & $(w\,e_c)^2 \, g_{n_2-2}$ 
 & $(w\,e_c)^2 \, g_{-(n_2+n_3+3)}$ & $(w\,e_c)^3 \, g_{-4}$  
 \\
 $\Upsilon_{2;n_2,n_3}^\times$ 
 & $|w\,e_c|^2 \, g_{n_3-1}$ 
 & $|w\,e_c|^2 \, g_{n_2-2}$ 
 & $(w\,e_c)^2 \, g_{-(n_2+n_3+5)}$ & $(w\,e_c)^2(w\,e_c)^* \, g_{-4}$
 \\
 $\Upsilon_{3;n_2,n_3}^\times$ 
 & $|w\,e_c|^2 \, g_{n_3-1}$ 
 & $(w\,e_c)^2 \, g_{n_2-6}$ 
 & $|w\,e_c|^2 \, g_{-(n_2+n_3+1)}$ & $(w\,e_c)^2(w\,e_c)^* \, g_{-4}$
 \\
 $\Upsilon_{4;n_2,n_3}^\times$ 
 & $(w\,e_c)^2 \, g_{n_3-5}$ 
 & $|w\,e_c|^2 \, g_{n_2-2}$ 
 & $|w\,e_c|^2 \, g_{-(n_2+n_3+1)}$ & $(w\,e_c)^2(w\,e_c)^* \, g_{-4}$
 \\
 $\Upsilon_{5;n_2,n_3}^\times$ 
 & $|w\,e_c|^2 \, g_{n_3+1}$ 
 & $|w\,e_c|^2 \, g_{n_2+2}$ 
 & $(w\,e_c)^2 \, g_{-(n_2+n_3+3)}$ & $(w\,e_c)^2(w\,e_c)^* \, g_{0}$
 \\
 $\Upsilon_{6;n_2,n_3}^\times$ 
 & $|w\,e_c|^2 \, g_{n_3+1}$ 
 & $(w\,e_c)^2 \, g_{n_2-2}$ 
 & $|w\,e_c|^2 \, g_{-(n_2+n_3-1)}$ & $(w\,e_c)^2(w\,e_c)^* \, g_{0}$  \\
 $\Upsilon_{7;n_2,n_3}^\times$ 
 & $(w\,e_c)^2 \, g_{n_3-3}$ 
 & $|w\,e_c|^2 \, g_{n_2+2}$ 
 & $|w\,e_c|^2 \, g_{-(n_2+n_3-1)}$ & $(w\,e_c)^2(w\,e_c)^* \, g_{0}$  \\ 
 $\mathcal{N}_{n_2,n_3}$ 
 & $w^2 \, g_{n_3}$ 
 & $w^2 \, g_{n_2}$ 
 & $w^2 \, g_{-(n_2+n_3)}$ 
 & $w^3$ \\
 \hline
\end{tabular}
\label{tab:4PCFMCCorr}
\end{table*}

\begin{table*}[t] 
\centering
\captionsetup{}
\caption{Symmetries of the multipole components of the shear 4PCF estimator}
\begin{tabular}{|c|c|c|c|c|c|}
\hline
 $(\Theta_1,\Theta_2,\Theta_3)$
 & $(\Theta_2,\Theta_3,\Theta_1)$ 
 & $(\Theta_3,\Theta_1,\Theta_2)$
 & $(\Theta_1,\Theta_3,\Theta_2)$
 & $(\Theta_2,\Theta_1,\Theta_3)$
 & $(\Theta_3,\Theta_2,\Theta_1)$ \\
\hline
 $\Upsilon^\times_{0,(n_2,n_3)}$ 
 & $\Upsilon^\times_{0,(n_3+1,-n_2-n_3)}$ 
 & $\Upsilon^\times_{0,(-n_2-n_3+1,n_2-1)}$ 
 & $\Upsilon^\times_{0,(n_3+1,n_2-1)}$ 
 & $\Upsilon^\times_{0,(-n_2-n_3+1,n_3)}$ 
 & $\Upsilon^\times_{0,(n_2,-n_2-n_3)}$ 
 \\
 $\Upsilon^\times_{1,(n_2,n_3)}$ 
 & $\Upsilon^\times_{1,(n_3-1,-n_2-n_3)}$ 
 & $\Upsilon^\times_{1,(-n_2-n_3-1,n_2+1)}$ 
 & $\Upsilon^\times_{1,(n_3-1,n_2+1)}$ 
 & $\Upsilon^\times_{1,(-n_2-n_3-1,n_3)}$ 
 & $\Upsilon^\times_{1,(n_2,-n_2-n_3)}$  
 \\
 $\Upsilon^\times_{2,(n_2,n_3)}$ 
 & $\Upsilon^\times_{4,(n_3+1,-n_2-n_3)}$ 
 & $\Upsilon^\times_{3,(-n_2-n_3+1,n_2-1)}$ 
 & $\Upsilon^\times_{2,(n_3+1,n_2-1)}$ 
 & $\Upsilon^\times_{3,(-n_2-n_3+1,n_3)}$ 
 & $\Upsilon^\times_{4,(n_2,-n_2-n_3)}$ 
 \\
 $\Upsilon^\times_{3,(n_2,n_3)}$ 
 & $\Upsilon^\times_{2,(n_3+1,-n_2-n_3)}$ 
 & $\Upsilon^\times_{4,(-n_2-n_3+1,n_2-1)}$ 
 & $\Upsilon^\times_{4,(n_3+1,n_2-1)}$ 
 & $\Upsilon^\times_{2,(-n_2-n_3+1,n_3)}$ 
 & $\Upsilon^\times_{3,(n_2,-n_2-n_3)}$ 
 \\
 $\Upsilon^\times_{4,(n_2,n_3)}$ 
 & $\Upsilon^\times_{3,(n_3+1,-n_2-n_3)}$ 
 & $\Upsilon^\times_{2,(-n_2-n_3+1,n_2-1)}$ 
 & $\Upsilon^\times_{3,(n_3+1,n_2-1)}$ 
 & $\Upsilon^\times_{4,(-n_2-n_3+1,n_3)}$ 
 & $\Upsilon^\times_{2,(n_2,-n_2-n_3)}$ 
 \\
 $\Upsilon^\times_{5,(n_2,n_3)}$ 
 & $\Upsilon^\times_{7,(n_3-1,-n_2-n_3)}$ 
 & $\Upsilon^\times_{6,(-n_2-n_3-1,n_2+1)}$ 
 & $\Upsilon^\times_{5,(n_3-1,n_2+1)}$ 
 & $\Upsilon^\times_{6,(-n_2-n_3-1,n_3)}$ 
 & $\Upsilon^\times_{7,(n_2,-n_2-n_3)}$ 
 \\
 $\Upsilon^\times_{6,(n_2,n_3)}$ 
 & $\Upsilon^\times_{5,(n_3-1,-n_2-n_3)}$ 
 & $\Upsilon^\times_{7,(-n_2-n_3-1,n_2+1)}$ 
 & $\Upsilon^\times_{7,(n_3-1,n_2+1)}$ 
 & $\Upsilon^\times_{5,(-n_2-n_3-1,n_3)}$ 
 & $\Upsilon^\times_{6,(n_2,-n_2-n_3)}$ 
 \\
$\Upsilon^\times_{7,(n_2,n_3)}$ 
 & $\Upsilon^\times_{6,(n_3-1,-n_2-n_3)}$ 
 & $\Upsilon^\times_{5,(-n_2-n_3-1,n_2+1)}$ 
 & $\Upsilon^\times_{6,(n_3-1,n_2+1)}$ 
 & $\Upsilon^\times_{7,(-n_2-n_3-1,n_3)}$ 
 & $\Upsilon^\times_{5,(n_2,-n_2-n_3)}$ 
 \\
 $\mathcal{N}_{(n_2,n_3)}$ 
 & $\mathcal{N}_{(n_3,-n_2-n_3)}$ 
 & $\mathcal{N}_{(-n_2-n_3,n_2)}$ 
 & $\mathcal{N}_{(n_3,n_2)}$ 
 & $\mathcal{N}_{(-n_2-n_3,n_3)}$ 
 & $\mathcal{N}_{(n_2,-n_2-n_3)}$  
 \\ \hline
\end{tabular}
\label{tab:4PCFSymm}
\end{table*}

\begin{figure}
    \centering
    \includegraphics[width=\linewidth]{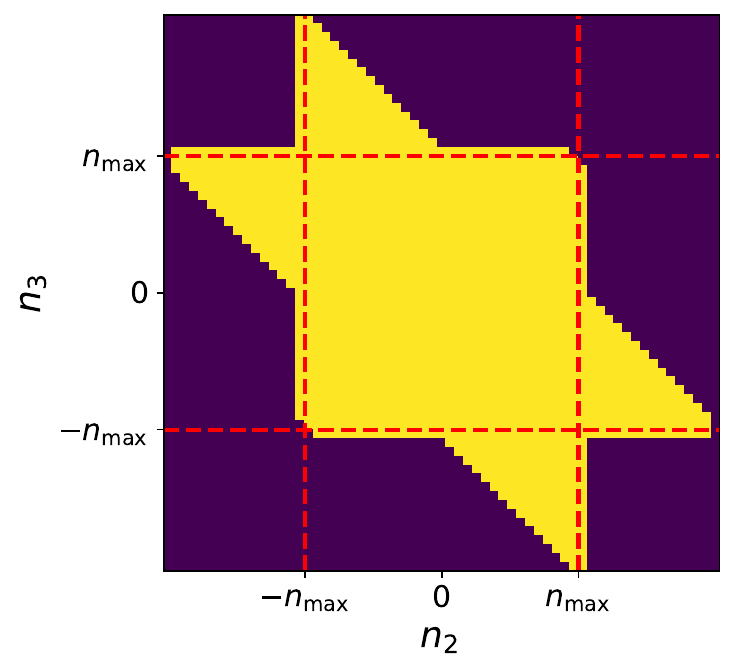}
    \captionsetup{width=\linewidth}
    \caption{Configurations of multipoles that are required to be allocated when making use of the symmetry properties in Table \ref{tab:4PCFSymm}. The inner region bounded by the red dashed lines corresponds to the configurations of multipoles that are required when one does not invoke any symmetry properties.}
    \label{fig:multipole_symmetries}
\end{figure}

\subsection{Low-memory implementation}\label{app:Est_LowMemImplementation}
Within a two-dimensional random field exhibiting statistical homogeneity and isotropy, the shape of an $N$-tuplet of points is characterised by $2N-3$ parameters such that the number of different $N$-tuplets for a fixed binning scheme with a comparable number of bins in each dimension increases quadratically with $N$. Due to this scaling, the memory footprint of a finely binned 4PCF, as it is required for the computation of the fourth-order aperture statistics, can exceed the available memory on state-of-the art compute nodes. To minimise the memory requirement at runtime while maintaining strong scaling throughout the computation, we devise the following implementation.  

Given the choice of $(N_{\Theta}, 2n_{\mathrm{max}}+1, N_{\Phi})$ bins for the radial components, the number of considered multipoles, and the angular components, as well as the number of available threads, $n_{\mathrm{threads}}$, we proceed as follows. First, we distribute the elements of the set 
$\{(\vartheta_i, \vartheta_j, \vartheta_k) \mid 1 \leq i \leq j \leq k \leq N_{\Theta}\}$ into $n_{\mathscr{s}} \geq n_{\mathrm{threads}}$ approximately equal-sized subsets, $\mathscr{s}_{\Theta,t}$. Each thread, $t$, then independently loops over the entire catalog. For each galaxy, in a first step, \circled{A}, it allocates all necessary elements of $G_n$, $W_n$, and the various multiple-counting corrections. Then, in a second step, \circled{B}, for each element in $\mathscr{s}_{\Theta,t}$, it updates $\Upsilon_\mathbf{n}$ and $\mathcal{N}_\mathbf{n}$ for the relevant $(n_2, n_3)$ pairs, see \figref{fig:multipole_symmetries} for a visualization. Once the full catalogue has been processed, each thread iterates over \(\mathscr{s}_{\Theta,t}\) again. For each element, it first applies the symmetry relations from Table \ref{tab:4PCFSymm} to reconstruct all possible permutations of radial bins. Subsequently, it computes the contribution of each permutation to the fourth-order aperture mass correlators and stores them in a shared array.

The potentially dominant contributions to the memory footprint per thread in this implementation scale as $\mathcal{O}\left(|\mathscr{s}_{\Theta,t}| \, n_{\mathrm{max}}^2 + N_{\Phi}^2\right)$, where the first term can be adjusted based on hardware constraints. However, progressively reducing the space complexity of the method may compromise strong scaling. To determine a suitable value for $n_\mathscr{s}$ that balances time and space complexity, we assume a constant tracer density, $\overline{n}$, across the survey and a maximum radial bin-edge, $\vartheta_{\rm{max}}$, for the 4PCF. Since, for realistic cosmic shear surveys, the time complexity is primarily dominated by steps \circled{A} and \circled{B} described above, we focus on analyzing these steps in more detail. The operations in \circled{A} scale as $\mathcal{C}_{\rm{A}} = \overline{n} \vartheta_{\rm{max}}^2  (c_1+c_2 \, n_{\rm{max}}) \approx \overline{n} \vartheta_{\rm{max}}^2 c_2 \, n_{\rm{max}}$, where $c_1$ represents the average complexity of assigning each tracer within the relevant range to its corresponding bin, and $c_2$ denotes the complexity of allocating one component to the relevant arrays in \circled{A}. By using tree-based methods, this complexity is further reduced, which we account for by introducing an effective tracer density, $\overline{n}_t$. In contrast, the steps in \circled{B} scale as $\mathcal{C}_{\rm{B}} = c_3 \, n_{\rm{max}}^2 \, |\mathscr{s}_{\Theta,t}|$. Since step \circled{A} is repeated for each thread, it introduces redundancy. Thus, we estimate the runtime of the low-memory implementation, $T_{\rm{lm}}$, as $T_{\rm{lm}} \approx\left(1+\frac{\mathcal{C}_{\rm{A}}}{\mathcal{C}_{\rm{B}}}\right)T_{\rm{rt}} \approx \left(1+\frac{c_2\, \overline{n}_t \, \vartheta_{\rm{max}}^2}{c_3 \, |\mathscr{s}_{\Theta,t}| \, n_{\rm{max}}}\right)T_{\rm{rt}}$, where $T_{\rm{rt}}$ represents the runtime of a runtime-optimised implementation that processes parts of the footprint in each thread and shares the full 4PCF across threads. For the analyses presented in this work, we selected values such that the memory footprint of the 4PCF-related computations per thread was limited to 2\,GB for which we found the runtime increase to be approximately 25\%.

\section{Dominant multipole components of the disconnected 4PCF}\label{app:DisconnectedMultipoles}
In this appendix we show that the multipole structure visible in the top right panel of \figref{fig:Gaussfield_4PCFcompare} follows from the assumption of a Gaussian field. We only reproduce the derivation of the sixth natural component, but note that the derivations for all the other $\Upsilon^\times_\mu$ go along the same lines.

Adapting the definition \Eq{eq:4PCFUpsilon0_Derivation} to the sixth multipole component, restricting ourselves to infinitely thin bins, a configuration with $\theta_1=\theta_2=\theta_3\equiv\theta$, and setting $\varphi_1 \equiv 0$ we have 
{\allowdisplaybreaks
\begin{align}
    \Upsilon^\times_{6,\mathbf{n}}(&\theta,\theta,\theta)  
    \nonumber \\ &\hspace{-0.6cm} =
    \int_{0}^{2\pi} \dd \phi_{12} \int_0^{2\pi} \dd \phi_{13}  \ \ee^{-\ii n_2 \phi_{12}} \ \ee^{-\ii n_3 \phi_{13}} 
    \nonumber \\ &\hspace{0.2cm}
    \times \ee^{-2\ii \left( -\zeta^\times_0+\zeta^\times_1-\zeta^\times_2+\zeta^\times_3 \right)} \ \left\langle \gammac^*(\mathbf{X}_0) \, \gammac(\mathbf{X}_1) \, \gammac^*(\mathbf{X}_2) \, \gammac(\mathbf{X}_3)\right\rangle 
    \nonumber \\ &\hspace{-0.6cm} =
    \xi_+(\theta) \int_{0}^{2\pi} \dd \phi_{12} \, \ee^{-\ii (n_2-2) \phi_{12}} \int_0^{2\pi} \dd \phi_{13} \, \ee^{-\ii (n_3+1) \phi_{13}} \, \xi_+(\theta_{23})
    \nonumber \\ &\hspace{-0.6cm} +
    \xi_-(\theta) \int_{0}^{2\pi} \dd \phi_{13} \, \ee^{-\ii (n_3-1) \phi_{13}} \, \xi_-(\theta_{13}) \int_0^{2\pi} \dd \phi_{12} \, \ee^{-\ii (n_2+2) \phi_{12}} 
    \nonumber \\ &\hspace{-0.6cm} +
    \xi_+(\theta) \int_{0}^{2\pi} \dd \phi_{12} \, \ee^{-\ii (n_2-2) \phi_{12}} \, \xi_+(\theta_{12}) \int_0^{2\pi} \dd \phi_{13} \, \ee^{-\ii (n_3+1) \phi_{13}} \ ,
\end{align}
}
where in the second step we expanded the correlator using Wick's theorem, rewrote all the the angles appearing in the different summands in terms of $\phi_{12}$ and $\phi_{13}$ and defined $\theta_{ij} \equiv |\boldsymbol{\theta}_j-\boldsymbol{\theta}_i|$.

Looking at the second term, we note that the integral vanishes unless $n_2 \equiv -2$, and when additionally choosing $n_3 \equiv 1$ the term obtains its maximum absolute value. The first condition describes the pronounced vertical line in \figref{fig:Gaussfield_4PCFcompare} while the joint condition is clearly visible as a peak on this line. A similar argument applies to the the other two terms giving the additional pronounced lines. In the case of the sixth natural component, both of the remaining terms have their maximum value located at $(n_2,n_3)=(2,-1)$ and this feature can be seen as the largest peak in  \figref{fig:Gaussfield_4PCFcompare}.

\section{The 4PCF in the SLICS ensemble}\label{app:SLICS_4PCF}

\begin{figure*}
  \centering
  \includegraphics[width=.999\textwidth,valign=t]{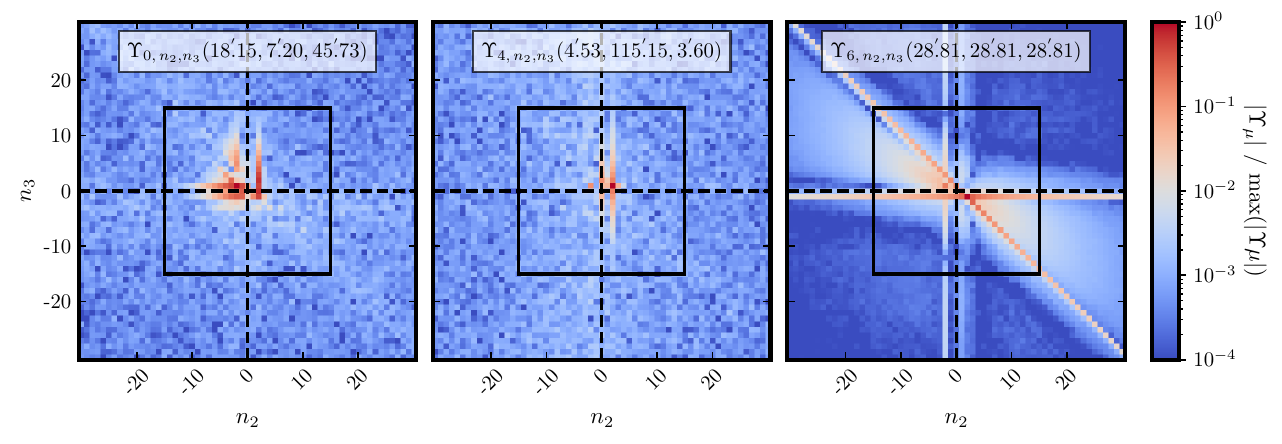}\\
  \includegraphics[width=.999\textwidth,valign=t]{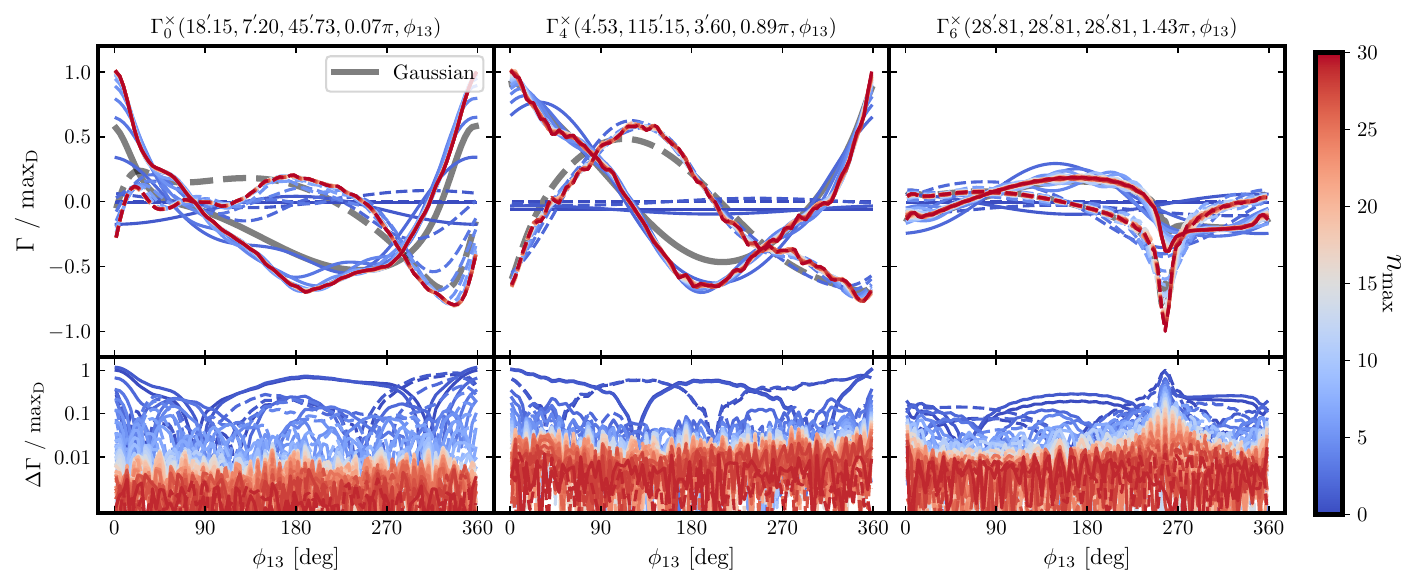}\\
  \includegraphics[width=.999\textwidth,valign=t]{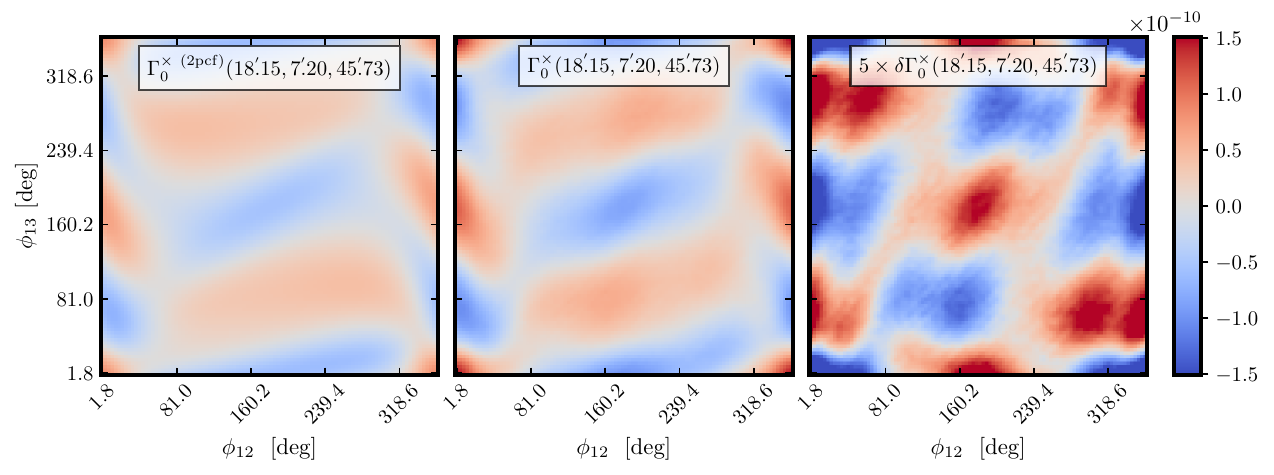}
\captionsetup{width=\linewidth}
\caption{Estimator validation using the shear 4PCF on the SLICS ensemble. The panels mirror the setup described in \figref{fig:Gaussfield_4PCFcompare}. We note that the discrepancies in the middle row and in the lower right panel are expected due to the non-vanishing connected 4PCF contribution present in the SLICS ensemble.}
\label{fig:slicsensemble_4PCFcompare}
\end{figure*}

\begin{figure*}
  \centering
  \includegraphics[width=.999\textwidth,valign=t]{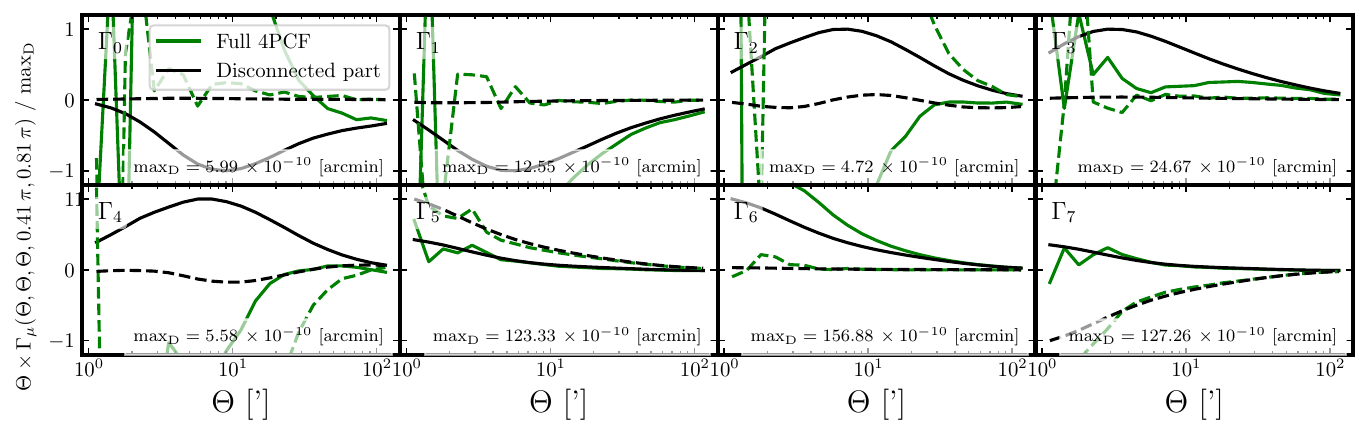}\\
  \includegraphics[width=.999\textwidth,valign=t]{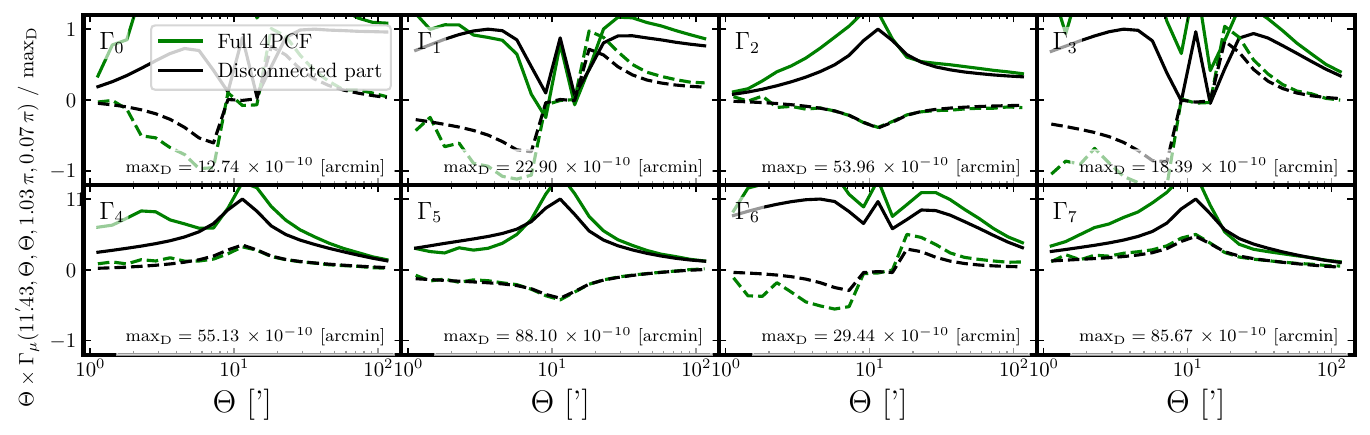}\\
  \includegraphics[width=.999\textwidth,valign=t]{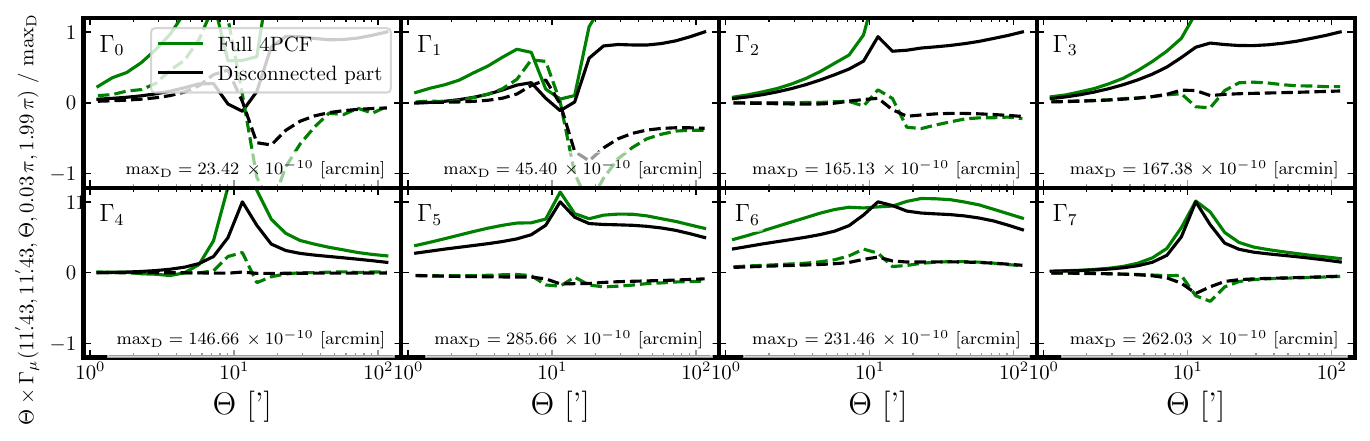} \\
  \includegraphics[width=.999\textwidth,valign=t]{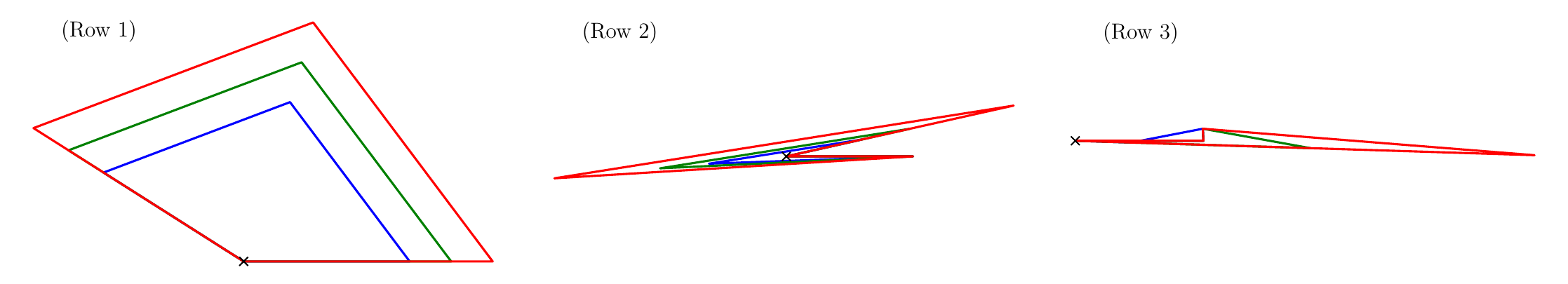}
\captionsetup{width=\linewidth}
\caption{Scaled configurations of the shear 4PCF in the SLICS ensemble. Solid lines indicate the real part while dashed lines display the imaginary part. In each panel the lines are normalised by the largest absolute value, $\rm{max}_D$,  the disconnected 4PCF takes one for each configuration. In the bottom panel we show three different scaled quadrilaterals for each of the above configurations with the postion of $\mathbf{X_0}$ being indicated by the black cross.}
\label{fig:slicsensemble_4PCFscaledconfigurations}
\end{figure*}
To assess the shape dependence of the connected 4PCF we repeated the analysis described in \secref{ssec:Gaussfield_4PCFcompare} on the SLICS ensemble, for which the data was reduced as described in \secref{ssec:EstimatorCompare} and for which we set the shape noise to zero. In \figref{fig:slicsensemble_4PCFcompare} we show the
measurements of the same components of the 4PCF  as in \figref{fig:Gaussfield_4PCFcompare}. In the top panel one can clearly see that additional multipole components are nonzero and in the lower panels how this affects the 4PCF in the real-space basis. While clear differences are visible we note that the overall complexity of the Gaussian and the full 4PCF appears to be similar, which motivates that the expected integration accuracy for the transformation equations to the aperture mass statistics can be estimated by only considering the disconnected 4PCF.  We further note that convergence is reached at a similar multipole order of about 15, but that the level of the remaining scatter is different from the results presented in \secref{ssec:Gaussfield_4PCFcompare}. We speculate that this is related to the lower $\overline{n}$ in the SLICS mocks, as the difference in the remaining scatter is largest for those configurations for which we expect the least number of quadruplet configurations and therefore the smallest level of angular resolution.

In \figref{fig:slicsensemble_4PCFscaledconfigurations} we show various scaled configurations of the full 4PCF in the SLICS ensemble and compare them with their connected parts. In the first row we show quadrilaterals of a kite shape and we see that when fixing the two angles to non-extreme values, as expected, the connected 4PCF asymptotes to zero for large radial separations.\footnote{We note that the components, for which visually the convergence has not been achieved, are the ones of the lowest overall amplitude as indicated by the value of $\mathrm{max}_D$.} The second row shows approximately dart-like configurations for which we observe a peak when all radial bins are equal and the configuration is closest to being of a dart shape. When the two free radial bins become large, we again observe the connected 4PCF to vanish. The third row shows complicated configurations of fairly flattened quadrilaterals with two radial bins fixed to intermediate scales. For those configurations we observe a non-vanishing connected 4PCF even when allowing the free radial bin to become very large.

\section{Sampling distribution of higher-order aperture mass moments in the T17 mocks.}
\label{app:SamplingDistribution}
\begin{figure*}
    \centering
    \includegraphics[width=.99\textwidth]{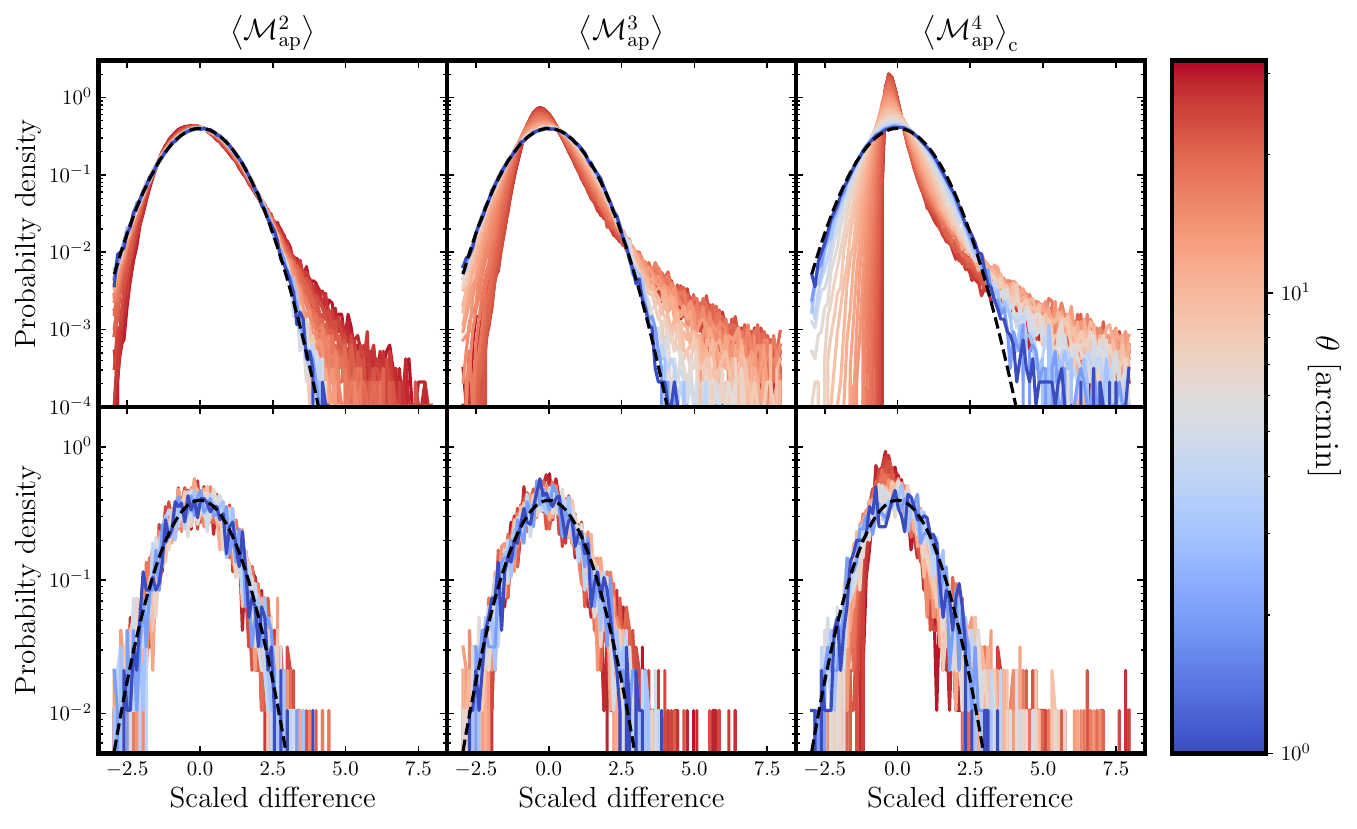}
    \captionsetup{width=\linewidth}
    \caption{Sampling distribution of the standardised higher-order aperture mass cumulants in the T17 mocks. The upper (lower) row shows the results for the individual patches (footprints). The lines are colour-coded by aperture radius. In each panel, the black dashed line displays the pdf of a standard normal distribution. For visual purposes we cut-off extreme tails outside the interval $[-3\sigma, 8\sigma]$. }
\label{fig:t17mocks_mapnsamplingdistribution_notomo}
\end{figure*}

In \figref{fig:t17mocks_mapnsamplingdistribution_notomo} we show the sampling distribution of the higher-order aperture mass cumulants in the T17 mocks for both the measurements on the individual 86\,400 individual patches, as well as for the 864 \desyt\!-like footprints. 

For both rows, we observe that while for aperture scales which are shapenoise-dominated, the sampling distribution approaches a Gaussian, this is not the case for larger aperture scales, for which the distribution becomes gradually more skewed. The level of skewness further increases for higher cumulants. These effects can be qualitatively understood when interpreting the aperture mass cumulants as the spatial average of the cumulants of the convergence field, smoothed with a compensated filter. For increasing smoothing scales, the aperture mass field becomes more susceptible to only the largest peaks in the underlying convergence field, which can be produced e.g. by high-mass haloes; taking cumulants of such fields enhances this dependence. 

On the other hand, compared to the distribution on the patches, we find the sampling distribution of the footprints to be more Gaussian. This can be motivated by the Central Limit Theorem, as taking the spatial averaging over more aperture mass cumulants is expected to reduce the level of non-Gaussianity. We note that this observation also serves as motivation to use mocks with an appropriate area when assessing the level of Gaussianity of the distribution of a summary statistic, or when performing a simulation-based inference in which the shape of the likelihood is implicitly learned. 

\section{Impact of PSF residuals}
\label{app:PSFTests}
\begin{figure*}
    \centering
    \includegraphics[width=.99\textwidth]{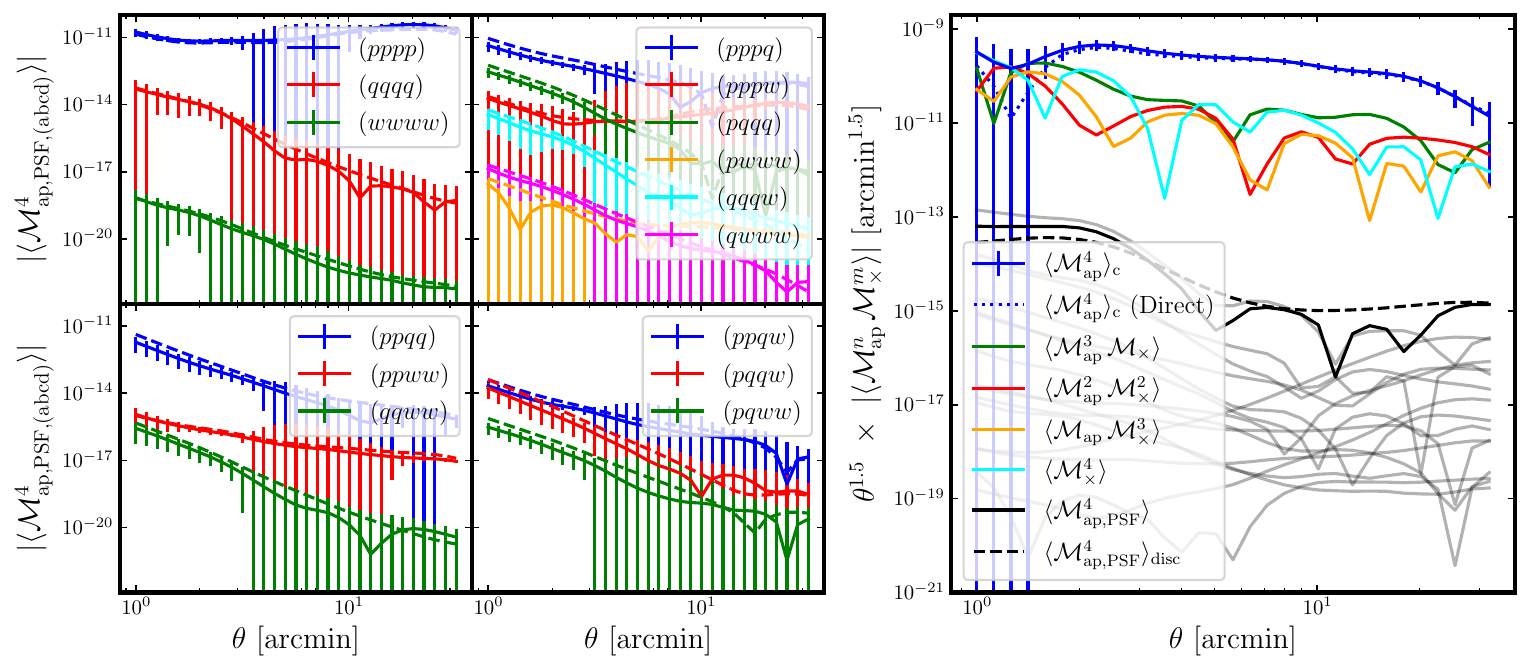}
    \captionsetup{width=\linewidth}
    \caption{Impact of additive PSF residuals on the fourth-order aperture statistics. In the panels on the left we show the aperture statistics derived from the 15  components of the fourth-order $\rho$-statistics. Solid lines denote the total statistics while dashed lines correspond to their disconnected parts. In the right panel we compare the amplitude of each contribution to the amplitude of the shear-based analysis, using our conservative choices for $\alpha, \, \beta, \, \eta$ and the prefactors of the trinomial expansion in \Eq{eq:RhoStatsMap4}. In addition to the fourth-order aperture statistics inferred from the shear 4PCF we show $\mapfourcens$ when measured using the direct estimator as the dotted blue line.}
    \label{fig:map4_psfresiduals}
\end{figure*}
To estimate the importance of an additive PSF systematics we made use of the $\rho$-statistics \citep{Rowe2010}. In this framework, one splits the measured ellipticities as 
\begin{align}
    e^{\rm{obs}} &= e + \delta e^{\rm{PSF}}
     \ , \label{eq:Rhostatsa} \\
    \delta e^{\rm{PSF}} 
    &= \alpha e_{\rm{model}} + \beta \left(e_* -  e_{\rm{model}} \right) + \eta \left(e_* \frac{T_*-T_{\rm{model}}}{T_*}\right)
    \nonumber \\
    &\equiv \alpha p + \beta q + \eta w \ , \label{eq:Rhostatsb}
\end{align}
where $T_{\rm{model}}$ and $e_{\rm{model}}$ are the modeled PSF size and PSF ellipticity while $T_*$ and $e_*$ are the same quantities directly measured from a set of reserve stars that were not included in fitting the PSF model. While a non-vanishing value for $\alpha$ could arise from an imperfect PSF deconvolution from galaxy images, the magnitude of other two parameters, $\beta, \, \eta$, relates to errors in the PSF model of sizes and ellipticities.

Instead of propagating the decomposition \Eq{eq:Rhostatsa} into the expression of the fourth-order aperture measures in terms of the 4PCF, we made use of an equivalent formulation that resembles the shape of the direct estimator \Eq{eq:MapDirectEstimator},
\begin{align}\label{eq:RhoStatsMap4}
    \left\langle\mathcal{M}^4_{\mathrm{ap,obs}}\right\rangle(\theta) 
    &= \int \dd^2 \bvartheta \, \int \dd^2 \bvartheta_1 \,  \int \dd^2 \bvartheta_2 \,  \int \dd^3 \bvartheta_3 \,  \int \dd^2 \bvartheta_4 
    \nonumber \\ &\hspace{-1cm} \times \prod_{i=1}^4 Q_\theta\left(|\bvartheta_i|\right) \ \left[e_\mathrm{t}(\bvartheta+\bvartheta_i;\varphi_i) + \delta e^{\mathrm{PSF}}_\mathrm{t}(\bvartheta+\bvartheta_i;\varphi_i) \right]
    \nonumber \\ &\equiv 
    \mapfourens(\theta) + \langle\mathcal{M}^4_{\mathrm{ap,PSF}}\rangle(\theta) + \text{cross terms} \ ,
\end{align}
and we assumed the terms mixing cosmic shear with PSF errors to vanish. Similar to the analysis of \citet{Seccoetal2022} we do not fit $\alpha$, $\beta$ and $\eta$ to fourth-order observations, but instead use conservative bounds based on the best-fit values that \citet{Gattietal2021} obtained for second-order shear statistics. In particular, we chose $\alpha=0.016$, $\beta=1.3$, $\eta=2.0$, which are each $3\sigma$ above the best-fit values of \citet{Gattietal2021}.

After having retrieved the values for $e_*$, $T_*$, $e_{\rm{model}}$ and $T_{\rm{model}}$ for the reserved stars, we used the direct estimator\footnote{We used the direct estimator due to computational constraints given that the multipole-based shear 4PCF estimator scales quartic with the number of different components, here three. Although the direct estimator is known to be biased due to a the nontrivial geometry of the \desyt footprint, we reduced this effect by only including apertures that are masked by at most 20 per cent. While we expect some bias to remain (see i.e. the blue dotted line in the right panel in \figref{fig:map4_psfresiduals}), this is at the level of ten percent and is therefore sufficient to assess the relevance of PSF effects which is based on order-of-magnitude arguments.} to compute the 15 contributions to $\langle\mathcal{M}^4_{\mathrm{ap,PSF}}\rangle$. In \figref{fig:map4_psfresiduals} we show the fourth-order aperture statistics derived from $p,q,w$. Using a patch-based covariance matrix we see that for most scales of interest the results are noise dominated. By comparing the fourth-order measures against their disconnected part we find both to be consistent with each other for most measures, implying that the connected fourth-order contribution is negligible. The strongest evidence for a connected fourth-order contribution arises for small scales and for statistics containing at least one contribution of $q$. The latter observation is somewhat expected as it implies that neither the PSF model nor the measured ellipticities around stars carry fourth-order information while this might not be the case for the PSF residuals, i.e. the true PSF. We found the full $\langle\mathcal{M}^4_{\mathrm{ap,PSF}}\rangle$ term to be multiple orders of magnitude smaller than any aperture measure derived from the shape catalog.
We have additionally verified that each of the connected parts of the 19 cross terms mixing correlations of the reserved stars with the source galaxies were either dominated by noise or had an amplitude of less than $1\%$ of the $E$-mode signal; we found the combined contribution to be noise dominated with an amplitude at the level of the cross- and mixed-mode aperture measures\footnote{The dominant contributions arise from the $(\gamma \, \gamma \, q \, q)$ and the  $(\gamma \, \gamma \, \gamma \, q)$ correlators.} that is not sufficient not explain the low amplitude of the $E$-mode signal.

\onecolumn
{\allowdisplaybreaks
\section{Multipole-based estimators for $N$th-order shear
statistics}
As we have seen in the \secref{sec:Estimator}, the construction of the estimator for fourth-order statistics is conceptually straightforward, but the expressions that need to be implemented are rather numerous and lengthy - and therefore very susceptible to hard-to-find typos. It is therefore useful to aim for a less error-prone implementation that generates all expressions on-the-fly by making use of dynamically defined methods. Here we will derive those relations.

\subsection{Notation}
We enumerate the natural components of the shear \npcf \, as follows:
\begin{enumerate}
    \item Start with $\Gamma_0 \equiv \left\langle\gammac \, \gammac \cdots \gammac \, \gammac\right\rangle$ 
    \item Enumerate through the list of $N$ entries with a single complex conjugate shear (cc) by starting from the left and moving the cc towards the right,
    $\{\Gamma_1, \cdots, \Gamma_{N}\} \equiv \{\left\langle\gammac^* \, \gammac \cdots \gammac \, \gammac\right\rangle \ \ , \ \ \left\langle\gammac \, \gammac^* \cdots \gammac \, \gammac\right\rangle \ \ , \ \ \cdots \ \ , \ \  \left\langle\gammac \, \gammac \cdots \gammac \, \gammac^*\right\rangle\}$
    \item When we have $p$ cc components, start with both of them on the left and then gradually move the most outer one to the right. Once finished, move the next cc one step to the right and the previous one right next to it. Repeat this procedure until all cc components are at the right. 
\end{enumerate}
Repeating the above algorithm until $p=\lfloor N/2 \rfloor$, $\binom{N}{p}$ components are generated for each $p$.\footnote{We note for even $N$ we only need the first half of the components for the $p=N/2$ case as the second half can be obtained by a complex conjugation of the first half.} In total, this yields a set of $2^{N-1}$ complex-valued natural components fully describing the shear \npcf. The location of the cc'd elements can equivalently be encoded in a $2^{N-1} \times N$ matrix with components $\beta^\mu_\ell \in \{-1,1\}$ where $\beta^\mu_\ell\equiv-1$ if the shear in the $\ell$th position of $\Gamma_\mu$ is cc'd, else $\beta^\mu_\ell\equiv 1$.
    
\subsection{Multipole-based estimator for the shear \npcf}\label{app:NthOrderEstimator}
In order for the multiploles to decouple we use a generalised version of the $\times$-projection in which $\zeta^\times_0 = \frac{1}{2}\left(\varphi_1+\varphi_{N-1}\right)$ and $\zeta^\times_i = \varphi_i \ (i>0)$. We further define the multipole decomposition as
\begin{align}
    \Upsilon^{\times}_{\mu}&\left(\Theta_1, \cdots, \Theta_{N-1},\phi_{1 \, 2},\cdots,\phi_{1 \, N-1}\right) 
     \equiv
     \frac{1}{(2\pi)^{N-2}} \sum_{\mathbf{n}=-\infty}^\infty \Upsilon^{\times}_{\mu,\mathbf{n}}(\Theta_1,\cdots, \Theta_{N-1}) \ \ee^{\ii n_{2}\phi_{1 \, 2}} \cdots \ee^{\ii n_{N-1}\phi_{1 \, N-1}} \ ,
\end{align}
where $\mathbf{n} \equiv \left(n_2, \cdots, n_{N-1}\right)$. Writing down the argument of the exponent in the second step in \Eq{eq:4PCFUpsilon0_Derivation}, but this time for the $\mu$th \npcf \ component, we find
\begin{align}
    \mathrm{Arg}_\mu &= 
    -\sum_{\ell=2}^{N-1} n_{\ell} \, \phi_{1\ell} \ -\beta^{\mu}_1 \left( \varphi_1 + \varphi_{N-1} \right) \ - \ 2 \sum_{\ell=1}^{N-1} \beta^{\mu}_{\ell+1} \, \varphi_\ell
    \nonumber \\
    &= 
    \varphi_1 \left( \sum_{\ell=2}^{N-1} n_{\ell} \ \ -\beta^{\mu}_1 -2 \beta^{\mu}_2 \right) + \sum_{\ell=2}^{N-2} \varphi_\ell \left( -n_{\ell} - 2 \beta^{\mu}_{\ell+1} \right) + \varphi_{N-1} \left( -n_{N-1} -\beta^{\mu}_1 -2 \beta^{\mu}_{N} \right) \ ,
    \nonumber \\
\end{align}
where the final line shows that the the $\times$-projection always decouples the multipole components -- as an example we would get for the $\mu_p$th component in which the first $2\leq p < N$ elements in the correlator are cc'd the following:
\begin{align}
\Upsilon^\times_{\mu_p,\mathbf{n}_{N-2}}\left(\Theta_1,\cdots,\Theta_{N-1}\right)  
    &= \sum_i^{\Ngal} w_i\gammaci{,i}^* \ \left(G^{\mathrm{disc}}_{-\left(\sum_{k=2}^{N-1}n_{k}+3\right)}(\btheta_i,\Theta_1) \right)^*
    \nonumber \\ &\hspace{-1cm}\times
    \left[\prod_{\ell=3}^p \left(G^{\mathrm{disc}}_{n_{\ell-1}-2} (\btheta_i,\Theta_{\ell-1})\right)^*\right] \ 
    \left[\prod_{\ell'=p+1}^{N-1} \left(G^{\mathrm{disc}}_{-\left(n_{\ell'-1}+2\right)} (\btheta_i,\Theta_{\ell'-1})\right)\right] \ 
    G^{\mathrm{disc}}_{-\left(n_{N-1}+1\right)} (\btheta_i,\Theta_{N-1})
    \ .
\end{align} 

Once a radial binning scheme and a largest multipole, $n_{\rm max}$, is chosen, all natural components of the shear \npcf \ can be dynamically allocated using the matrix $\beta$. As for the case of the lower-order \npcf s, the time complexity consists of a quadratic contribution (allocation of the $G_n$ blocks) and a linear contribution (allocation of the $\Upsilon^\times_{\mu_p,\mathbf{n}}$). While the time complexity of the quadratic contribution grows linearly with the order $N$ of the statistics, the linear one grows proportional to $n_{\rm max}^{N-2}$ for each $(\Theta_1, \cdots, \Theta_{N-1})$ element.\footnote{More explicitly, we can approximate the number of operations for the linearly scaling part for each natural component as $N_{\rm op,multipol} \sim N_{\rm gal} \ (2 n_{\rm max}+1)^{N-2} \ n_{\Theta}^{N-1}$. In contrast, using a brute-fore estimator up until some radial scale $\theta_{\rm max}$ would require $N_{\rm op,brute} \sim N_{\rm gal} \left(\overline{n} \pi \theta_{\rm max}\right)^{N-1}$ operations. Taking the ratio of the two we find that $\frac{N_{\rm op,brute}}{N_{\rm op,multipol}} \sim (2 n_{\rm max}+1) \ \left(\frac{N_{\rm{gal},\theta_{\rm max}}}{N_{\rm bins,eff}}\right)^{N-1}$, where we defined $N_{\rm{gal},\theta_{\rm max}} \equiv \overline{n} \pi \theta_{\rm max}$ and $N_{\rm bins,eff} \equiv (2 n_{\rm max}+1) n_{\Theta}$. This means that for any order $N$ the multipole-based estimator is favored over a brute-force implementation whenever the effective number of bins per order is smaller than the expected number of tracers within a circle of radius $\theta_{\rm max}$. In case one only requires a fraction, $f_{\Theta}$, of radial bin combinations, the time complexity of the multipole-based estimator reduces by a factor $f_{\Theta}^{-1}$ while the runtime of the brute-force estimator does not reduce significantly.}
}

\subsection{Algorithmic implementation}\label{app:NthOrderImplementation}
While the procedure outlined in \secref{app:NthOrderEstimator} is sufficient to estimate the $N$PCF in quadratic time complexity, it may still be unfeasible in case of a high number density of tracers or a very fine binning scheme. In this subsection we generalise the `combined estimator' formalism outlined in \citetalias{Porthetal2024} to arbitrary order and present an implementation in which the additional memory footprint introduced by using the approximation schemes is expected to be subdominant compared to the required memory for storing the $N$PCF. In the following, to homogenise the nomenclature with the \textsc{orpheus} code, we will refer to the `combined estimator' as the `DoubleTree' estimator.

We present the approximation schemes used within the \textsc{orpheus} code for a hypothetical \npcf \, correlator, $\mathscr{C}$, that, in its multipole basis, has the structure
\begin{align}
\label{eq:NPCF_baseestimator}
    \mathscr{C}_{\mathbf{n}_{N-2}}(\Theta_1, \cdots, \Theta_{N-1}) 
    &\sim
    \sum_{i=1}^{N_{\rm{disc}}} x\left(\vec{\vartheta_i}\right)  \ X_{n'_2}^{\rm{disc}} \left( \Theta_1; \vec{\vartheta_i}\right) \ \cdots \ X_{n'_{N}}^{\rm{disc}} \left( \Theta_{N-1}; \vec{\vartheta_i}\right) \\ 
    &\approx
\label{eq:NPCF_treeestimator}
    \sum_{i=1}^{N_{\rm{disc}}} x\left(\vec{\vartheta_i}\right)  \ X_{n'_2}^{\left(\left\{\Delta^{\mathrm{L}}\right\}\right)} \left( \Theta_1; \vec{\vartheta_i}\right) \ \cdots \ X_{n'_{N}}^{\left(\left\{\Delta^{\mathrm{L}}\right\}\right)} \left( \Theta_{N-1}; \vec{\vartheta_i}\right) \ ,
\end{align}
where in the first line $x$ denotes the value of the tracer in question (i.e. $w$ for number counts or $we_\mathrm{c}$ for ellipticities), the $X_{n'_k}^{\rm disc}$ are the building blocks (i.e. $W_n$ for number counts or $G_n$ for ellipticities) and the $n'_k$ are a linear combination of the multipole components $n_k, \ k\in\{2,\cdots,N-1\}$. In going to the second line we defined the 'Tree'-approximation that computes the $X_{n'_k}$ by using a hierarchy of `reduced' catalogues with resolutions $\left\{\Delta^{\mathrm{L}}\right\}$. Such an approximation significantly reduces the allocation time of the $X_n$ and therefore the quadratically scaling contribution of the estimator. We used this approximation scheme for measuring the 4PCF in the \desyt data as it can straightforwardly be adapted to the low-memory implementation of the estimator described in \secref{app:Est_LowMemImplementation}.

To reduce the linearly scaling contribution we additionally allow for a coarser sampling of the catalogue at position $\boldsymbol{X}_0$ using a (potentially different) hierarchy of reduced catalogues  with resolutions $\left\{\Delta^{\mathrm{B}}\right\}$. We coin this approximation `BaseTree' and dub its combination with the `Tree'-approximation the `DoubleTree'-approximation. Assuming that $\Theta_1 \leq \Theta_2 \leq \cdots \leq \Theta_{N-1}$, which can always be achieved by evaluating the  $\mathscr{C}_{\mathbf{n}_{N-2}}$ for a larger set of multipoles, we devise the recursion
\begin{align}
    \mathscr{C}_{\mathbf{n}_{N-2}}(\Theta_1, \cdots, \Theta_{N-1}) 
    &\approx
    \sum_{i=1}^{N_{\Delta^{\mathrm{B}}_{\Theta_1}}}
    \left( \sum_{j \in p_i^{\left(\Delta^{\mathrm{B}}_{\Theta_1}|\Delta^{\mathrm{B}}_0\right)}}x\left(\vec{\vartheta_j}\right)\right)  \ X_{n'_2}^{\left(\left\{\Delta^{\mathrm{L}}\right\}\right)} \left( \Theta_1; \vec{\vartheta}_i^{\left(\Delta^{\mathrm{B}}_{\Theta_1}\right)}\right) \ X_{n'_3}^{\left(\left\{\Delta^{\mathrm{L}}\right\}\right)} \left( \Theta_2; \vec{\vartheta}_i^{\left(\Delta^{\mathrm{B}}_{\Theta_1}\right)}\right) \ \cdots \ X_{n'_{N}}^{\left(\left\{\Delta^{\mathrm{L}}\right\}\right)} \left( \Theta_{N-1}; \vec{\vartheta}_i^{\left(\Delta^{\mathrm{B}}_{\Theta_1}\right)}\right) 
    \nonumber \\ &\equiv 
    \sum_{i=1}^{N_{\Delta^{\mathrm{B}}_{\Theta_1}}}
    \left( xX^1\right)^{\left(\left\{\Delta^{\mathrm{L}}\right\}\right)}_{n'_2}\left(\Theta_1;\vec{\vartheta}_i^{\left(\Delta^{\mathrm{B}}_{\Theta_1}\right)}\right)  \ \cdot \  X_{n'_3}^{\left(\left\{\Delta^{\mathrm{L}}\right\}\right)} \left( \Theta_2; \vec{\vartheta}_i^{\left(\Delta^{\mathrm{B}}_{\Theta_1}\right)}\right) \ \cdots \ X_{n'_{N}}^{\left(\left\{\Delta^{\mathrm{L}}\right\}\right)} \left( \Theta_{N-1}; \vec{\vartheta}_i^{\left(\Delta^{\mathrm{B}}_{\Theta_1}\right)}\right)
    \nonumber \\ &\approx 
    \sum_{i=1}^{N_{\Delta^{\mathrm{B}}_{\Theta_2}}}
    \left( \sum_{j \in p_i^{\left(\Delta^{\mathrm{B}}_{\Theta_2}|\Delta^{\mathrm{B}}_{\Theta_1}\right)}} \left( xX^1\right)^{\left(\left\{\Delta^{\mathrm{L}}\right\}\right)}_{n'_2}\left(\Theta_1;\vec{\vartheta}_j^{\left(\Delta^{\mathrm{B}}_{\Theta_1}\right)}\right)  \right)  \ X_{n'_3}^{\left(\left\{\Delta^{\mathrm{L}}\right\}\right)} \left( \Theta_2; \vec{\vartheta}_i^{\left(\Delta^{\mathrm{B}}_{\Theta_2}\right)}\right) \ \cdots \ X_{n'_{N}}^{\left(\left\{\Delta^{\mathrm{L}}\right\}\right)} \left( \Theta_{N-1}; \vec{\vartheta}_i^{\left(\Delta^{\mathrm{B}}_{\Theta_2}\right)}\right)
    \nonumber \\ &\equiv 
    \sum_{i=1}^{N_{\Delta^{\mathrm{B}}_{\Theta_2}}}
    \left( xX^2\right)^{\left(\left\{\Delta^{\mathrm{L}}\right\}\right)}_{n'_2,n'_3}\left(\Theta_1,\Theta_2;\vec{\vartheta}_i^{\left(\Delta^{\mathrm{B}}_{\Theta_2}\right)}\right)  \ \cdot \  X_{n'_4}^{\left(\left\{\Delta^{\mathrm{L}}\right\}\right)} \left( \Theta_3; \vec{\vartheta}_i^{\left(\Delta^{\mathrm{B}}_{\Theta_2}\right)}\right) \ \cdots \ X_{n'_{N}}^{\left(\left\{\Delta^{\mathrm{L}}\right\}\right)} \left( \Theta_{N-1}; \vec{\vartheta}_i^{\left(\Delta^{\mathrm{B}}_{\Theta_2}\right)}\right)
    \nonumber \\ &\approx \cdots \nonumber \\ &\approx
    \sum_{i=1}^{N_{\Delta^{\mathrm{B}}_{\Theta_{N-1}}}}
    \left( xX^{N-1}\right)^{\left(\left\{\Delta^{\mathrm{L}}\right\}\right)}_{n'_2,\cdots,n'_{N}}\left(\Theta_1,\cdots,\Theta_{N-1};\vec{\vartheta}_i^{\left(\Delta^{\mathrm{B}}_{\Theta_{N-1}}\right)}\right) \ ,
\end{align}
where $\vec{\vartheta}_i^{\left(\Delta^{\mathrm{B}}_k\right)}$ denotes the position of the $i$th tracer in the reduced catalogue of resolution $\Delta^{\mathrm{B}}_k$ and $p_i^{\left(\Delta^{\mathrm{B}}_k|\Delta^{\mathrm{B}}_{k'}\right)}$ is the set of cells of the reduced catalogue of resolution $\Delta^{\mathrm{B}}_{k'}$ contained within the $i$th cell of the reduced catalogue of resolution $\Delta^{\mathrm{B}}_k$. From the recursion one sees that for any $(\Theta_1,\cdots,\Theta_{N-1})$-tuple it is possible to evaluate the linearly scaling part at resolution $\Delta^{\mathrm{B}}_{\Theta_{N-1}}$ while maintaining some information from the higher-resolution calculations. In case of a runtime-optimised implementation one allocates the various multipole components in a $(N-1)$-fold nested loop in which at each level the $(xX^k)$ contributions can be cached for all coarser resolutions. This implies that in the innermost loop not only the effective spatial resolution is highly reduced, but also that all but one of the $N$ complex multiplications in \Eq{eq:NPCF_baseestimator} have already been cached. 

To apply the above recursion to a $N$PCF one needs to specify the basis $(xX^k)$ and the relations $n'_k\left(\mathbf{n}_{N-2}\right)$. While in principle it is possible to couple this scheme with a low-memory implementation, its effectiveness might be strongly reduced. Currently, we employ the DoubleTree-approximation within the \textsc{orpheus} code for second- and third-order correlation functions for which the memory footprint of a runtime-optimised implementation is manageable.

\twocolumn
\end{appendix}

\end{document}